\documentclass[prd,aps,a4paper,floatfix,amsmath,amssymb,twocolumn,nofootinbib]{revtex4-1}
\usepackage{amssymb,amsmath,amsthm,xcolor,graphicx,slashed}
\usepackage[margin=2cm]{geometry}
\usepackage[pdftex,breaklinks,colorlinks,
linkcolor=blue,
citecolor=teal,
anchorcolor=red,
urlcolor=cyan]{hyperref}

\begin{document}


\title{Simulations of gravitational collapse in null coordinates:
  I. Formulation and weak-field tests in generalised Bondi gauges}


\author{Carsten Gundlach}
\affiliation{Mathematical Sciences, University of Southampton,
  Southampton SO17 1BJ, United Kingdom} 
\author{David Hilditch}
\affiliation{CENTRA, Departamento de F\'isica, Instituto Superior
  T\'ecnico IST, Universidade de Lisboa UL, Avenida Rovisco Pais 1,
  1049 Lisboa, Portugal} 
\author{Thomas W. Baumgarte}
\affiliation{Department of Physics and Astronomy, Bowdoin College,
  Brunswick, ME 04011, USA} 

\date{23 April 2024}


\begin{abstract}
  We present a code for numerical simulations of the collapse of  regular initial data to a black hole in null coordinates. We  restrict to twist-free axisymmetry with scalar field matter. Our  coordinates are $(u,x,\theta,\varphi)$, where the retarded time $u$  labels outgoing null cones emerging from a regular central  worldline, the angles $(\theta,\varphi)$ label the null generators  of each null cone, and the radial coordinate $x$ labels points along  these generators. We focus on a class of generalised Bondi radial  coordinates $x$ with the twin properties that $x=0$ is the central  world line and that the numerical domain $(u\ge0$,  $0\le x\le x_\text{max})$ is a subset of the domain of dependence of  the initial data on $(u=0$, $0\le x\le x_\text{max}$). In critical  collapse, an appropriate choice of these coordinates can be made to  zoom in on the accumulation point of scale echos of the critical  solution, without the need for explicit mesh refinement. We  introduce a novel numerical scheme that in effect reduces the  angular resolution at small radius, such that the time step  $\Delta u$ for an explicit numerical scheme is limited by the radial  resolution $\Delta x$, rather than $\Delta x(\Delta\theta)^2$. We  present convergence tests in the weak-field regime, where we have  exact solutions to the linearised scalar and gravitational-wave  equations.
\end{abstract}

\maketitle
\tableofcontents


\section{Introduction}


\subsection{Motivation}


Formulations of the Einstein equations on null surfaces are attractive
in both mathematical and numerical relativity for several reasons.

As is generally done in the literature, we will here consider null
coordinates where the surfaces of constant coordinate $u$ are null
hypersurfaces, and the lines of constant $(u,\theta,\varphi)$ are
their null generators \cite{BondiBurgMetzner1962,Sachs1962}. Such
coordinates are sometimes called ``Bondi-like'', not to be confused
with Bondi coordinates, where in addition the radial coordinate $x$ is
chosen to be the area radius.

As we will review, the Einstein equations in Bondi-like coordinates
can be formulated in a way that makes them maximally constrained, in
the following sense. In affine gauge or Bondi gauge, one solves two
evolution equations, representing two polarisations of gravitational
wave (or one in twist-free axisymmetry). On each null slice of
constant $u$, the metric is completely determined by solving partial
differential equations (PDEs) with derivatives only in the slice.
Moreover these can be solved by explicit integration along the null
cone generators. In double-null coordinates there is one additional
evolution equation for the area radius $R$.

This is useful in numerical relativity in two ways. First, the
evolution cannot drift away from a consistent state on each time slice
through numerical error, in the sense that the hypersurface equations
are solved on each time slice, not just the initial one, and so only
free data are evolved. 

Secondly, the equations can easily be discretised in a way that is
compatible with causality. This makes it straightforward to evolve on
the domain of dependence of the initial data, or to impose boundary
conditions on timelike inner and outer boundaries, or to extend the
numerical domain to future null infinity.

A third reason for using Bondi-like coordinates in numerical
relativity is that any results obtained in them have immediate
geometric significance, in contrast to, say, harmonic coordinates or
``puncture'' coordinates. For all these reasons, null coordinates are
often the method of choice in spherical symmetry (see
Sec.~\ref{section:nullnumerics} for references).

Beyond spherical symmetry, there is an obvious problem with null
coordinates: null surfaces generically form caustics. However, we
expect the expansion of outgoing null cones to prevent caustics in
spacetimes that are sufficiently close to being either flat or
spherically symmetric, with the origin of the null cones near the
centre of approximate spherical symmetry. In a companion paper
\cite{paper2} (from now, Paper~II) we will consider axisymmetric
spacetimes with an additional reflection symmetry through the
equatorial plane, so that there is a preferred worldline fixed by this
symmetry.

This motivates us to investigate {\em when} null coordinates can be
used to simulate non-spherical gravitational collapse, and how best to
do this. We are not, in fact, aware of {\em any} use of null
coordinates in the numerical evolution of regular initial data to a
black hole (gravitational collapse) beyond spherical symmetry. Our
ultimate motivation for this is vacuum critical collapse, see
\cite{GundlachLRR} for a general review of critical collapse and
\cite{Vacuum_Collapse} for what we consider to be the state of
the art, at the time of writing, in vacuum critical collapse.

The present paper is concerned with general considerations, the
derivation of a number of possible radial gauges adapted to critical
collapse, spacetime diagnostics, the presentation of our numerical
methods, and weak-field convergence tests. In Paper~II we apply these
methods to axisymmetric scalar field critical collapse. In
another companion paper \cite{paper3}, we consider issues of
hyperbolicity and well-posedness.


\subsection{Previous numerical work in null coordinates}
\label{section:nullnumerics}


A spacetime coordinate $u$ is called null if the surfaces of constant
$u$ are null or, in terms of the spacetime metric, $g^{uu}=0$. We
speak of null coordinates when one of the coordinates is null, and of
double null coordinates when two of them (usually called $u$ and $v$)
are null. In our terminology, $u$ will always be an outgoing null
coordinate, also called retarded time. For general review papers on
the use of such (single) null coordinates in numerical relativity see
\cite{Winicour2005,BishopRezzolla2016}.

One natural choice of null surfaces is the set of null cones emerging
from a regular central worldline. The cones are labelled by the
retarded time $u$ and their null generators by $(u,\theta,\varphi)$.
The fourth coordinate, which we generically call $x$, then
labels points on each generator.

This formulation has been implemented in twist-free axisymmetry, in
Bondi gauge, where $x$ is the area radius $R$, compactified at future
null infinity, both in vacuum \cite{GomezPapadopoulosWinicour1994} and
with perfect fluid matter
\cite{SiebelFontMuellerPapaodopoulos2002}. Here, the null cones
emanate from a regular centre. This brings about a severe limitation
of the time step to $\Delta u\sim \Delta x\,(\Delta\theta)^2$, see
also Appendix~\ref{appendix:timestepproblem}.

Without symmetry restrictions, Bondi gauge in vacuum has been
implemented in \cite{BishopGomezLehnerMaharajWinicour1997}, and with
perfect fluid matter in \cite{BishopGomezLehnerMaharajWinicour1999},
both using stereographic coordinates on the 2-spheres of constant
$(u,x)$. The vacuum case has also been implemented using angular
coordinates on the 2-spheres in
\cite{ReisswigBishopPollney2013}. These papers are focused on
gravitational wave extraction, and so their null cones emanate from a
regular timelike world cylinder, on which boundary data must be
given. This also avoids the time step problem at the origin.

A formulation where the radial coordinate $x$ is the affine parameter
$\lambda$ along the null generators has been implemented in spherical
symmetry in a cosmological setting with fluid matter in
\cite{VanderWaltBishop2012}, and in an application to spherical scalar
field critical collapse in \cite{CrespoOliveiraWinicour2019}, and to
vacuum in spherical symmetry with initial data on two intersecting
null cones in \cite{Maedler2019}. Affine
gauge in vacuum without symmetries was formulated in
\cite{Winicour2013}, but not implemented in a code.

Affine gauge has also been used in a number of papers in the context
of asymptotically anti-de~Sitter spacetimes, on {\em ingoing} null
surfaces emanating from the timelike infinity and terminating inside a
black hole apparent horizon \cite{CheslerYaffe2014}. Mentioning only the
two applications most relevant for us, in \cite{CheslerYaffe2011} this was
done in 4+1 dimensions with two commuting translation symmetries (and
therefore mathematically similar to the twist-free axisymmetric case
in 3+1 dimensions), and in 3+1 dimensions without symmetries in
\cite{CheslerLowe2019,Chesler2022}.

As far as we know, no attempt has been made to simulate the collapse
of regular initial data to a black hole on null cones emanating from a
regular centre, except in spherical symmetry. From among the many
successful applications in spherical symmetry, we review here only the
application to the gravitational collapse of a spherically symmetric
massless scalar field. An early study of scalar field collapse in
Bondi coordinates was \cite{GoldwirthPiran1987}. Essentially the same
algorithm was used in \cite{GundlachPricePullin1994} for the study of
power-law tails and quasinormal modes in scalar field collapse.
Double-null coordinates were used for the study of spherical scalar
field critical collapse in \cite{HamadeStewart1996}, Bondi
coordinates compactified at future null infinity in
\cite{PuerrerHusaAichelburg}, and (as already mentioned) affine
coordinates in \cite{CrespoOliveiraWinicour2019}.

``Type II'' critical collapse is characterised by an arbitrarily large
range of spacetime scales, and hence typically requires adaptive mesh
refinement in numerical simulations. In fact, the pioneering paper
\cite{Choptuik1993} was made possible only by the first use of
adaptive mesh refinement in numerical relativity. However, with the
benefit of hindsight, the required mesh refinement zooms in on a
single spacetime point. Once that point has been identified, a much
simpler refinement scheme is possible, such as nested boxes in
Cartesian coordinates. A fixed grid in polar-radial coordinates
centred on this point and spaced logarithmically in radius also
provides the required spatial resolution, but at the cost of a time
step everywhere set by the smallest radial grid spacing.

A fixed grid providing the required mesh refinement in space and time
for spherically symmetric critical collapse was implemented by
Garfinkle \cite{Garfinkle1995} using double-null coordinates. The
numerical domain is $(u\ge0,x\le x_0$), with $x$ an ingoing null
coordinate. Hence the numerical domain is exactly the domain of
dependence of the initial data. Its outer boundary is the ingoing null
cone $x=x_0$, which converges to a point. In the evolution of
near-critical initial data, an appropriate choice of $x_0$ then puts
the apex of the numerical domain near the accumulation point of scale
echos in near-critical solutions. Whenever half the $x$-grid points
have fallen into the centre, the resolution is doubled by regridding,
with the time step adjusted accordingly. The numerical grid thus
``zooms in'' on the (approximate) critical solution as efficiently as
possible, without the considerable complications of standard adaptive
mesh refinement schemes.

Beyond spherical symmetry, ingoing null cones generically develop
caustics, rather than refocus on a central world line. (This is not a
concern for a sufficiently short time in a setup with initial data on
two null cones $u=0$ and $v=0$ that intersect in a spacelike
two-sphere, one often used in mathematical relavity for the study
of black-hole spacetimes, see for example
\cite{DafermosHolzegelRodnianskiTaylor2021}).

However, we can rescue the key idea of \cite{Garfinkle1995} if we
choose $x$ in such a way that $x=0$ is the regular centre while
$x=x_0$ is ingoing null (possible in spherical symmetry only) or future
spacelike (ingoing faster than light). We first implemented this in spherical
symmetry \cite{PortoGundlach2022}, and found that it provides mesh
refinement for critical collapse as efficiently as the algorithm of
\cite{Garfinkle1995}.

In hindsight this is similar to using an ingoing radial shift with
spacelike time slices to make the outer boundary ingoing null or
future spacelike. This had already been used for spherical critical
collapse on spacelike time slices in \cite{Rinne2020}.


\subsection{Plan of this paper}


A fresh approach to critical collapse using null coordinates looks
promising for the following three reasons, already mentioned
above. First, because of their geometric rigidity, the use of null
cones give us any approximate critical solution we find in coordinates
already adapted to discrete self-similarity. Second, we do not expect
problems with constraint violations. Finally, in a suitable
discretisation the outer boundary (assumed future spacelike) can be
treated exactly like the interior points.

A natural stepping stone from spherical symmetry to vacuum collapse is
non-spherical scalar field collapse, which can be examined with an
arbitrary degree of non-sphericity, whereas vacuum collapse is
necessarily very non-spherical. We shall present our results for
scalar field critical collapse in twist-free axisymmetry in Paper~II.

The structure of the present paper is as follows. In order to
highlight the general mathematical structure of the Einstein equations
in null coordinates, in Sec.~\ref{sec:nullanydim} we review them in
$n+2$ spacetime dimensions, without symmetry assumptions.

Starting from Sec.~\ref{sec:polaxi3+1}, we restrict to twist-free
axisymmetry in the usual 3+1 dimensions, and add a massless scalar
field as matter. We review standard gauge choices, and propose several
new ones for critical collapse. We also discuss diagnostic quantities
such as the Hawking mass, and how one can hope to identify apparent
and event horizons. 

Sec.~\ref{sec:numericalmethods} describes our numerical methods, and
in particular a novel method for completely overcoming the time step
problem mentioned above, such that $\Delta u\sim \Delta x$.

Sec.~\ref{sec:almostlin} describes convergence tests of the full nonlinear
equations in a small data regime where the linearisation of the
equations about Minkowski is a good approximation.

We conclude with a summary and outlook in Sec.~\ref{sec:conclusions}.
 
A number of appendixes give details of
(\ref{appendix:timestepproblem}) the time step problem, null
coordinates in (\ref{appendix:Minkowski}) Minkowski spacetime and
(\ref{appendix:sphericalsymmetry}) spherical symmetry,
(\ref{appendix:linperts}) exact solutions of the linearised equations
that we use as testbeds in the small data regime,
(\ref{appendix:residualgaugefreedom}) the residual gauge freedom in
our coordinate choice, and (\ref{appendix:regularity}) regularity
conditions for the metric at the origin and on the symmetry axis.


\section{Null coordinates on generic spacetimes of arbitrary
  dimension}
\label{sec:nullanydim}


\subsection{Metric ansatz}


Throughout this paper, $a,b,\dots$ are abstract tensor indices on the
full spacetime ($n$+2-dimensional in this section, and
2+2-dimensional in the rest of the paper), and $\nabla_a$ is the
covariant derivative with respect to the spacetime metric
$g_{ab}$. $i,j,k=1...n$ are angular coordinate indices, and
$\mu,\nu,...=1...n+2$ are spacetime coordinate indices, with $n \ge 2$ 
in this section only, and $n=2$ in the remainder of the paper. 

In $n+2$ spacetime dimensions, we define null coordinates
$(u,x,\theta^i)$, with $i=1...n$, by demanding that the hypersurfaces
of constant $u$ are null, in the sense that
$g^{ab}\nabla_au\nabla_bu=g^{uu}=0$. The vector field
\begin{equation}
\label{Udef}
U^a:=-\nabla^au
\end{equation}
is obviously null, and obeys
\begin{equation}
\label{guucalculation}
U^a\nabla_aU_b=U^a\nabla_bU_a={1\over 2}\nabla_b(U_aU^a)=0.
\end{equation}
Hence $U^a$ is the tangent vector to the affinely parameterised null
geodesics ruling the null surfaces of constant $u$.  It is easy to verify
that the most general metric obeying $g^{uu}=0$ can be written in $n+2$
form as
\begin{eqnarray}
  ds^2&=&-2G\,du\,dx-H\,du^2 \nonumber \\ &&
  +\tilde\gamma_{kl}
(d\tilde\theta^k+\hat\beta^k\,du+\tilde\beta^k\,dx)
  (d\tilde\theta^l+\hat\beta^l\,du+\tilde\beta^l\,dx). \nonumber \\
\end{eqnarray}
Note there are
\begin{equation}
  2+{n(n+1)\over 2}+2n={(n+3)(n+2)\over 2}-1
\end{equation}
metric coefficients
$(G,H,\tilde\gamma_{kl},\tilde\beta^k,\hat\beta^k)$, which is the full
number of independent metric coefficients in $n+2$ spacetime
dimensions, minus the one coordinate condition $g^{uu}=0$.  At this
point we still have two independent $n$-dimensional angular ``shift
vectors'' $\tilde\beta^k$ and $\hat\beta^k$.

We now make the gauge transformation from $(u,x,\tilde\theta^k)$ to
$(u,x,\theta^i)$ where $\theta^i(u,x,\tilde\theta^k)$ is given
implicitly by a solution of the system
\begin{equation}
\tilde\theta^k_{,x}(u,x,\theta^i)=-\tilde\beta^k[u,x,\tilde\theta^l(u,x,\theta^j)]
\end{equation}
of $n$ coupled first-order ordinary differential equations
(ODEs) in $x$ for the functions $\tilde\theta^k(u,x,\theta^i)$.
In the new coordinates the metric then takes the form
\begin{eqnarray}
  ds^2&=&-2G\,du\,dx-H\,du^2 
\nonumber \\ &&
\label{mymetric}
+\gamma_{ij}(d\theta^i+\beta^i\,du)
(d\theta^j+\beta^j\,du),
\end{eqnarray}
where
\begin{eqnarray}
\gamma_{ij}&=&\tilde\gamma_{kl}\tilde\theta^k_{,i}\tilde\theta^l_{,j},
\\
\beta^i&=&\theta^i_{,k}(\tilde\theta^k_{,u}+\hat\beta^k),
\end{eqnarray}
and $\theta^i_{,k}$ is the matrix inverse $\tilde\theta^k_{,i}$.

With the coordinates in the order $x^\mu:=(u,x,\theta^i)$, the
metric tensor can be written in matrix form as
\begin{equation}
\label{munumetric}
g_{\mu\nu}=\left(\begin{array}{ccc} 
-H+{\gamma_{ij}\beta^i\beta^j} & -G &  \gamma_{jk}\beta^k \\
-G & 0 & 0\\
\gamma_{ik}\beta^k & 0 & \gamma_{ij} \\
\end{array}\right).
\end{equation}
Note that the induced metric on the surfaces of constant $u$, given by
the bottom right sub-matrix of (\ref{munumetric}), is
degenerate with signature $0++\dots+$, as one would expect. 

The inverse metric is
\begin{equation}
\label{munuinvmetric}
g^{\mu\nu}={
\left(\begin{array}{ccc} 
0 & -{1\over G} & 0\\
-{1\over G} & {H\over G^2} & {\beta^j\over G} \\
0 & {\beta^i\over G} & \gamma^{ij} \\
\end{array}\right)},
\end{equation}
where $\gamma^{ij}$ is the the inverse of $\gamma_{ij}$. We see that
$G=0$ would be a coordinate singularity where the metric has no
inverse. Hence we assume that $G>0$. There is no such restriction on
$H$.

We see from (\ref{munuinvmetric}) that in our coordinates the
vector field $U^a$ takes the simple form
\begin{equation}
\label{Ugeneral}
U={1\over G}\partial_x,
\end{equation}
and so the outgoing null geodesics that rule each surface of constant
$u$ (its generators) are simply the lines of constant
$(u,\theta^i)$. (We have already mentioned that such coordinates are
sometimes called ``Bondi-like''). With $G>0$, $x$ is strictly
increasing with the affine parameter of the null cone generators.

\begin{figure}
\includegraphics[width=0.4\textwidth, angle=0]{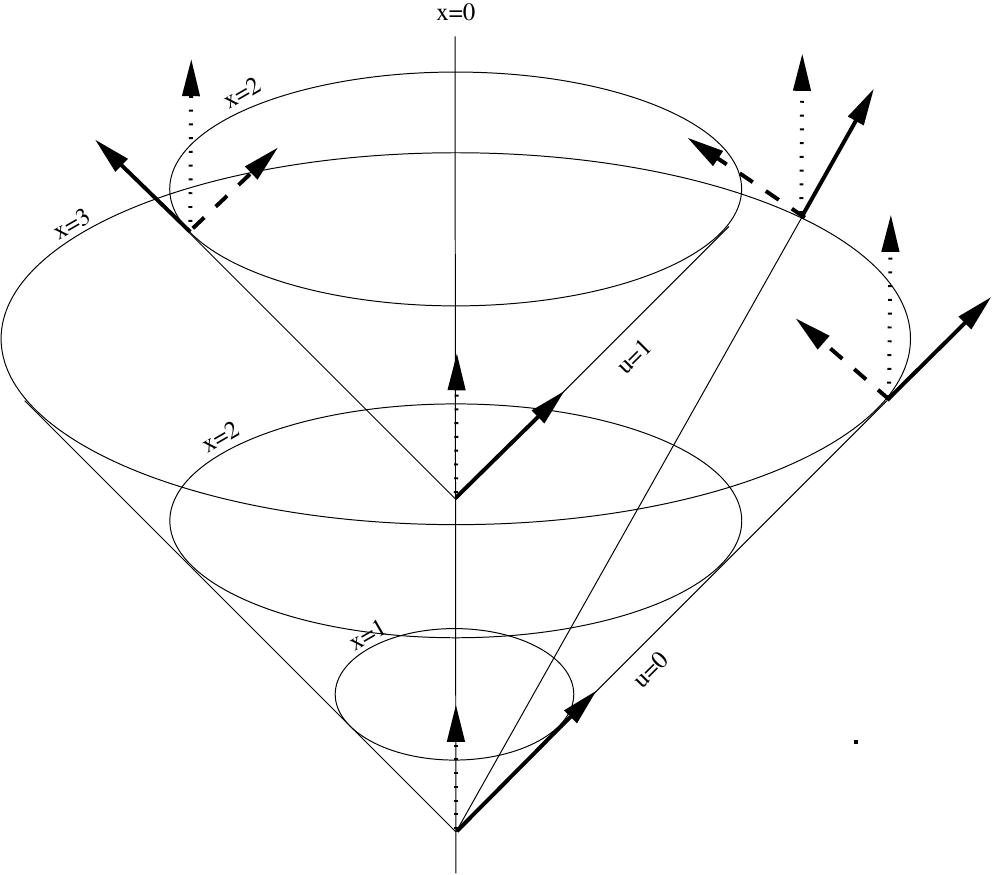} 
\caption{Schematic spacetime picture showing our coordinates and basis
  vectors. Shown are two null cones of constant $u$, four spacelike
  closed 2-surfaces of constant $(u,x)$, the central worldline $x=0$,
  and outgoing null rays of constant $(u,y,\varphi)$ (the angular
  coordinate $y=-\cos\theta$ is suppressed here). Solid arrows
  represent the outgoing null vector $U\propto \partial_x$, dashed
  arrows the ingoing null vector $\Xi$, and dotted arrows the timelike
  vector $\partial_u$ (which are timelike near the origin but may tip
  inwards to become spacelike further out).}
\label{fig:lightcones}
\end{figure}

For the following discussions, we denote by ${\cal N}^+_u$ the
outgoing $n+1$-dimensional null hypersurfaces of constant $u$, and by
${\cal L}^+_{u,\theta^i}$ the outgoing affinely parameterised null
geodesics that rule each ${\cal N}^+_u$ [lines of constant
  $(u,\theta^i)$]. We denote by ${\cal S}_{u,x}$ the $n$-dimensional
spacelike surfaces of constant $u$ and $x$, by ${\cal N}^-_{u,x}$ the
ingoing null surface that emerges from each ${\cal S}_{u,x}$, and by
${\cal L}^-_{u,x,\theta^i}$ the affinely parameterised null geodesics
that rule it. Note that on ${\cal N}^-_{u,x}$ and ${\cal
  L}^-_{u,x,\theta^i}$ none of the coordinates are constant: the
coordinate values that label them are only starting values. Our
coordinates and basis vectors are sketched in
Fig.~\ref{fig:lightcones}.

\subsection{Standard radial gauge choices}
\label{section:standardgauges}


We see from (\ref{munumetric}) that $g_{xi}=0$ for $i=1\dots n$, and
from (\ref{munuinvmetric}) that $g^{uu}=0$, both by construction. We
have used up $n+1$ of the possible $n+2$ coordinate conditions, with
one remaining to be imposed. 

From an $n+2$ perspective, this final gauge condition should not
single out any spatial coordinate $\theta^i$, and so should involve
only $G$, $H$ and $\det \gamma_{ij}$. If we think of this condition as
fixing, for example, the metric coefficient $H$, we have
\begin{equation}
{(n+3)(n+2)\over 2}-(n+2)= 1+{n(n+1)\over 2}+n
\end{equation}
independent metric coefficients: on the left-hand side the number of
algebraically independent metric coefficients $g_{\mu\nu}$ in $n+2$
spacetime dimensions without symmetry, minus $n+2$ coordinate
conditions, and on the right-hand side the number of components in
$(G,\gamma_{ij},\beta^k)$. We are aware of three such conditions in
the literature.


\subsubsection{Bondi} 


One may be able to assume that the surfaces
${\cal S}_{u,x}$ are $n$-spheres and that their volume increases
monotonically with $x$. Then Bondi coordinates are defined by the
coordinate condition $\det \gamma_{ij}=x^2\det\bar\gamma_{ij}$, where
$\bar\gamma_{ij}$ is the unit round metric on $S^n$ in the coordinates
$\theta^i$. $x$ is then called the area radius and is usually, and in
this paper, denoted by $r$.


\subsubsection{Double-null} 


Double-null coordinates are defined by
$g^{xx}=0$, which, from (\ref{munuinvmetric}), is equivalent to
$H=0$. $x$ is then a second null coordinate, and is usually, and in
this paper, denoted by $v$. The ${\cal N}^-_{u,v}$ are now surfaces of
constant $v$. The affinely parameterised ${\cal L}^-_{u,v,\theta^i}$
have the tangent vector
\begin{equation}
\label{Vdef}
V^a:=-\nabla^a v,
\end{equation}
which in coordinates takes the form
\begin{equation}
\label{Vdefbis}
V={1\over G}\left(\partial_u-\beta^i\partial_i\right).
\end{equation}
As already mentioned, we are not aware of a numerical application of
double null coordinates beyond spherical symmetry. This may be because
we expect ingoing null cones to develop caustics generically, rather
than converge to a point.


\subsubsection{Affine} 


Finally, one sees from (\ref{Ugeneral}) that $x$
is an affine parameter along the outgoing null geodesics if and only
if the coordinate condition $G_{,x}=0$ holds. $x$ is then often called
$\lambda$. $\lambda$ on each outgoing null geodesic is fixed up to an
additive constant by fixing the function $G(u,\theta^i)$. The most
common choice is $G=1$.


\subsection{The hierarchy of Einstein equations}


In order to have nontrivial spacetimes with a regular centre even in
the limit of spherical symmetry, we add a massless minimally coupled
scalar field $\psi$ that obeys the wave equation
\begin{equation}
\nabla^a\nabla_a\psi=0.
\end{equation}
The Einstein equations with scalar field matter {in any spacetime
  dimension} can be written, in trace-reversed form, as
\begin{equation}
E_{ab}:=R_{ab}-8\pi\nabla_a\psi\nabla_b\psi=0.
\end{equation}
$R_{ab}$ is the spacetime Ricci tensor, and we use gravitational units
where $c=G=1$.

We define the ingoing null vector field
\begin{equation}
\label{Xidefgen}
\Xi:=\partial_u-{H\over 2G}\partial_x-\beta^i\partial_i.
\end{equation}
Together $\Xi^a$ and $U^a$ span the normal space to each ${\cal
  S}_{u,x}$. They are normalised so that $\Xi^aU_a=-1$. $U^a$, given
in (\ref{Ugeneral}), is tangent to the affinely parameterised
generators of ${\cal N}_u^+$, while $\Xi^a$ is tangent to the
generators of ${\cal N}_{u,x}^-$ where they emerge from ${\cal
  S}_{u,x}$.

Given only the metric components $\gamma_{ij}$ on a surface of
constant $u$, and boundary values for $G$, $\beta^i$ and
${\beta^i}_{,x}$ on any past boundary $x=x_{\rm min}(u,\theta^i)$ of
that surface, we can solve the Einstein equation $E_{xx}=0$ for $G$,
${E_x}^i=0$ for $\beta^i$, and $E_{ij}=0$ for the null derivatives
$\Xi \gamma_{ij}$, in this order, by explicit integration in $x$. In
the terminology of \cite{Sachs1962} these are the ``main
equations''. We will call them the ``hierarchy equations''. They
contain $H$ only in the combination $\Xi$. Explicit expressions in
twist-free axisymmetry in 3+1 spacetime dimensions will be given in
Sec.~\ref{sec:polaxi3+1} below. Given the $\Xi \gamma_{ij}$, one gauge
condition is required to fix $H$ and so find the ``time'' derivatives
$\gamma_{ij,u}$. We can then advance $\gamma_{ij}$ in $u$ and repeat.

The remaining $n+2$ Einstein equations $E_{ux}=0$ (the ``trivial
equation'' \cite{Sachs1962}), $E_{uu}=0$ and ${E_u}^i=0$ (the
``supplementary conditions'' \cite{Sachs1962}) contain higher
$u$-derivatives than the hierarchy equations and are redundant modulo
the $n+2$ contracted Bianchi identities. The supplementary conditions
also act as constraints on the boundary data imposed at
$x=x_\text{min}$. We do not discuss them here because
$x=x_\text{min}=0$ will always be a regular central world line here
and in Paper II, so no free data can be imposed there. Finally, the trivial
equation is an algebraic consequence of imposing all other equations.

With the shorthand
\begin{equation}
B:={H\over 2G},
\end{equation}
we can write (\ref{Xidefgen}) as
\begin{equation}
\label{partialufromXi}
\partial_u=\Xi+B\partial_x+\beta^i\partial_i.
\end{equation}
This suggests that we consider $B$ and $\beta^i$ as the $x$ and
$\theta^i$ components of a ``shift vector'' representing the
difference between the coordinate time direction $\partial_u$ and the
null vector $\Xi$.  However, while the future-pointing unit vector
$n^a$ normal to a spacelike hypersurface $\Sigma$ is unique, the null
vector $\Xi^a$ depends not only on the null hypersurface ${\cal
  N}_u^+$, but also on its foliation by $n$-surfaces ${\cal S}_{u,x}$.


\section{Twist-free axisymmetry in 3+1 with a scalar field}
\label{sec:polaxi3+1}


\subsection{Metric ansatz and matter field}


We now restrict to 3+1 spacetime dimensions in spherical polar null
coordinates $(u,x,\theta,\varphi)$. We assume axisymmetry with Killing
vector $K:=\partial_\varphi$, meaning that $g_{\mu\nu,\varphi}=0$ in
those coordinates. In addition, we assume a reflection symmetry
$\varphi\to-\varphi$. This is a consistent truncation, in the sense
that if we impose $\gamma_{\theta\varphi}=0$ on the initial data, and
the boundary conditions $\beta^\varphi=\beta^\varphi_{,x}=0$ at
$x=x_\text{min}$ to start up the integration, the hierarchy equations
give us $\beta^\varphi=0$ and
$\gamma_{\theta\varphi,u}=0$. Geometrically, the twist vector
$\epsilon^{abcd}K_a\nabla_bK_c$ vanishes, hence the name twist-free
axisymmetry. Physically, this symmetry removes one of the two
gravitational wave degrees of freedom, hence the alternative name
polarized axisymmetry.

We identify a worldline on the symmetry axis world sheet, calling it
the central worldline, or centre for short. (There is a preferred
choice for this if the spacetime has a reflection symmetry $z\to-z$,
or $\theta\to\pi-\theta$, but in general the choice is arbitrary.  For
now we stay in the general case.) The null cones of constant $u$ are
assumed to have a regular vertex on the central worldline.

We parameterise the metric under these conditions as
\begin{eqnarray}
\label{metricuvtheta}
  ds^2&=&-2G\,du\,dx-H\,du^2\nonumber \\ &&
+R^2\left[e^{2F}(d\theta+\beta\,du)^2+e^{-2F}\sin^2\theta\,d\varphi^2\right],
\nonumber \\
\end{eqnarray}
where the metric coefficients $(G,H,R,F,\beta)$ depend on the
coordinates $(u,x,\theta)$ only. $R$ is the area radius, in the sense
that $\det\gamma_{ij}=R^2\sin^2\theta$, and so the area of ${\cal
  S}_{u,x}$ is $4\pi R^2$.  The central worldline is at $R=0$.


\subsection{Regularisation of the axis}


The field equations can be regularised on the symmetry axis by
reparameterising \cite{GomezPapadopoulosWinicour1994}
\begin{equation}
\beta=:\sin\theta\,b, \qquad
F=:\sin^2\theta\,f
\end{equation}
and replacing the coordinate $\theta$ by 
\begin{equation}
y:=-\cos\theta, \qquad -1\le y\le 1.
\end{equation}
Intuitively, $b$ is the $z$-component of the shift vector
$\beta\partial_\theta$, and therefore regular on the
symmetry axis. The metric becomes
\begin{eqnarray}
\label{ymetric}
  ds^2&=&-2G\,du\,dx-H\,du^2 \nonumber \\
&&+R^2\left[e^{2S f}S^{-1}(dy+S\,b\,du)^2
+e^{-2S f}S\,d\varphi^2\right], \nonumber \\ 
\end{eqnarray}
where we have defined the shorthand 
\begin{equation}
S:=1-y^2=\sin^2\theta.
\end{equation}
We also have
\begin{equation}
\label{Xidef}
\Xi=\partial_u -B\partial_x-bS \partial_y.
\end{equation}
Even though the metric now has a division by $S$, the hierarchy
equations for $(G,b,H,\Xi R,\Xi f,\Xi\psi)$ are regular on the axis in
the sense that they contain neither square roots nor divisions by $y$
or $S$, see Eqs.~(\ref{Geqn}-\ref{Spsi}) below.

We note in passing the identities
\begin{eqnarray}
\label{psithetaeqn}
\psi_{,\theta}&=&\sqrt{1-y^2}\psi_{,y}, \\
\label{psithetathetaeqn}
\psi_{,\theta\theta}&=&(1-y^2)\psi_{,yy}-y\psi_{,y}. 
\end{eqnarray}
From (\ref{psithetaeqn}) we see that the usual regularity condition
for scalars on the symmetry axis, $\psi_{,\theta}=0$ at $\theta=0$ and
$\pi$, corresponding to $y=-1$ and $y=1$, does not impose a condition
on $\psi_{,y}$ there. Rather, we see from (\ref{psithetathetaeqn})
that $\psi_{,y}=\pm \psi_{,\theta\theta}$ there, which is
unconstrained by regularity.


\subsection{The equation hierarchy}


In twist-free axisymmetry, the Einstein equations
have seven algebraically independent components $E_{\mu\nu}$. It is
convenient to define the two linear combinations
\begin{equation}
E_\pm:=e^{-2F}SE_{yy}\pm e^{2F}S^{-1}E_{\varphi\varphi}.
\end{equation}
 
Geometric free data on an outgoing null cone of constant $u$ consist of $f$ and
$\psi$ as functions of $R$ and $y$. In double-null gauge, we also need
to specify $R$ as a function of $x$ and $y$ there, considering this as
fixing a gauge freedom in the initial data.

The two Einstein equations $E_{xx}=0$ and $E_{xy}=0$ do not contain
any $u$-derivatives. They can, respectively, be written as
\begin{eqnarray}
\label{Geqn}
\left(\ln {G\over R_{,x}}\right)_{,x}
&=&S_G[R,f,\psi] \\ 
\label{dbdxeqn}
\left({R^4e^{2S f}b_{,x}\over G}\right)_{,x}&=&S_b[R,f,\psi,G].
\end{eqnarray}
The Einstein equations
$E_+=0$ and $E_-=0$ and the scalar wave equation all contain
$u$-derivatives. They can, respectively, be written as
\begin{eqnarray}
\label{Reqn}
\left(R\,\Xi R\right)_{,x}&=&S_R[R,f,\psi,G,b], \\
\label{feqn}
\left(R\,\Xi f\right)_{,x}&=&S_f[R,f,\psi,G,b]-(\Xi R)f_{,x},\\
\label{psieqn}
\left(R\,\Xi\psi\right)_{,x}&=&S_\psi[R,f,\psi,G,b]-(\Xi R)\psi_{,x},
\end{eqnarray}
where $\Xi$ is the derivative operator defined in (\ref{Xidef}), and
$H$ only appears as part of $\Xi$. The right-hand sides $S[f,\dots]$
contain the derivatives $f_{,x}$, $f_{,y}$, $f_{,xy}$ and $f_{,yy}$
(but not $f_{,xx}$), and the same derivatives of $R$, $G$, $b$ and
$\psi$, with the exceptions of $\psi_{,xy}$ and $b_{,yy}$.

The remaining algebraically independent Einstein equations are
$E_{uu}$, $E_{ux}$ and $E_{yy}$, and contain $u$-derivatives other
than those already appearing in the hierarchy equations. In
particular, $E_{uu}$ (only) contains $R_{,uu}$; $E_{uu}$ and $E_{uy}$
contain $R_{,uy}$, $f_{,uy}$ and $G_{,uy}$; $E_{uu}$ and $E_{ux}$
contain $G_{,ux}$; and all three contain $G_{,u}$. We do not
investigate here what relevance these have as constraints on the data
at an inner boundary $x=x_\text{min}$.

The full right-hand sides of the hierarchy equations are as follows,
beginning with the principal terms on a separate line ($S_G$ is all
non-principal), and the scalar field stress-energy terms at the end:
\begin{widetext}
\begin{eqnarray}
\label{SG}
S_G&=&{R\over R_{,x}}\left(S^2f_{,x}^2+4\pi
\psi_{,x}^2\right) ,
\\
S_b&=&  2 R^2 S f_{,{xy}}-\frac{R^2 G_{,{xy}}}{G}-2 R
R_{,{xy}}
\nonumber \\ &&
+f_{,{x}} \left(-4 R^2 S^2 f_{,{y}}+8 R^2 y (f S-1)+4 R S R_{,{y}}\right)+\frac{R^2
   G_{,{x}} G_{,{y}}}{G^2}+\frac{2 R G_{,{y}}
  R_{,{x}}}{G}+2 R_{,{x}} R_{,{y}}
-16 \pi  R^2 \psi_{,{x}} \psi _{,{y}} ,
\\
S_R&=& \frac{1}{4} R^2 S b_{,{xy}}+\frac{G S e^{-2 f S} R_{,{yy}}}{2 R}-\frac{1}{2} G S^2 e^{-2 f
   S} f_{,{yy}}+\frac{1}{4} S e^{-2 f S} G_{,{yy}}
\nonumber \\ &&
+\frac{R^4 S b_{,{x}}^2 e^{2 f S}}{8 G}+b_{,{x}} \left(-\frac{R^2 y}{2}-\frac{1}{2} R S
   R_{,{y}}\right)+R S b_{,{y}} R_{,{x}}-2 b R y
   R_{,{x}}
\nonumber \\ && 
+e^{-2 f S}
   \Biggl[f_{,{y}} \left(-4 G S y (f S-1)-\frac{G S^2 R_{,{y}}}{R}\right)+\frac{G y (2 f
   S-1) R_{,{y}}}{R}+G S^3 f_{,{y}}^2-\frac{S G_{,{y}}^2}{8 G}
\nonumber \\ &&
+G_{,{y}} \left(-\frac{1}{2} S^2 f_{,{y}}+f S
   y-\frac{y}{2}\right)+\frac{1}{2} G \left(2 f \left(S (5-4 f (S-1))-4\right)-1\right)-\frac{G S R_{,{y}}^2}{2
   R^2}+2 \pi  G S \psi _{,{y}}^2\Biggr], 
\label {SR} \\
S_f&=&\frac{R b_{,{xy}}}{4}+\frac{e^{-2 f S} G_{,{yy}}}{4 R}
\nonumber \\ &&
-3 b R yf_{,{x}}+\frac{R^3 b_{,{x}}^2 e^{2 f S}}{8 G}+b_{,{x}} \left(-\frac{1}{2}
   R S f_{,{y}}-f R y\right)+b_{,{y}} \left(\frac{1}{2} R S
   f_{,{x}}+\frac{R_{,{x}}}{2}\right)
\nonumber \\ &&
-2 b f y R_{,{x}}
+e^{-2 f S}
   \left(-\frac{G_{,{y}} R_{,{y}}}{2 R^2}-\frac{G_{,{y}}^2}{8 G R}
+\frac{2 \pi  G \psi_{,{y}}^2}{R}\right),
 \\
S_\psi&=& \frac{G S e^{-2 f S} \psi _{,{yy}}}{2 R} 
\nonumber \\ &&
+\frac{1}{2} R S b_{,{y}} \psi _{,{x}}-\frac{1}{2} R S
   b_{,{x}} \psi _{,{y}}-b R y\psi _{,{x}} 
+e^{-2 f S}
   \left(-\frac{G S^2 f_{,{y}} }{R}+\frac{G y (2 f S-1) }{R}+\frac{S
     G_{,{y}} }{2 R}\right)\psi _{,{y}}.
\label{Spsi}
\end{eqnarray}
\end{widetext}


\subsection{Formulation in terms of $\gamma:=\ln (G/R_{,x})$ 
and $R$-derivatives}


Note that $S_G$ given in (\ref{SG}) has a division by $R_{,x}$.  If we
assume that $R_{,x}>0$ everywhere, we can reparameterise the metric
variable $G$ by the new variable
\begin{equation}
g:={G\over R_{,x}} \quad \Rightarrow \quad G=R_{,x}g.
\end{equation}
Then all $x$-derivatives in the hierarchy equations can be eliminated in
favour of the derivative
\begin{equation}
\label{calDdef}
{\cal D}:={1\over R_{,x}}\partial_x,
\end{equation}
that is, the derivative with respect to $R$ at constant $u$ and
$y$. Moreover, $g$ is invariant under reparameterising $x$, and in
particular is simply $1$ in flat spacetime. Obviously, ${\cal D}$ does not
commute with $\partial_u$ and $\partial_y$, so we have to specify the
order of mixed derivatives. As a convention, we apply ${\cal D}$ last. If
we further replace $g$ by
\begin{equation}
\gamma:=\ln g,
\end{equation}
the equations simplify a little further, and $\gamma$ appears
undifferentiated only in the two combinations
$\exp\pm(2Sf-\gamma)$. Hence in our final numerical formulation we
treat $\gamma$ as the primary variable.

The hierarchy equations now take the form 
\begin{eqnarray}
\label{gammaeqn}
{\cal D}\gamma&=&\bar S_\gamma[R,f,\psi] \\ 
\label{blhs1eqn}
{\cal D}\left(R^4e^{2S f-\gamma}{\cal D} b\right)&=&\bar
S_b[R,f,\psi,\gamma], \\
\label{newReqn}
{\cal D}\left(R\,\Xi R\right)&=&\bar S_R[R,f,\psi,\gamma,b], \\
\label{newfeqn}
{\cal D}\left(R\,\Xi f\right)&=&\bar S_f[R,f,\psi,\gamma,b]-(\Xi
R){\cal D} f,
\nonumber \\ \\
\label{newpsieqn}
{\cal D}\left(R\,\Xi\psi\right)&=&\bar
S_\psi[R,f,\psi,\gamma,b]-(\Xi R){\cal D}\psi, \nonumber \\
\end{eqnarray}
where $\bar S=S/R_{,x}$. With $G=gR_{,x}$, third derivatives of $R$
appear in the Einstein equations when we write them in terms of $g$ or
$\gamma$ . In the hierarchy equations, only $R_{,xxy}$ and $R_{,xyy}$
appear. The former, which appears only in the equation for $b$, is
problematic numerically, but it can be eliminated by writing
\begin{equation}
\label{barSbbyparts}
\bar S_b[R,f,\psi,\gamma]=-{\cal D}\left(R^2{\cal D}(R_{,y})\right)+\tilde
S_b[R,f,\psi,\gamma].
\end{equation}
The modified source $\tilde S_b$ no longer contains third derivatives,
and we can explicitly integrate the first term on the
right. 

The source terms are now
\begin{widetext}
\begin{eqnarray}
\bar S_\gamma&=&R \left(S^2 {\cal D}(f)^2+4 \pi  {\cal D}(\psi )^2\right), \\
\tilde S_b&=&R^2 \Biggl(-\Bigl(4 {\cal D}(f) \left(S^2 f_{,y}-2 f S
y+2y\right) 
-2 S {\cal D}\left(f_{,y}\right)+{\cal D}\left(\gamma _{,y}\right)
+16 \pi {\cal D}(\psi ) \psi _{,y}\Bigr)\Biggr) \nonumber \\ &&
+2 R \left(2 S {\cal D}(f) R_{,y}+{\cal D}\left(R_{,y}\right)
+\gamma_{,y}\right)+2 R_{,y}, \\
\bar S_R&=&\frac{1}{8} e^{2 f S-\gamma } R^4 S {\cal D}(b)^2 
-\frac{1}{4} R \Bigl(-S \left(R {\cal D}\left(b_{,y}\right)+4
b_{,y}\right) 
+2 {\cal D}(b) \left(SR_{,y}+R y\right)+8 b y\Bigr) 
+\frac{1}{8} e^{\gamma -2 f S} \Biggl(32 f^2 S y^2 \nonumber \\ &&
-4 {\cal D}\left(R_{,y}\right) \left(S^2 f_{,y}-2 f S y+y\right) 
-\frac{8R_{,y}}{R}  \left(S^2 f_{,y}-2 f S y+y\right)
+8 S^3 f_{,y}^2-8 f \left(4S^2 y f_{,y}+5 y^2-1\right) \nonumber \\ &&
+2 S \gamma _{,y} {\cal D}\left(R_{,y}\right)
-S {\cal D}\left(R_{,y}\right){}^2+2 S{\cal D}\left(R_{,yy}\right)
+\frac{4 S R_{,yy}}{R}-S \gamma _{,y}^2+2 S
\left(\gamma _{,y}^2+\gamma _{,yy}\right) - 4\nonumber \\ &&
-4 \gamma _{,y} \left(S^2 f_{,y}-2 f S y+y\right)
+32 S y f_{,y}-4 S^2 f_{,yy}
-\frac{4 S R_{,y}^2}{R^2} +16 \pi S\psi _{,y}^2
\Biggr), \\
\bar S_f&=&\frac{1}{8}e^{2 f S-\gamma } R^3 {\cal D}(b)^2 
+\frac{1}{4} \Bigl(R \left(2 S b_{,y}
   {\cal D}(f)+2 {\cal D}(b) \left(-S f_{,y}-2 f
   y\right)+{\cal D}\left(b_{,y}\right)\right)-4 b y (3 R {\cal D}(f)+2 f)+2
   b_{,y}\Bigr)\nonumber \\ && 
+\frac{e^{\gamma -2 f S}}{8 R^2} \left(-4
   R_{,y} \gamma _{,y}+\left(2 R \gamma _{,y}-4 R_{,y}\right)
   {\cal D}\left(R_{,y}\right)-R {\cal D}\left(R_{,y}\right){}^2+R \left(2
   {\cal D}\left(R_{,yy}\right)+\gamma _{,y}^2+2
   \gamma _{,yy}+16 \pi  \psi _{,y}^2\right)\right), \\
\bar S_\psi&=&\frac{1}{2} \left({\cal D}(\psi ) \left(R \left(S b_{,y}-2 b y\right)\right)-R S {\cal D}(b) \psi _{,y}\right)+\frac{e^{\gamma -2 f S}}{2 R}
   \left(\psi _{,y} \left(-2 S^2 f_{,y}+4 f S y+S
   {\cal D} \left(R_{,y}\right)+S \gamma _{,y}-2 y\right)+S \psi
   _{,yy}\right). \nonumber \\
\end{eqnarray}
\end{widetext}


\subsection{Regular centre}
\label{section:regular centre}


When the null surfaces are cones emanating from a regular central
world line, null geodesics leaving the central worldline at the same
time must carry the same $u$, and those leaving at different times are
naturally identified as setting off in the same direction
$(y,\varphi)$ via parallel transport along the central wordline. The
only remaining gauge freedom is to relabel $x$ by $\tilde x(u,x,y)$,
and $u$ by $\tilde u(u)$.

Obviously, all metric coefficients must be single-valued on the
central worldline, that is, at $x=0$ they must be independent of $y$
for all $u$. We stress this by writing $G(u,0,y)=G(u,{\bf 0})$, and so
for all other quantities that are single-valued at the centre.

We show in Appendix~\ref{appendix:Minkowski} that when the centre
$R=0$ is regular, geodesic, and has coordinate location $x=0$, in
our gauge we have $b(u,{\bf 0})=f(u,{\bf 0})=0$ and
\begin{eqnarray}
g(u,{\bf 0})&=&U'(u),\\
H(u,{\bf 0})&=&U'(u)^2,
\end{eqnarray}
where $U(u)$ is proper time along the central worldline. Choosing $u$
itself to be proper time, we have $g(u,{\bf 0})=H(u,{\bf
  0})=1$. Because $g$ and $G$ are single-valued at the origin,
$R_{,x}$ at the origin must also be single-valued, and we denote it by
$R_{,x}(u,{\bf 0})$.


\subsection{Standard radial gauge choices}


We review here the standard radial gauge choices already outlined for
arbitrary spacetime dimension above.


\subsubsection{Bondi} 


The condition $R=x=:r$ defines Bondi coordinates
$(u,r,y,\varphi)$. In particular, the definition
\begin{equation}
\label{XiRdef}
\Xi R:=R_{,u}-B R_{,x}-bS R_{,y}, 
\end{equation}
reduces in Bondi gauge, where $R_{,u}=R_{,y}=0$, to
\begin{equation}
\label{Bondishift}
B=B_\text{Bondi}:=-{\Xi R\over R_{,x}}
\end{equation}
Given $\psi$ and $f$ as functions of $(x,y)$ on a surface of constant
$u$, such as $u=0$, and the gauge initial data $R(0,x,y)=x$, the
hierarchy equations are solved for $(\gamma,b,\Xi R,\Xi f,\Xi\Psi)$ by
integration. We then find $B$ from $\Xi R$ and (\ref{Bondishift}).


\subsubsection{Double-null} 


The condition $H=0$ and hence $B=0$ defines double-null coordinates
$(u,v,y,\varphi)$, where $x=:v$ becomes the second null
coordinate. Given $\psi$, $f$ and $R$ as functions of $(x,y)$ on a
surface of constant $u$, the hierarchy equations are again solved for
$(\gamma,b,\Xi R,\Xi f,\Xi\psi)$ by integration.  Here $R(0,x,y)$ can
be thought of as fixing a gauge freedom. As already mentioned, this
formulation is only almost-maximally constrained because the evolution
equation for $R$ does not relate to a physical degree of freedom but
propagates this initial gauge choice.


\subsubsection{Affine} 


The coordinate condition $G_{,x}=0$ defines $x=:\lambda$ to
be an affine parameter along the null geodesic generators of our time
slices. In this case, the first hierarchy equation
(\ref{Geqn},\ref{SG}) becomes
\begin{equation}
\label{RaffineODE}
R_{,xx}+\left(S^2f_{,x}^2+4\pi \psi_{,x}^2\right)R=0,
\end{equation}
and so does not contain $G$ at all, but instead becomes an ODE in $x$
(at constant $u$ and $y$) for $R$, given $f$ and $\psi$.  
$\Xi R$ is now used for finding $H$, rather than for evolving $R$. By
taking an $x$-derivative of the hierarchy equation (\ref{Reqn}) for $\Xi R$ and a
$u$-derivative of (\ref{RaffineODE}), and eliminating $R_{,uxx}$
between them, we find an equation of the form
\begin{equation}
H_{,xx}=S_H[R,b,f,\psi,\Xi R,\Xi f,\Xi\psi],
\end{equation}
which can be integrated twice to find $H$.

We stress that any null gauge in which we solve the Raychaudhuri
equation (\ref{Geqn}) for $G$ or $\gamma$, such as Bondi or double
null gauge, breaks down where $R_{,x}=0$. But, as we will see later,
the divergence $\rho_+$ of the null generators of our ${\cal N}_u^+$
is proportional to $R_{,x}$, so this will generically happen in strong
gravity. The only alternative is to solve (\ref{Geqn}) for $R$, for
example in affine gauge. We will consider this elsewhere.


\subsection{Choices of radial gauge for critical collapse}


We now present possible choices of radial gauge adapted to critical
collapse that generalise the ideas of \cite{Garfinkle1995} beyond
spherical symmetry. In these coordinates, we want $x=0$ to be the
regular centre $R=0$ and the outer boundary $x=x_\text{max}$ to be
future spacelike or null. We therefore evolve on the domain of
dependence of the initial data, without the need for an explicit outer
boundary condition, and the numerical domain shrinks with time. We can
then hope to control this shrinking in such a way that resolution of
the critical solution is maintained without the need for adaptive mesh
refinement.

The domain of dependence of the data on any $u=u_0$,
$0\le x\le x_\text{max}$ is bounded by the ingoing null surface
${\cal N}^{-}_{u_0,x_\text{max}}$, whose null tangent vector at
$S_{u_0,x_\text{max}}$ is $\Xi$. Hence at $x=x_\text{max}$, $x$ should
not decrease along $\Xi$, or
\begin{equation}
\label{outerboundaryacausal}
\Xi x=-{H\over 2G}\ge0 \text{ at } x=x_{\rm max},
\end{equation}
and this must hold for all $u$. An equivalent requirement is that the
surface $x=x_\text{max}$ is null or spacelike, that is
\begin{equation}
\label{outerboundaryacausalbis}
|\nabla x|^2={H\over G^2}\le0 \text{ at } x=x_{\rm max}.
\end{equation}
In short, $H$ and therefore $B$ must be non-positive at
$x=x_\text{max}$.  This also means that numerically we can
consistently upwind the advection term $B\partial_x$ in the time
evolution equations at the outer boundary, with the one-sided stencil pointing
into the numerical domain.

At the inner boundary $x=0$, the condition that $R=0$ remains at $x=0$
fixes $B$ to be given by (\ref{Bondishift}). This is positive, and so
again we can upwind consistently with the one-sided stencil pointing
into the numerical domain.

For applications to critical collapse we impose a condition at some
$0<x_0\le x_\text{max}$ that makes $x=x_0$ approximately null:
approximately in the sense that $B=0$ there in some average
sense, or that $B\le 0$ with equality at one or more values of
$y$. If a spacetime approaches a self-similar critical solution, and
$x_0$ is chosen appropriately, the past lightcone of the critical
solution will then be at $x\simeq x_0$. This gauge condition should
then also imply $B\le 0$ at the outer boundary.

We now present a few possible gauges that obey these two boundary
conditions at $x=x_0$ and $x=0$.


\subsubsection{Shifted double null gauge}


Consider first the choice
\begin{equation}
\label{sdnshift}
B=B_\text{sdn}:=
\left(1-{x\over x_0}\right)B_\text{Bondi}({\bf
  0}),
\end{equation}
for some constant parameter $x_0\le x_\text{max}$. The Bondi radial
shift was defined in (\ref{Bondishift}). In other words,
$B$ decreases linearly in $x$ from its value in Bondi gauge at
the centre to zero at the some $x_0$. Then $x=0$ remains at
$R=0$, and $x=x_0$ is an ingoing null surface.

Our convention that $u$ is proper time along the central worldline
implies that $\Xi R=-1/2$ there, and so (\ref{Bondishift}) gives
\begin{equation}
\label{sdnshift2}
B_\text{sdn}=\left(1-{x\over x_0}\right){1\over
  2R_{,x}(u,{\bf 0})}.
\end{equation}
We shall call this gauge choice shifted double null (from now on, sdn)
gauge because it simply rescales the ingoing null coordinate $v$.
More precisely, if we choose $x=v$ at $u=0$, then at fixed $u$ the new
coordinate $x$ is a linear function of $v$ defined by $v=v_0$ at $x=0$
and $R=0$ at $x=0$ \cite{PortoGundlach2022}.

Therefore, sdn gauge is a continuous version of the repeated
regridding of the double-null coordinate $v$ in
\cite{Garfinkle1995}. We have previously implemented it in spherical
symmetry in \cite{PortoGundlach2022}, and found that it works exactly
as well as our earlier implementation, in \cite{ymscalar}, of the
original Garfinkle regridding algorithm. The resulting numerical
domains are identical for $x_0=x_{\rm max}$, but we found in
\cite{ymscalar,PortoGundlach2022} that choosing $x_{\rm max}>x_0$,
which gives us a spacelike buffer zone, has the advantage of revealing
more of the critical solution spacetime.

On a regular spacetime beyond spherical symmetry, we expect sdn gauge
to fail because ingoing null cones, and in particular $x=x_0$, do not
reconverge on the central worldline. It is, however, extremely useful
in spherical symmetry. We will now discuss alternatives that work
better beyond spherical symmetry, but reduce to sdn gauge in spherical
symmetry.

We have also not been able to find a stable discretization of the
equations in sdn gauge beyond spherical symmetry, even in the limit of
weak fields. The instability {\em looks} like a purely numerical
problem at the origin, but we cannot exclude formation of caustics or
ill-posedness as problems in the continuum. We have not explored this
further.


\subsubsection{Global shifted Bondi gauge}


An alternative starting point is to demand that
\begin{equation}
\label{sdef}
R(u,x,y)=s(u)\,x,
\end{equation}
where $s(u)$ is a function to be specified. This simplifies the
hierarchy equations in the same way as Bondi gauge does, and in
particular $G=g/s(u)$. Substituting this into the equation defining
$\Xi R$, we have
\begin{equation}
\label{gsBshift}
B=B_\text{gsB}:={s'(u)x-\Xi R\over s(u)}. 
\end{equation}
We shall this class of gauges global shifted Bondi (from now on, gsB)
gauge.

To avoid potential numerical instabilities from evolving $R$ as a
dynamical variable, we update it directly with $R_{,u}=s'(u)x$, rather
than the generic expression based on $\Xi R$ plus shift terms, and
also use (\ref{sdef}) in simplifying other derivatives. This gauge
does not suffer from the same numerical instabilities as sdn gauge.

To fix $s(u)$, we demand that the surface $x=x_0$ is ingoing
null or future spacelike, in the sense that $H(u,x_0,y)\le
0$. This gives
\begin{equation}
\label{gsBx0min}
s'(u)={1\over x_0}\min_y\Xi R(u,x_0,y).
\end{equation}
As we shall see in Paper~II, gsB gauge behaves very different from sdn
gauge in strong fields, already in spherical symmetry, and does not
seem to be a good choice for critical collapse.


\subsubsection{Local shifted Bondi gauge}


Yet another starting point is the observation that the instability we
observe in our implementation of sdn gauge seems to be connected to
the $y$-dependence of $R$, while the choice (\ref{sdef}) is too
restricted. Hence we can attempt the more general gauge
\begin{equation}
R(u,x,y)=\bar R(u,x), 
\end{equation}
by setting 
\begin{equation}
B=B_\text{lsB}
:={\bar R_{,u}-\Xi R\over \bar R_{,x}},
\end{equation}
where $\bar R_{,u}(u,x)$ and hence $\bar R(u,x)$ is yet to be specified.
We call this local shifted Bondi (from now on, lsB) gauge. We require
\begin{equation}
\bar R(u,0)=0
\end{equation}
to keep the origin regular at $x=0$ and 
\begin{equation}
\bar R_{,u}(u,x_\text{max})\le\min_y\Xi R(u,x_\text{max},y)
\end{equation}
to make $B\le 0$ at the outer boundary. 

We restrict the possible choices of $\bar R_{,u}$ by demanding that in
spherical symmetry we revert to sdn gauge. An obvious possibility is
to start from the Bondi shift (\ref{Bondishift}) with its spherical
part subtracted (``non-spherical Bondi'', from now on nsB),
\begin{equation}
\label{nsB}
B_\text{nsB}:=-{\Xi
  R-(\Xi R)_{l=0}\over \bar R_{,x}},
\end{equation}
where the suffix $l=0$ denotes the spherical part, and add the (purely
spherical) sdn shift:
\begin{equation}
\label{lsb-XiRmean}
B=B_\text{lsB1}:=B_\text{nsB}+B_\text{sdn}.
\end{equation}
In this gauge $x_0$ is a null surface ``on average''.  

We can add a further term that makes the shift non-positive everywhere
at $x_0$, giving
\begin{equation}
\label{lsb-XiRmeanminx0}
B_\text{lsB2}:=B_\text{lsB1}
+{x\over x_0}{(\min_y\Xi R)-(\Xi R)_{l=0}\over R_{,x}}(x_0). 
\end{equation}
In this gauge, $B_{,u}$ is discontinuous at such values of $u$ where
the location in $y$ of $\min_y\Xi R(u,x_0,y)$ changes
discontinuously. In either lsB1 or lsB2, there is no guarantee that
$H<0$ at the outer boundary to make it future spacelike.

Another possibility is to subtract, at {\em every} $(u,x)$, the global
maximum over $y$ of the Bondi radial shift instead of its spherical part
(``non-negative Bondi'')
\begin{equation}
\label{nnB}
B_\text{nnB}:=-{\Xi R-\min_y(\Xi R)\over\bar R_{,x}},
\end{equation}
and again add the sdn shift,
\begin{equation}
\label{lsb-XiRmin}
B_\text{lsB4}:=B_\text{nnB}+B_\text{sdn}.
\end{equation}
This version has the property that every surface of constant
$x\ge x_0$, and so the outer boundary in particular, is null or future
spacelike. It has the disadvantage that $B_{,x}$ and $B_{,u}$ are
discontinuous at all values of $(u,x)$ where the location in $y$ of
$\min_y\Xi R(u,x,y)$ changes discontinuously.

We can also make a transition from lsb gauge, with the spherical part
given by sdn, near the centre, to full sdn at the outer boundary, so
that the outer boundary is null everywhere. In other words, we
consider
\begin{eqnarray}
\label{lsBtosdn}
B_\text{lsBtosdn}&:=&B_\text{sdn}\nonumber \\ && 
\hspace{-2.2cm}+\left[1-K_{01}\left({x-\lambda x_0\over x_0-\lambda
    x_0}\right)\right]
\left[
\left({\Xi R\over R_{,x}}\right)_{l=0}
-{\Xi R\over R_{,x}}
\right],
\end{eqnarray}
where $K_{01}(x)$ is a sufficiently smooth switching function with
$K_{01}(x\le 0)=0$ and $K_{01}(x\ge 1)=1$, and $0<\lambda\le 1$ is a
parameter.  We then have pure lsb gauge for $0\le x\le \lambda x_0$,
and a transition to pure sdn over the interval $\lambda x_0\le x\le
x_0$.  For $K_{01}$ on the transition range $0\le x\le 1$ we have
settled on the 9th order polynomial defined by the first four
derivatives vanishing at $x=0$ and $x=1$, and the symmetry condition
$K_{01}(1/2)=1/2$.


\subsection{Spacetime diagnostics}



\subsubsection{Affine parameter} 


In this section we look at a number of ways of extracting geometric
information from our simulations. After fixing the central worldline,
a completely geometric coordinate system is given by
$(u,\lambda,y,\varphi)$, where $u$ labels the null cones emanating
from the central worldline, and $(y,\varphi)$ label the generators of
these null cones. $u$ is fixed to be the proper time along the central
worldline. The same $(y,\varphi)$ at different $u$ are identified by
parallel transport of the vector $\partial_u$ along the central
worldline. $\lambda$ is the affine parameter along the generators,
with origin $\lambda=0$ at $R=0$ and normalisation $\lambda\simeq R$
near the origin.

The tangent vector to the affinely parameterised generators of our
null cones is $U$ given by (\ref{Ugeneral}). We have
$dx/d\lambda:=U^a\nabla_ax=G^{-1}$ and $\lambda=0$ on the central
worldline $R=x=0$. Integration then gives us
\begin{equation}
\lambda(u,x,y)=\int_0^x G(u,x',y)\,dx'=\int_0^R g(u,x',y)\,dR'.
\end{equation}
for the affine parameter. Recall that in our convention $g=1$ at the
origin, so we have the required normalisation.


\subsubsection{Redshift} 


Let $\tau$ be the proper time measured by a timelike observer at
coordinate location $(u,x,y,\varphi)$ and with 4-velocity $u^a$
(normalised to $u^au_a=-1$). The redshift of photons emitted from the
central world line at $(u,{\bf 0})$, measured by this observer, is
\begin{equation}
\label{Zdef}
Z:={d\tau\over du}={1\over u^a\nabla_au}
\end{equation}
where we have used our convention that $d\tau/du=1$ along the central
world line.

One natural choice of a family of timelike observers puts them at constant
$R$, $y$ and $\varphi$. The ansatz 
\begin{equation}
u^a=Z^{-1}\left((\partial_u)^a-{R_{,u}\over R_{,x}}(\partial_x)^a\right)
\end{equation}
for the corresponding $u^a$ gives
$u^a\nabla_aR=u^a\nabla_ay=u^a\nabla_a\varphi=0$ and $u^a\nabla_a
u=Z^{-1}$ as required. From $u_au^a=-1$ we then find that $Z$ is given
by
\begin{equation}
Z=\left(-2g(\Xi R+S bR_{,y})-R^2e^{2S f}S
b^2\right)^{1\over 2}.
\end{equation}


\subsubsection{Hawking mass and Hawking compactness}


Following \cite{SzabadosLRR}, let ${\cal S}$ be a smooth closed
spacelike 2-surface. Let $l^a$ and $n^a$ be a pair of future-pointing
null vectors normal to ${\cal S}$, normalised such that
$l^an_a=-1$. Let $n^a$ be outgoing and $l_a$ be ingoing. This is
unique up to multiplying $n^a$ by a positive function $e^\Lambda$ on
${\cal S}$ and multiplying $l_a$ by $e^{-\Lambda}$. The projection
operator onto the tangent space of ${\cal S}$ is
\begin{equation}
\pi_{ab}:=g_{ab}+l_an_b+n_al_b,
\end{equation}
and is unique. We then define the null congruence expansions
\begin{equation}
\label{rhorhoprimedef}
\rho_+:={1\over 2}\pi^{ab}\nabla_an_b, \qquad \rho_-:={1\over
  2}\pi^{ab}\nabla_al_b,
\end{equation}
It is easy to see that
\begin{equation}
\pi^{ab}\nabla_a(e^\Lambda n_b)=e^\Lambda(\nabla_an^a+n^bl^a\nabla_a
l_b)=e^\Lambda\nabla_an^a,
\end{equation}
where the first equality holds when $n^a$ is null and the second when
$l^a$ is an affinely parameterised geodesic. A similar result then
holds for $e^{-\Lambda}l_a$. Without loss of generality we now define
$n^a$ and $l^a$ to be continued off ${\cal S}$ as affinely
parameterised geodesics. Then the product
\begin{equation}\rho_+\rho_-={1\over 4}(\nabla_al^a)(\nabla_bn^b)
\end{equation}
is independent of the normalisation $e^\Lambda$, and therefore
uniquely determined by the spacetime geometry and ${\cal S}$.

From $\rho_+\rho_-$, the Hawking mass $M$ of ${\cal S}$ is now defined
by
\begin{eqnarray}
\label{MHaw}
M({\cal S})&:=&{1\over 2}\sqrt{A({\cal S})\over 4\pi}\,C({\cal S}), \\
\label{CHaw}
C({\cal S})&:=&1+{1\over 2\pi}\int_{\cal S} \rho_+\rho_-\,dS,
\end{eqnarray}
where $A:=\int_{\cal S} dS$ is the area of ${\cal S}$. We have defined
the ``Hawking compactness'' $C$ as an intermediate quantity that is of
interest in its own right. In particular, a marginally outer-trapped
surface, defined by $\rho_+=0$ and $\rho_-<0$ at each point, has
Hawking compactness $C=1$, and an outer-trapped surface, defined by
$\rho_+<0$ and $\rho_-<0$ at each point, has Hawking compactness
$C>1$, but the converses are not true.

We now restrict attention to spacelike 2-surfaces ${\cal S}_0$ that
lie within a single coordinate null cone. We define such surfaces by
\begin{equation}
u-u_0=0, \qquad x-x_0(y)=0.
\end{equation}
A basis of tangent vectors to ${\cal S}_0$ is given by
\begin{equation}
t_{(y)}:=\partial_y+x_0'\, \partial_x,\qquad t_{(\varphi)}:=\partial_\varphi. 
\end{equation}
The outgoing and ingoing null vectors orthogonal to ${\cal S}_0$,
normalised so that $n^al_a=-1$ and $l^a\nabla_a u=1$, are
\begin{eqnarray}
n&=&U, \\ l&=&\Xi+{e^{-2S f}S gR_{,x}\over R^2}
\left(x_0'\,\partial_y+{x_0'^2\over 2}\partial_x\right). 
\end{eqnarray}
$n^a=U^a$ holds because the outgoing null surface that emanates from
${\cal S}_0$ is simply the part of ${\cal N}^+_u$ that lies to the
future of ${\cal S}_0$. If and only if $x_0(y)$ is constant, we also
have $l^a=\Xi^a$.

The corresponding null geodesic expansions are
\begin{eqnarray}
\label{rhoscalarnew}
R\rho_+&=&{1\over g}, \\
\label{myrhoscalarnew}
2R\rho_-&=&K_0+K_1\,x_0'+K_2\,x_0'^2+K_3\,x_0''.
\end{eqnarray}
The $K_i$ are evaluated at $(u_0,x_0(y),y)$, but do not contain
derivatives of $x_0(y)$. Here we only give
\begin{equation}
K_0=2\Xi R-R(S b)_{,y}
\end{equation}
for later reference.

The induced metric on ${\cal S}_0$ is
\begin{equation}
h_{ij}=x^\mu_{,i}x^\nu_{,j}h_{\mu\nu}=x^\mu_{,i}x^\nu_{,j}g_{\mu\nu}=g_{ij},
\end{equation}
where $x^i:=(y,\varphi)$ are the coordinates on ${\cal S}_0$. Its
determinant is therefore simply $R^4$, independent of $x_0(y)$, so the
volume element on ${\cal S}_0$ is
\begin{equation}
\label{dSnew}
dS=R^2\,dy\,d\varphi.
\end{equation}
We therefore have
\begin{equation}
\sqrt{A({\cal S}_0)\over 4\pi}=\left({1\over 2}\int_{-1}^1
R^2\,dy\right)^{1/2}
\end{equation}
and 
\begin{equation}
\label{CHawbis}
C({\cal S}_0)=1+{1\over 2}\int_{-1}^1 2\rho_+\rho_-R^2\,dy.
\end{equation}
Note that the integral ${1\over2}\int_{-1}^1 ...dy$ is equal to the
$l=0$ component of its integrand, and that $2R^2\rho_+\rho_-=-1$ for
light cones in flat spacetime.

After an integration by parts to eliminate the $x_0''$ term,
$C$ becomes
\begin{equation}
\label{Cx0y}
C({\cal S}_0)=1+{1\over 2}\int_{-1}^1 
L(x_0,x_0')\,dy,
\end{equation}
where
\begin{eqnarray}
L&=&L_0+L_1\,x_0'+L_2\,x_0'^2, \\
L_0&:=&{K_0\over g}={2\Xi R -R(S b)_{,y}\over g}, \\
L_1&:=&{S\left(e^{-2S f}
R_{,x}(R g)_{,y}-R^3b_{,x}\right)\over gR^2}, \\
L_2&:=&{e^{-2S f}S g_{,x}R_{,x}\over gR}.
\end{eqnarray}

If we now further restrict to surfaces ${\cal S}_{u,x}$ of
constant $u$ and $x$, (\ref{myrhoscalarnew}) simplifies to 
\begin{equation}
2R\rho_-=K_0,
\end{equation}
and (\ref{Cx0y}) simplifies to
\begin{equation}
\label{CHawux}
C(u,x):=C({\cal S}_{u,x})=1+{1\over 2}\int_{-1}^1 L_0\,dy.
\end{equation}

Beyond spherical symmetry, $M({\cal S}_{u,x})$ need not be monotonic
in $x$ or non-negative. However, in the lsB class of gauges, one can
show that it is non-decreasing with $x$, and non-negative. To prove
this, note that for $R=R(u,x)$, the integral for $A$ is trivial,
giving $\sqrt{A/4\pi}=R$. We can pull this factor of $R$ into the
integral for $C(u,x)$ to obtain
\begin{equation}
\label{MlsB}
M(u,x):=M({\cal S}_{u,x})={1\over 4}\int_{-1}^1 R(1 +2\rho_+\rho_-R^2)\,dy.
\end{equation}
To evaluate $M_{,x}({\cal S}_{u,x})$, one can pull $\partial_x$ into
the integral and use integration by parts in $y$ to express
it as 
\begin{eqnarray}
M(u,x)_{,x}&=&{R_{,x}\over 4}\int_{-1}^1\Bigl\{
{e^{-2S f}S\over 4g^2}
\left(g_{,y}-R^2e^{2S f}{\cal D} b\right)^2 \nonumber \\ &&
-2R^4\rho_+\rho_-\left[S^2({\cal D} f)^2+4\pi({\cal
    D}\psi)^2\right] \nonumber \\ &&
+4\pi e^{-2S f}S\psi_{,y}^2\Bigr\}\,dy.
\label{DMlsB}
\end{eqnarray}
Hence a sufficient condition for ${\cal D}M:=M_{,x}/R_{,x}$ to be
non-negative is that $\rho_+\rho_-<0$. (This result is a special case of
Eardley's observation \cite{EardleyBattelle} that the Hawking mass
increases on a foliation of an outgoing null surface by luminosity
distance, as long as $\rho_+\rho_-<0$ and the matter obeys the dominant
energy condition.)  From ${\cal D} M\ge 0$ and $M=0$ along the central
worldline we then obtain $M\ge 0$.

In lsB gauge, where 
\begin{equation}
\label{MCRlsB}
M(u,x)={R(u,x)\over 2}C(u,x)
\end{equation}
we can therefore either compute $M$ from (\ref{MCRlsB}) with $C$ given
by (\ref{Cx0y}), or by integrating (\ref{DMlsB}) as
\begin{equation}
\label{tildeMdef}
\tilde M(u,x,):=\int_0^xM_{,x}(u,x')\,dx',
\end{equation}
with
\begin{equation}
\tilde C(u,x):={2\tilde M(u,x)\over R(u,x)}
\end{equation}
then defined from $\tilde M$. In the continuum, $C=\tilde C$ and $M=\tilde
M$, but their discretizations are very different, to the extent that
their agreement is a highly non-trivial test of the accuracy of our
discretization.


\subsubsection{No MOTS on coordinate lightcones $u=\text{const}$}
\label{section:MOTS}


A marginally outer-trapped surface (MOTS) in 4-dimensional spacetime
is a smooth closed 2-dimensional spacelike surface in the spacetime,
such that the generators of its outgoing null cone have zero
divergence. We now ask if a MOTS ${\cal S}_0$ can exist that lies in a
single null time slice, defined by $\rho_+=0$ on the surface $u=u_0$,
$x=x_0(y)$. 

A first problem with this is that in any gauge where $R$ is specified
as free data and evolved, including sdn, gsB and lsB gauge, the
Raychaudhuri equation will necessarily involve division by
$R_{,x}$. To see this, write it as
\begin{equation}
\left(\ln {G\over R_{,x}}\right)_{,x}={R\over R_{,x}}\tilde S_G,
\qquad \tilde S_G:=S^2f_{,x}^2+4\pi\psi_{,x}^2,
\end{equation}
where $\tilde S_G$ is now regular even when $R_{,x}=0$. It is now easy
to check that if we write this as a first-order ODE in $x$ for $G$,
$g:=G/R_{,x}$ or $\rho_+=R_{,x}/(RG)$, or as a second-order ODE in $x$
for the affine parameter $\lambda$ with $\lambda_{,x}=G$, these ODEs
involve a division by $R_{,x}$, and so become singular where the null
expansion $\rho_+$ changes sign. Where $R_{,x}>0$, we can absorb it
into ${\cal D}:=d/dR$ (as we have done), but ${\cal D}$ is not defined
where $R_{,x}=0$. This division by $R_{,x}$ is purely a gauge problem:
in any gauge where we specify $G$ a priori and solve the Raychaudhuri
equation for $R$, for example in affine gauge, $R_{,x}$ and therefore
$\rho_+$ can change sign without a problem.

A second, purely geometrical, problem is that when we trace the
future-outgoing null rays that emerge from the MOTS backwards in time
we generically do not expect them to converge to a point, but rather
to form caustics, except in spherical symmetry. In particular this
means that this backwards null surface cannot in general be a {\em
  coordinate} null cone with regular vertex.

We now ask if one can find a {\em non}-marginally outer-trapped
surface (OTS) on $u=u_0$, one where $\rho_+\le 0$ everywhere. The
geometric non-genericity argument then does not apply, so we expect to
be able to find OTS (but not MOTS) in affine gauge.

But in twist-free vacuum axisymmetry even this is not possible because
of a third, again purely geometrical, problem: the shear along the
symmetry axis is zero, (as we see from $S=0$ there), so in vacuum the
Raychaudhuri equation on the symmetry axis on the symmetry axis
reduces to $g_{,x}=0$ or $R_{,xx}=0$, and so $\rho_+=1/(gR)=R_{,x}/G$
cannot become zero or change sign. There may of course be OTS and MOTS in
such a spacetime, but they cannot be embedded in an outgoing null cone
with regular vertex.


\subsubsection{Surfaces of maximal Hawking compactness}
\label{section:Hawkingcompactness}


In spite of the obstacles set out in the last subsection, in spherical
symmetry we and other authors have happily identified MOTS, and used
their Hawking mass as an estimate of the initial black hole mass
while working in double-null, Bondi or sdn gauge. How did this work?

Quite naively, we were led by what one does in spherical symmetry on
Cauchy surfaces, for example in polar-radial coordinates $(t,r)$ (see
Appendix~\ref{appendix:sphericalsymmetry}): these coordinates become
singular on a MOTS, but one simply identifies the
first appearance of a local maximum in $r$ of $C(t,r)$ with
$C\ge 0.99$, say, as an approximate MOTS, and estimates $M\simeq r/2$
at its location.

Similarly, in null coordinates we called a surface ${\cal S}_{u,x}$ an
approximate MOTS when it was a local maximum in $x$ of $C(u,x)$ with
$C\ge 0.99$, say \cite{ymscalar,PortoGundlach2022}. Unlike
$\rho_+(u,x)$, which is often a decreasing function of $x$ (while
remaining strictly positive), $C(u,x)$ typically does have one or more
local maxima in $x$. This singled out a specific ${\cal S}_{u,x}$ as
our MOTS candidate. The same approach was also taken by the authors of
\cite{Garfinkle1995,PuerrerHusaAichelburg}.

How then can we generalise this procedure in spherical symmetry to the
non-spherical case? The obvious difference is that we no longer have a
preferred foliation of our coordinate null cones into 2-surfaces.

Ideally, we would look for surfaces $x=x_0(y)$ in $u=u_0$ that
maximise $C[u_0,x_0(y)]$ given by (\ref{Cx0y}). The resulting
Euler-Lagrange equation is a quasilinear second-order ODE for
$x_0(y)$. The boundary term usually obtained in deriving the
Euler-Lagrange equation vanishes, and so the variation is
unconstrained, consistent with our earlier observation following
Eq.~(\ref{psithetathetaeqn}) that a regular scalar need not have
vanishing $y$-derivative on the symmetry axis.

In Paper~II, we shall for simplicity limit ourselves to finding the
local maxima in $x$ of $C(u,x)$ given by (\ref{Cx0y}) for the
compactness of the {\em coordinate} 2-spheres ${\cal S}_{u,x}$, as we
did in spherical symmetry. It turns out that with a smaller threshold
value, say $C\ge 0.8$, our heuristic criterion consistently
distinguishes collapsing and dispersing solutions. Recall, however,
our earlier observation following Eq.~(\ref{CHaw}) that beyond
spherical symmetry $C>1$ is necessary but not sufficient for a
spacelike 2-surface to be outer-trapped.


\subsubsection{Event horizons and coordinate null cones}


We now show that if a spacetime admits an event horizon, this has at
least one generator in common with each of at least two coordinate
null cones.

The intersection ${\cal H}\cap\Sigma$ of an event horizon ${\cal H}$
with a Cauchy surface $\Sigma$ with topology ${\Bbb R}^{3}$ (which
excludes external black holes with two spatial infinities) at
sufficiently late time is a 2-sphere (or consists of disconnected
2-spheres). Hence on ${\cal H}\cap\Sigma$ the coordinate $u$ attains a
global minimum and global maximum. If the intersection is at least
$C^2$ in our coordinate system, they are also local extrema. At any
such local extremum $u_*$, ${\cal H}\cap\Sigma$ and
$\{u=u_*\}\cap\Sigma$ have the same (2-dimensional, spacelike) tangent
space $V$. It follows that the unique future outgoing null geodesic
through that point and normal to $V$ is a generator of both ${\cal H}$
and $u=u_*$.

Independently, in axisymmetry there will be (at least) two local
extrema $u_*$ of $u$ on ${\cal H}\cap\Sigma$ located at the poles
$y=\pm 1$.  If the spacetime has an additional $y\to-y$ symmetry
(reflection through the equatorial plane), and ${\cal H}\cap\Sigma$
(or one of its components) straddles the equatorial plane, there is a
third local extremum on the equator $y=0$, while the other two are
related by the reflection symmetry.

It seems unlikely that we can identify any horizon generators, even if
they are also coordinate null cone generators.


\section{Numerical methods}
\label{sec:numericalmethods}



\subsection{Legendre pseudospectral method in $y$}



\subsubsection{Choice of basis functions, and synthesis matrices}


In the physical scenarios we are interested in, it is natural to
maintain constant angular resolution, and to expect the solution to be
smooth. We therefore use a pseudospectral method in the angle
$\theta$, or $y$. In this subsection, we write $N$ for $N_y$, the
number of grid points in $y$.

We represent $\psi$ as
\begin{equation}
\label{psilexpansion}
\psi=\sum_{l=0}^{N-1} \psi_l(u,x)P_l(y),
\end{equation}
where $P_l$ is the Legendre polynomial of order $l$, proportional to
the spherical harmonic $Y_{l0}$. We represent any other quantity that
transforms as a scalar under coordinate changes on the 2-spheres
${\cal S}_{u,x}$, including the metric components $R$, $G$ and $H$, in
the same way.

By contrast, $b$ and $f$ are not scalars but components of a vector
and symmetric 2-tensor on $S^2$, respectively. We show in
Appendix~\ref{appendix:linperts} that if we represent them as
\begin{eqnarray}
\label{blexpansion}
b&=&\sum_{l=1}^N b_l(u,x)P_l'(y), \\
\label{flexpansion}
f&=&\sum_{l=2}^{N+1} f_l(u,x)P_l''(y),
\end{eqnarray}
then in the linearisation of the Einstein and wave equations about
spherical symmetry the different $l$ decouple. We choose this spectral
representation for $b$, $f$ and the scalars also in the nonlinear
case.

In any pseudospectral method, one goes backwards and forwards between a
finite number of coefficients in Fourier space and an equal number of
carefully chosen collocation points in real space, carrying out
differentiation in Fourier space and nonlinear algebraic operations in
real space.

Here we transform between collocation points $y_i$ and spherical
harmonic components $l$ by full matrix multiplication. This is clearly
inefficient for large $N$, in contrast to the fast Fourier transform
available for a Fourier series or Chebyshev polynomials.

We call the matrices that take variables from Fourier space (with
index $l$) to real space (with index $i$) synthesis matrices. For all
variables that transform as a scalar under coordinate changes on the
coordinate 2-spheres, we use the synthesis matrix
$S_{il}^{(0)}:=P_l(y_i)$, where $y_i$ are the collocation points (to
be determined below) and $l$ is the spectral index. In other words,
each {\em column} of $S^{(0)}$ represents one $P_l$. For $b$ we use
$S_{il}^{(1)}:=P_l'(y_i)$, and for $f$ we use
$S_{il}^{(2)}:=P_l''(y_i)$. Note this means that with $i=1,\dots N$,
we can represent $l=0,\dots N-1$ for scalars, $l=1,\dots N$ for $b$
and $l=2,\dots N+1$ for $f$.


\subsubsection{Differentiation in $y$ and choice of collocation points}


Rather than transforming back to Fourier space in order to
differentiate in $y$, we differentiate directly in real space. Either
method requires full-matrix multiplication. Making differentiation
exact for certain functions of $y$ then fixes the collocation points.

We implement first $y$-derivatives through the Legendre-Gauss-Lobatto
differentiation matrix \cite{Boyd}
\begin{equation}
D_{ij}=\begin{cases}
{P_{N-1}(y_i)\over (y_i-y_j)P_{N-1}(y_j)}, & i\ne j, \\
-{N(N-1)\over 4}, & i=j=1, \\
{N(N-1)\over 4}, & i=j=N, \\
0, & \text{otherwise},
\end{cases}
\end{equation}
where the grid points $y_2,\dots,y_{N-1}$ are the zeros of
$P_{N-1}'(y)$ in increasing order, $y_1=-1$ and $y_N=1$. For second
derivatives, we use the matrix square of $D$. This discretisation is
known to be exact for polynomials up to order $2N-3$.  We choose
Legendre-Gauss-Lobatto because we want the north and south poles
$y=\pm 1$ to be on the grid. In order to also have the equator on the
grid, we then choose $N$ to be odd, typically $2^K+1$.


\subsubsection{Analysis matrices}


An analysis matrix takes grid functions from physical to Fourier
space. One might compute the analysis matrix $A_{li}^{(0)}$ using the
formulas for Legendre-Gauss-Lobatto quadrature and the fact that
$\int P_mP_n\,dy=(2n+1)\delta_{mn}/2$. However, with $A^{(0)}$ defined
this way, the product $A^{(0)}S^{(0)}$ differs from the expected unit
matrix in that its bottom right element is $2+1/(N-1)$. This is due to
the fact that Legendre-Gauss-Lobatto quadrature is exact for
polynomials in $y$ up to order $2N-3$, whereas this bottom right
element requires integration of $P_{N-1}P_{N-1}$, which is a
polynomial of order $2N-2$.

Therefore we use as our analysis matrix $A_{li}^{(0)}$ the matrix
inverse of $S^{(0)}_{il}$, which differs from the naive analysis matrix
only in its last row. (Having to make this choice could have been
avoided by using Gauss-Legendre quadrature, which is exact for
polynomials up to order $2N-1$.) We similarly define $A_{li}^{(1)}$
and $A_{li}^{(2)}$ as the matrix inverses of $S_{il}^{(1)}$
and $S_{il}^{(2)}$.


\subsubsection{Discrete versions of continuum identities and
  consistent truncation}


We define the $N\times N$ matrices
\begin{equation}
\Delta^{(s)}:=(1-Y^2)D^2-2(s+1)YD,
\end{equation}
where $Y$ denotes the diagonal matrix
\begin{equation}
Y:={\rm diag}\{y_i\}_{i=1}^N,
\end{equation}
and we have defined the diagonal matrices
\begin{equation}
\Lambda^{(s)}:={\rm diag}\left\{\lambda^{(s)}_l\right\}_{l=s}^{N-1+s},
\end{equation}
where
\begin{equation}
\lambda^{(s)}_l:=-(l+s+1)(l-s), \quad l\ge s,
\end{equation}
are the eigenvalues of the Laplace operator on $S^2$ for the ``spins''
$s=0,1,2$.

Our synthesis and differentiation matrices obey discrete equivalents
of the continuum identities (\ref{L2Y}), (\ref{Peqn}) and (\ref{Qeqn})
between Legendre polynomials given in
Appendix~\ref{appendix:linperts}, which in this notation can be
written concisely as
\begin{equation}
\Delta^{(s)}S^{(s)}=S^{(s)}\Lambda^{(s)},
\end{equation}
for $s=0,1,2$.
We also have 
\begin{eqnarray}
DS^{(0)}&=&S^{(1)}I_+, \\
\label{S1toS2}
DS^{(1)}&=&S^{(2)}I_+, \\
\Rightarrow \quad D^2S^{(0)}&=&S^{(2)}I_+^2,
\end{eqnarray}
relating the different spins. Here $I_+$ denotes the matrix with ones
in the super-diagonal. We also define $I_-$ as the matrix with ones in
the sub-diagonal, and $I_0$ as the matrix with ones in the diagonal,
except for a zero in the bottom right element. These obey
$I_+I_-=I_0$. In particular, Eq.~(\ref{S1toS2}) is a discrete version
of the identity (\ref{Pp2Q}), which we need for the linearised
hierarchy equation for $f$, Eq.~(\ref{deltafBondi}), to be discretised
exactly in $y$. 

With the definition
\begin{equation}
\slashed{\Delta}^{(1)}:=(1-Y^2)D-4Y \quad\Rightarrow\quad
\slashed{\Delta}^{(1)}D=\Delta^{(1)},
\end{equation}
we can derive the identity 
\begin{equation}
\slashed{\Delta}^{(1)}S^{(2)}I_0= S^{(1)}\Lambda^{(1)}I_-.
\label{Qp2Pdiscrete}
\end{equation}
Except for the presence of the factor $I_0$ on the left,
(\ref{Qp2Pdiscrete}) is the discrete version of the identity
(\ref{Qp2P}), which is needed for the linearised hierarchy equation
for $b$, Eq.~(\ref{deltabBondi}). Without the right factor of $I_0$,
the left-hand side of (\ref{Qp2Pdiscrete}) would have non-zero entries
for $l=N-1$ and $N-3$ in its last column. With $N$ grid points we can
represent $f_{N+1}$ but not $b_{N+1}$, so one can think of this as an
aliasing error arising from the spectral truncation. To avoid this
error, which would lead to an instability, we need to suppress the
unpartnered $f_{N+1}$.

In the linearised field equations on a general spherically symmetric
background, or in a gauge other than Bondi gauge, $f_l$ and $b_l$
couple not only to each other but also to $\psi_l$, $G_l$, $R_l$ and
$H_l$. Therefore we must suppress $f_{N+1}$, $f_N$ and $b_N$, as all
three have no scalar partners.


\subsubsection{Half-range spectral method}


Most papers on axisymmetric gravitational collapse assume an
additional reflection symmetry, $z\to-z$ in cylindrical coordinates,
or $y\to-y$ in our spherical coordinates. In such situations, we can
save computing time by only representing the even $l$ in Fourier
space, or the values $-1\le y\le 0$ in real space. Let $\bar
N:=(N+1)/2$ denote the number of grid points on the half-range
equivalent to $N$ grid points on the full range. We define
reduced analysis and synthesis matrices as
\begin{eqnarray}
\bar A^{(0,2)}&:=&Q_+A^{(0,2)}V_+^T, \\
\bar S^{(0,2)}&:=&XS^{(0,2)}Q_+^T, \\
\bar A^{(1)}&:=&Q_-A^{(1)}V_-^T, \\
\bar S^{(1)}&:=&XS^{(1)}Q_-^T,
\end{eqnarray}
where we have defined (with $\bar N=3\Leftrightarrow N=5$ serving as a
prototype)
\begin{eqnarray}
Q_+:&=&\left(\begin{array}{cccccc}
1 & 0 & & & \\
& 0 & 1 & 0 & \\
 & & & 0 & 1 \\
\end{array}\right), \\
Q_-:&=&\left(\begin{array}{cccccc}
0 & 1 & 0 & & \\
& & 0 & 1 & 0 \\
& & & & 0 \\
\end{array}\right), \\
X:&=&\left(\begin{array}{cccccc}
1 & & & & \\
& 1 & & & \\
& & 1 & 0 & 0 \\
\end{array}\right), \\
V_+:&=&\left(\begin{array}{cccccc}
1 & & & & 1\\
& 1 & & 1 & \\
& & 1 & &  \\
\end{array}\right), \\
V_-:&=&\left(\begin{array}{cccccc}
1 & & & & -1\\
& 1 & & -1 & \\
& & 0 & &  \\
\end{array}\right).
\end{eqnarray}
Note that $\bar S$ with $N=5$ or $\bar N=3$ is a $3\times3$ matrix,
etc.  These obey
\begin{eqnarray}
\bar A^{(0,2)}\bar S^{(0,2)}&=&I, \\
\bar A^{(1)}\bar S^{(1)}&=&I_0.
\end{eqnarray}
To understand this, consider again the case $\bar N=3$: we can
represent $P_l$ for $l=0,2,4$, and $P_l^{(2)}$ for $l=2,4,6$, but
$P_l^{(1)}$ only for $l=2,4$.  $Q_-$ has row rank $\bar N-1$, and
therefore $S^{(1)}$ and $A^{(1)}$ have rank $\bar N-1$.

We also define separate derivative matrices acting on even and odd
functions,
\begin{eqnarray}
D_\pm&:=&XDV_\pm^T, \\
D_\pm^2&:=&D_\mp D_\pm.
\end{eqnarray}
We have then the same identities as already discussed, again with the
proviso that $P^{(2)}_{2\bar N}=P^{(2)}_{N+1}$ can be represented but
must be suppressed for consistency.


\subsubsection{Tests at finite numerical precision}
\label{sec:lgltests}


We numerically calculate the synthesis, analysis and differentiation
matrices for selected values of $N$ in Mathematica. The collocation
points (zeros of the Legendre polynomials) need to be determined
numerically, and we do this with precision $10^{-50}$. The matrix
calculations are then carried out with the same precision. We then
save the resulting expressions for the collocation points $y_i$ and
the synthesis, analysis and differentiation matrices result into ascii
files in slightly more than double precision, and read this into the
F90 code at runtime. 

As as test of the numerical error of the pseudo-spectral method, we
have explicitly evaluated the following matrices, which all vanish in
the continuum, numerically in the F90 code (in double precision).
\begin{eqnarray}
T_0&:=& A^{(0)}S^{(0)}-I, \\
T_1&:=& A^{(1)}S^{(1)}-\begin{cases}I\\I_0
  \end{cases}, \\
T_2&:=& A^{(2)}S^{(2)}-I, \\
T_3&:=&\Delta^{(0)}_+S^{(0)}-S^{(0)}\Lambda^{(0)}, \\
T_4&:=&\Delta^{(1)}_-S^{(1)}-S^{(1)}\Lambda^{(1)}, \\
T_5&:=&\Delta^{(2)}_+S^{(2)}-S^{(2)}\Lambda^{(2)}, \\
T_6&:=& D_+S^{(0)}-S^{(1)}I_+, \\
T_7&:=& D_-S^{(1)}-S^{(2)}\begin{cases}I_+\\I_0
  \end{cases}, \\
T_8&:=& D_+^2S^{(0)}-S^{(2)}\begin{cases}I_+^2\\I_+
  \end{cases}, \\
T_9&:=& \slashed{\Delta}_+^{(1)}S^{(2)}\begin{cases}I_+\\I_0
  \end{cases} -S^{(1)}\Lambda^{(1)}.
\end{eqnarray}
The matrices $A$, $S$, $\Delta$ and $\Lambda$ are understood as either
half-range or full-range. Where there is a case distinction, the upper
case applies to the full-range setup and the lower case to the
half-range setup. In the full-range case $D_+=D_-:=D$, and similarly
for $\Delta$ and $\slashed{\Delta}$.

To avoid duplication of code, in the numerical code we use the
half-range notation throughout, and simply set both $D_+$ and $D_-$ to
$D$ in the full-range case, and similarly $D_+^2$ and $D_-^2$ to
$D^2$. By contrast, the combinations $\Delta$ and $\slashed{\Delta}$
do not explicitly appear in the nonlinear equations and so are not
stored as matrices, but are used here as shorthands for their
definitions.

We also check that each row of $D$ or $D_+$ adds up to zero, as the
differencing of a constant grid function should give zero. However, no
such test applies to $D_-$, which acts only on odd functions of
$y$. Hence we define another test
\begin{equation}
T_{10}:=D_+{\bf 1},
\end{equation}
where ${\bf 1}$ is the column vector with 1 in every row.

At $N=3$ or $\bar N=2$, the error in all tests is zero or at machine
precision. At all higher resolutions, the largest error occurs in
$T_5$. All errors at the two equivalent resolutions $N$ and
$\bar N=(N+1)/2$ are approximately the same, as we would expect.  At
$N=5$, $\bar N=3$, this error is $\simeq 10^{-13}$. At each doubling
of resolution, it increases by a factor of $\sim 100$, up to
$4\cdot 10^{-5}$ at $N=65$, $\bar N=33$, and $8\cdot 10^{-3}$ at
$N=129$, $\bar N=65$.

We believe that the observed error in these tests is essentially
round-off error in double precision computations in the F90 code, and
that it increases so rapidly with $N$ because the synthesis and
analysis matrices become increasingly ill-conditioned with both
resolution $N$ and spin $s$. 

We note that while the error in $T_0$ at $N=65$, evaluated within
Mathematica, is $10^{-49}$ as expected, in $T_1$ and $T_2$ it is
already $10^{-23}$. We can avoid this by noting that $P'_l$ is a
linear combination with integer coefficients of $P_{l-1}$, $P_{l-3}$
and so on. Hence we can write $S^{(1)}=S^{(0)}T^{(01)}$ and
$S^{(2)}=S^{(0)}T^{(02)}$, where $T^{(01)}$ and $T^{(02)}$ are lower
diagonal matrices with integer coefficients, and so can be inverted
exactly. $T_1$ and $T_2$ then have a much smaller internal error of
$10^{-44}$. The maximum difference between $A^{(1)}$ or $A^{(2)}$
obtained in the two ways is still only $10^{-24}$ at $N=65$, so they
are identical when reduced to double precision in Fortran
code. However, for $N=129$, Mathematica cannot find $A^{(1)}$ and
$A^{(2)}$ by direct matrix inversion, but we can still find them from
$A^{(0)}$ and the inverses of $T^{(01)}$ and $T^{(02)}$.


\subsubsection{High-frequency filtering}
\label{sec:filtering}


As in any pseudo-spectral code, nonlinear terms spuriously excite
high-$l$ modes through aliasing, and {without care} this eventually
leads to numerical instabilities, even if the linearised code is
stable. For high-frequency filtering of a grid function in real space,
we therefore multiply by $F:=S\hat FA$, where $\hat F$ is a diagonal
matrix where only the entries corresponding to $l\le l_\text{max}$
diagonal entries are one, and those for $l>l_\text{max}$ are zero. (A
smooth transition from one to zero would also be possible, but we have
not tried this).

When the Einstein equations are linearised around any spherically
symmetric solution, $\psi_l$, $f_l$, $b_l$, $R_l$ and $\gamma_l$
couple only for the same $l$. This suggests that for consistency we
truncate at the same $l_\text{max}$ for all variables, rather than the
same degree of polynomial in $y$, using the appropriate analysis and
synthesis matrices. Define $\hat F(k)$ to be the
diagonal matrix with 1 in its first $0\le k\le N$ entries, and 0 in
the remaining ones. We then filter the scalars with $\bar F^{(0)}:=S^{(0)}\hat
F(l_\text{max}+1)A^{(0)}$, the variable $b$ with $\bar F^{(1)}:=S^{(1)}\hat
F(l_\text{max})A^{(1)}$, and the variable $f$ with $\bar F^{(2)}:=S^{(2)}\hat
F(l_\text{max}-1)A^{(2)}$. These matrices have therefore different rank.


\subsection {Finite differencing in $x$ of the hierarchy equations}


In this subsection, we write $N$ for $N_x$. As we have seen, the
hierarchy equations take the form
\begin{equation}
\label{dphiconsdx}
\phi_{{\rm int},x}=S(y,\phi,\phi_{,x},\phi_{,y},\phi_{,xy},\phi_{,yy})
\end{equation}
in terms of the intermediate quantities
\begin{equation}
\label{phiconstildedef}
\phi_{\rm int}:=\left(\gamma, R^4e^{2S f-\gamma}{\cal D}
b,b,R\Xi R,R\Xi f,R\Xi\psi\right),
\end{equation}
and the basic variables (metric coefficients and matter field)
\begin{equation}
\label{phidef}
\phi:=(\gamma,R, b,f,\psi).
\end{equation}
These equations can be solved by integration as
\begin{equation}
\label{phiintegration}
\phi_{\rm int}(u,x,y)=\phi_{\rm int}(u,0,y)+\int_0^xS(u,x',y)\,dx',
\end{equation}
for the intermediate quantities, and hence the constrained variables
\begin{equation}
\label{phiconsdef}
\phi_{\rm cons}:=(\gamma,b,\Xi R,\Xi f,\Xi\psi),
\end{equation}
in this order. The evolved variables
\begin{equation}
\label{phievol}
\phi_{\rm evol}:=(f,R,\psi)
\end{equation}
are specified freely at $u=0$ and evolved in $u$ using their
$\Xi$-derivatives.

In double-null gauge no influence propagates from larger to smaller
$x$. In a general gauge, this happens onlythrough the shift term
$B\partial_x$. This suggests to us that the hierarchy equations should
be solved by an integration scheme in $x$ that does not evaluate the
integrand to the right of the point where the integral is being
approximated. Then for $H\le 0$, strictly no information flows to
larger $x$. In the following, we present a simple second-order
accurate scheme that has these properties.

We use a grid equally spaced in $x$, $x_i=i\Delta x$, with $x_0=0$ and
$x_{N}=x_\text{max}$, so $\Delta x=x_\text{max}/N$. In the code, array
storage for fields at the mid-point $x_{i+1/2}$ is labelled by the
array index $i$, that is by the grid point to its left. We do not
store field values at the first grid point $x_0=0$ and first mid-point
$x_{1/2}=\Delta x/2$. Therefore arrays representing grid points have
array index ranging from $1$ to $N$, and arrays representing
mid-points from $1$ to $N-1$, corresponding to $i=3/2$ to
$N-1/2$. With this convention in mind, we define the shorthands
\begin{eqnarray}
\Delta_i \phi&:=&\phi_{i+1}-\phi_i, \\
\bar \phi_i&:=&{\phi_{i+1}+\phi_i\over 2}.
\end{eqnarray}
Here we write only the index $i$ corresponding to the coordinate $x$,
but not the index $j$ corresponding to the coordinate $y$.

For sufficiently smooth functions $\phi$ we then have
\begin{eqnarray}
\label{dfdxmidapprox}
(\phi_{,x})_{i+1/2}&=&{\Delta_i \phi\over \Delta x}+O(\Delta x^2), \\
\label{fmidapprox}
\phi_{i+1/2}&=&\bar \phi_i+O(\Delta x^2).
\end{eqnarray}
We then discretize the integrations (\ref{phiintegration}) with the
midpoint rule, that is
\begin{equation}
\label{discretehierarchy}
\phi_{{\rm int},i+1}=\phi_{{\rm int},i}+S_{i+{1\over 2}}\Delta x,
\end{equation}
using the discretisations (\ref{dfdxmidapprox},\ref{fmidapprox}), the
discretizations of $\phi_{,y}$ and $\phi_{,yy}$ using $D$ and $D^2$,
and the discretization of $\phi_{,xy}$ and  $\phi_{,xyy}$ resulting from
combining (\ref{dfdxmidapprox}) with $D$. The resulting $\phi_{{\rm
    int},i+1}$ is second-order accurate but depends only on
$\phi_{i+1}$ and $\phi_i$, thus respecting causality. By contrast, the
trapezoid rule would require left derivatives at the right-hand
gridpoint to respect causality. 

In the formulation where we use ${\cal D}$ defined in (\ref{calDdef})
rather than $\partial_x$, we differentiate
\begin{equation}
\label{dfdRmidapprox}
({\cal D} \phi)_{i+1/2}\simeq{(\phi_{,x})_{i+1/2}\over
  (R_{,x})_{i+1/2}} = {\Delta_i \phi\over \Delta_i R}
\end{equation}
and integrate 
\begin{equation}
\label{discretehierarchyR}
\phi_{{\rm int},i+1}=\phi_{{\rm int},i}+\bar S_{i+{1\over 2}}\Delta_i
R. 
\end{equation}


\subsection{Shift terms}


For the evolved variables (\ref{phievol}), the definition of $\Xi$
gives
\begin{equation}
\label{dphidu}
\phi_{{\rm evol},u}=\Xi \phi_{\rm evol}+B\phi_{{\rm evol},x}+Sb\phi_{{\rm evol},y}
\end{equation}
The analogy of $B$ with the $x$-component of a shift vector suggests
that, in contrast to the $x$-derivatives in the right-hand side of
(\ref{dphiconsdx}), $\phi_{{\rm evol} ,x}$ in (\ref{dphidu}) should be
upwinded: as right derivatives $\phi_{,x}^+$ for $B>0$, and left
derivatives $\phi_{,x}^-$ for $B<0$. In particular, no numerical
boundary condition is then required at an outer boundary $x=x_{\rm
  max}$ as long as $B\le 0$ there, or at the innner boundary where
$B=B_\text{Bondi}>0$. We do not upwind the shift term in the $y$
direction, as we take all $y$-derivatives spectrally. 

The upwind $x$-derivatives are evaluated with the second-order accurate
three-point formulas on grid points, namely
\begin{eqnarray}
\label{phixplus}
(\phi_{,x})^+_{i}&:=&{2\phi_{i+1}-{3\over 2}\phi_{i}-{1\over 2}\phi_{i+2}\over \Delta
    x}, \\
\label{phixminus}
(\phi_{,x})^-_{i}&:=&-{2\phi_{i-1}-{3\over 2}\phi_{i}-{1\over 2}\phi_{i-2}\over \Delta
    x}.
\end{eqnarray}

Where $R_{,x}(u,{\bf 0})$ is needed, we use $R_{0,j}=0$ in
(\ref{phixplus}) to evaluate the right derivative as
\begin{equation}
(R_{,x})^+_{0,j}={2R_{1,j}-{1\over 2}R_{2,j}\over \Delta x}, \\
\end{equation}
and then average over $y$ to obtain $R_{,x}(u,{\bf 0})$. 


\subsection{The time step problem, and a resolution}


An important, if somewhat vaguely defined, necessary condition for
numerical stability of any explicit time-evolution scheme is the
Courant-Friedrichs-Lewy (from now, CFL) condition. This says that the
outermost characteristic cone, in our case the light cone, is
contained in the spacetime numerical stencil. We consider this as a
heuristic guide to a stability limit on the time step, without
carrying out an actual discrete stability analysis. Unusually, in
spherical symmetry, in double-null coordinates $(u,v)$, this causality
condition does not impose any restriction on the time step $\Delta
u$. However, and not surprisingly, we found in \cite{ymscalar} that
even then a limit $\Delta u\sim \Delta x$ is required for
stability. Beyond spherical symmetry, however, the combination of
spherical polar coordinates and null coordinates imposes a severe
restriction, and we need to address this problem in order to make our
code efficient.

Explicit methods for hyperbolic problems using spacelike time slices
and Cartesian coordinates typically have a time step condition $\Delta
t\sim \Delta x$, where $\Delta x$ is the grid spacing of the Cartesian
spatial coordinates. By contrast, polar spatial coordinates on
spacelike slices give rise to a time step condition of $\Delta t\sim
\Delta r\Delta\theta$, which comes from evaluating the CFL condition
in the tangential direction, $\Delta t\sim r\Delta\theta$, near the
centre, that is at $r\sim \Delta r$. In Appendix
\ref{appendix:timestepproblem}, we show that in polar coordinates on
null cones, the stability criterion is an even worse $\Delta u\sim
\Delta r\Delta\theta^2$.

This is a problem not only when we use finite differencing in the
angle $\theta$. When we split the linearised wave equation into
spherical harmonics as in (\ref{psilexpansion}) and finite-difference
in $r$ for given $l$, we find empirically that the time step limit for
evolving $\psi_l(u,r)$ is $\Delta u\sim l^{-2} \Delta r$. There is no
differentiation in $y$ to give rise to a CFL condition in the
tangential direction, but instead the wave equation decomposed into
spherical harmonics now contains an $l(l+1)/r^2$ potential. It turns
out that this requires the same restriction on the time stesp as if we
had used finite differencing in $\theta$, with $\Delta\theta\sim
1/l_\text{max}$.

In the context of the wave equation on Minkowski spacetime, the
$l(l+1)/r^2$ barrier can be transformed into a $2(l+1)/r$ barrier for
a first-order reduction of the wave equation, and this was made
stable for $\Delta t\sim\Delta r$ by the introduction of a suitable
summation by parts (SBP) differencing scheme in $r$ \cite{SBP}. We
could not see how to do this in null coordinates, even for the
flat-space scalar wave equation. 

However, the spectral method in $y$ that we use suggests a possible
remedy to the time step restrictions in polar coordinates, both on
null slices and on the usual spacelike slices: we simply filter out
all spatial frequencies above $l\sim i$, where $l$ is the spherical
harmonic index, and $i$ the grid index in the radial coordinate
$x$. In other words, at radius $x$ only spherical harmonics up to
$l\sim x/(\Delta x)$ are represented. We discuss how this is done in
the code in the next subsection.

An intuitive understanding of why this boundary condition removes the
extra restrictions on the time step due to spherical polar coordinates
and null coordinates is that, with $\Delta\theta \sim
1/l_\text{max}\sim 1/i\sim \Delta x/x$ near
the centre, we now have an effective angular resolution such that
$x\Delta\theta\simeq \Delta x$, so that the effective ``grid cells''
have roughly equal sides, giving us the stability benefit of a
Cartesian grid. 

Outside the central region, for $i\gtrsim l_{\rm max}$, the angular
resolution is constant, giving us the physical benefits of a spherical
grid, namely efficient resolution of ingoing and outgoing waves of
finite $l$ and an outer boundary with spherical topology.

A similar filtering approach to overcoming the stricter CFL limit in
spherical polar coordinates has been presented on spacelike times
lices in \cite{Jietal2023}. Here the filtering uses fast Fourier
transforms in $\theta$ and $\varphi$.

These adjustments allow us to run the axisymmetric code with the time
step
\begin{equation}
\label{timestep}
\Delta u = \min_{i,j} \min(C_1\Delta u_1, C_2\Delta u_2),
\end{equation}
where we have defined the local time step criteria
\begin{eqnarray}
\label{Deltau1def}
\Delta u_1(u,x,y)&:=&\Delta x
\left|{R_{,x}\over \Xi R}\right|, \\
\label{Deltau2def}
\Delta u_2(u,x,y)&:=&\Delta x\left|B\right|^{-1}.
\end{eqnarray}
The parameters $C_1$ and $C_2$ are independent of $\Delta x$ and
$\Delta y$ (or $l_{\text max}$), and are of order one. As $B$ is the
shift in the $x$-direction, $\Delta u\le \Delta u_2$ is the standard
limit on the time step for any explicit finite differencing of an
advection equation. In double-null gauge, $B=0$, and so this criterion
is empty, but empirically the limit $\Delta u\le \Delta u_1$ on the
time step is required even then. The quantity $|R_{,x}/\Xi R|$ can be
understood in two ways: as $|R_{,x}|/|R_{,u}|$ in double-null gauge,
implying that $R$ always changes less per time step than per grid
point, or as the value taken by $B$ in Bondi gauge. Empirically, we
find that this term guarantees stability in double null and Bondi
gauges and their variants. Because of our gradual suppression of high
angular frequencies near the centre this holds independently of
$l_\text{max}$.


\subsection{Treatment of the central region}


\subsubsection{High-frequency filtering after each time step}


In order to allow for a time step $\Delta u\sim \Delta x$ in the way
just discussed, in the initial data and after each full time step we
apply a filter to the evolved variables $R$, $f$ and $\psi$ as
discussed in Sec.~\ref{sec:filtering}, but with a local $l_\text{max}$.
This sets
\begin{equation}
\phi_{\text{evol},l}=0 \text{ for } l>l_\text{max,local}(i)
\end{equation}
where 
\begin{equation}
l_\text{max,local}(i):=\min(\max(2,2i-2),l_\text{max,global}),
\end{equation}
which equals $2,2,4,6,...$ for $i=1,2,3,4...$
This filtering means that at $i=1,2$, only $l=0,1,2$ are present. At $i=3$,
$l=0,1,2,3,4$ are present, at $i=4$, $l=0...6$, and so on.
$l_\text{max,global}$ is even to treat the full and half-range
discretisation equally. We use the relevant analysis and synthesis
matrices for the scalars, $b$ and $f$, respectively.

The filter always removes at least the top two $l$-modes of $f$ on
the full-range grid, or the top mode on the half-range grid, as these
do not have a counterpart in $\gamma$, $H$, $R$, $\psi$ and $b$. This
means that $l_\text{max,global}\le N_y-1$ on the full $y$-range and
$2(\bar N_y-1)$ on the half-range. (We always choose $N_y$ and
$\bar N_y$ to be odd.)


\subsubsection{Expansion of the field equations about the origin}
\label{section:originexpansion}


The filtering near the origin that gets round the bad CFL condition de
facto imposes unphysical numerical boundary conditions, for example
$\psi_l(u,x_l)=0$, at some $x_l\sim l\Delta x$. This is compatible
with second (or higher) order numerical accuracy in $\Delta x$ for
large $l$, but not for the smallest $l$. For example, in a regular
continuum solution $\psi_l(u,x)\sim x^l$ near the centre, so imposing
$\psi_l(u,x_l)=0$ imposes an error of $O(\Delta x^l)$. If we want the
code to be second-order accurate, this is acceptable for $l\ge 2$, but
for $l=0$ and $l=1$ we need to impose a more accurate, nonzero,
boundary condition.

In the code, we impose boundary conditions at
an inner boundary (including an unphysical one) as integration
constants when we solve the hierarchy equations by integration in $x$,
for example the constant $c_l$ in
$(R\Xi\psi)_l(u,x)=c_l+\int_{x_l}^x...dx'$. We now obtain the values
of those integration constants that correspond to a regular centre (to
a given order of accuracy) by expanding the full hierarchy equations
in powers of $x$.

We shall see that for second-order accuracy we need nonzero
integration constants only for the spherical harmonic components
$b_{1,2}$, $(\Xi R)_{0,1,2}$, $(\Xi f)_2$ and $(\Xi\psi)_{0,1,2}$
(where the suffix denotes the value of $l$). The general argument
above applies also to $\gamma_{0,1,2}$ but their integration constants
turn out to be zero.

We assume that in a regular axisymmetric solution of the
Einstein-scalar equations, the scalar field $\psi(u,x,y)$ admits a
convergent expansion around the origin $x=0$ of the form
\begin{eqnarray}
\label{psiexpansion1}
\psi&=&\sum_{l=0}^\infty \sum_{k=l}^\infty
\psi_{l(k)}(u)\,x^k\,P_l(y) \\
\label{psiexpansion2}
&=&\sum_{k=0}^\infty \sum_{l=0}^k \dots,
\end{eqnarray}
where the re-ordering obviously requires convergence. The first suffix
in $\psi_{l(k)}$ denotes the spherical harmonic and the second one the
power of $x$.  Appendix~\ref{appendix:regularity} shows that this
[together with analyticity of the $\psi_{l(k)}(u)$] corresponds to
analyticity in suitable Cartesian coordinates $(t,\xi,\eta,z)$.

Expanding any hierarchy equation $F=0$ as
(\ref{psiexpansion2}), and truncating this expansion as
\begin{equation}
F\simeq \sum_{k=0}^{k_\text{max}} \sum_{l=0}^k F_{l(k)}(u)\,x^k\,P_l(y)
\end{equation}
the result is a polynomial of finite order $k_\text{max}$ in both $x$
and $y$. This observation guarantees that when, order by order in $x$,
we set the coefficients of all nonvanishing powers of $y$ to zero
separately to obtain a system of algebraic equations for the
$\psi_{l(k)}$, the expansion remains exact in $y$.

A priori, we expand $\gamma$ and $R$ in the same way as
$\psi$. However, we impose $\gamma(u,{\bf 0})=\gamma_{0(0)}=0$ in
order to make $u$ proper time at the origin, and so
\begin{equation}
\gamma= \sum_{k=1}^\infty \sum_{l=0}^k \gamma_{l(k)}(u)\,x^k\,P_l(y).
\end{equation}
We additionally impose $R(u,{\bf 0})=R_{0(0)}=0$ to locate the origin
$R=0$ at $x=0$ and $R_{1(1)}=0$ to make $R_{,x}$ single-valued at the
origin. Moreover, we show in Appendix~\ref{appendix:regularity} that
regularity requires $R_{l(l)}=0$ for all $l$, so that
\begin{equation}
\label{Rexpansion}
R =\sum_{k=1}^\infty \sum_{l=0}^{k-1}R_{l(k)}(u)\,x^k\,P_l(y).
\end{equation}
Generalising from the behaviour of analytic solutions to the Einstein
equations linearised about Minkowski spacetime (see
Appendix~\ref{appendix:gwtoscalar}), and consistent with the
regularity requirements in Appendix~\ref{appendix:regularity} we
expand
\begin{eqnarray}
f&=&\sum_{k=2}^\infty \sum_{l=2}^k f_{l(k)}(u)\,x^k\,P''_l(y), \\
b&=&\sum_{k=0}^\infty \sum_{l=1}^{k+1} b_{l(k)}(u)\,x^k\,P'_l(y).
\end{eqnarray}
Here $b_{1(0)}(u)$ is free, corresponding to a gauge choice, see
Appendix~\ref{appendix:residualgaugefreedom}.  Finally, from
$\Xi=\partial_u-B\partial x+...$ we expect that the expansions of
$\Xi$-derivatives start at one power of $x$ lower, that is
\begin{eqnarray}
\Xi R&=&\sum_{k=0}^\infty \sum_{l=0}^{k+1}(\Xi
R)_{l(k)}(u)\,x^k\,P_l(y),\\
\Xi f&=&\sum_{k=1}^\infty \sum_{l=2}^{k+1}(\Xi
f)_{l(k)}(u)\,x^k\,P''_l(y), \\
\Xi \psi&=&\sum_{k=0}^\infty \sum_{l=0}^{k+1}(\Xi \psi)_{l(k)}(u)\,x^k\,P_l(y).
\end{eqnarray}

As a test of consistency, we have explicitly expanded all fields to
$O(x^5)$. This allows us to consistently expand the hierarchy equation
for $\gamma$ to $O(x^2)$, for $b$ to $O(x^5)$ and for $\Xi R$, $\Xi f$
and $\Xi\psi$ to $O(x^3)$, and the resulting coefficient equations can
be solved for $(\gamma,b,\Xi R,\Xi f,\Xi\psi)$ to $O(x^3)$.

In the code, we only need the expansions to $O(x)$, as the error of
$O(x^2)$ corresponds to $O(\Delta x)^2$ for the innermost few grid
points. The nonvanishing terms then involve only spherical harmonics
up to $l=2$, and are
\begin{eqnarray}
\gamma&=& O(x^2), \\
b&=&\Biggl[b_{1(0)}+\Bigl( -4\pi
R_{0(1)}^{-1}\psi_{0(1)}\psi_{1(1)}-R_{0(1)}^{-2}R_{1(3)}\nonumber \\ 
&& +2R_{0(1)}^{-3}R_{0(2)}R_{1(2)} \Bigr)\,x\Biggr]P'_1(y) \nonumber \\
&&+\Biggl[-4R_{0(1)}^{-1}\left(f_{2(2)}+{\pi\over
  3}\psi_{1(1)}^2\right) -R_{0(1)}^{-2}R_{2(3)} \nonumber \\
&& +{2\over3}R_{0(1)}^{-3}R_{1(2)}^2\Biggr]\,x\,P'_2(y) + O(x^2), \\
\Xi R&=&-{1\over 2}P_0(y) \nonumber \\
&&+\left(-R_{0(1)}b_{1(0)}-R_{0(1)}^{-1}R_{1(2)}\right)
\,x\, P_1(y) \nonumber \\
&&+O(x^2), \\
\Xi f&=&-R_{0(1)}^{-1}f_{2(2)}\,x\,P_2''(y)+O(x^2), \\
\Xi\psi&=&\Biggl[{1\over 2}R_{0(1)}^{-1}\psi_{0(1)} 
+{1\over 2}\Bigl(R_{0(1)}^{-1}\psi_{0(2)} \nonumber \\
&&+R_{0(1)}^{-2}\left(R_{1(2)}\psi_{1(1)}-R_{0(2)}\psi_{0(1)}\right)\Bigr)\,x
\Biggr]P_0(y)\nonumber \\
&&-{1\over 2}R_{0(1)}^{-1}\psi_{1(1)}\,P_1(y) \nonumber \\
&&-R_{0(1)}^{-1}\psi_{2(2)}\,x\,P_2(y)+O(x^2).
\end{eqnarray}
Note that $P_0(y)=1$, $P_1(y)=y$, $P_2(y)=(3y^2-1)/2$, $P'_1(y)=1$,
$P'_2(y)=3y$, $P''_2(y)=3$ but we have not substituted these values
for clarity of exposition.

Assuming both $b_{10}=0$ (the origin is geodesic) and $R_{,y}=0$ (lsB
gauge), as we do here and in Paper~II, these equations simplify to 
\begin{eqnarray}
\gamma&=& O(x^2), \\
b&=&-4\pi
R_{0(1)}^{-1}\psi_{0(1)}\psi_{1(1)}\,x\,P'_1(y) \nonumber \\
&&-4R_{0(1)}^{-1}\left(f_{2(2)}+{\pi\over
  3}\psi_{1(1)}^2\right) \,x\,P'_2(y) \nonumber \\
&&+ O(x^2), \\
\Xi R&=&-{1\over 2}P_0(y)+O(x^2), \\
\Xi f&=&-R_{0(1)}^{-1}f_{2(2)}\,x\,P_2''(y)+O(x^2), \\
\Xi\psi&=&\Biggl[{1\over 2}R_{0(1)}^{-1}\psi_{0(1)} 
+{1\over 2}\Bigl(R_{0(1)}^{-1}\psi_{0(2)} \nonumber \\
&&-R_{0(1)}^{-2}R_{0(2)}\psi_{0(1)}\Bigr)\,x
\Biggr]P_0(y)\nonumber \\
&&-{1\over 2}R_{0(1)}^{-1}\psi_{1(1)}\,P_1(y) \nonumber \\
&&-R_{0(1)}^{-1}\psi_{2(2)}\,x\,P_2(y)+O(x^2).
\end{eqnarray}

This means that we need to fit only the following coefficients from
the evolved variables $R$, $f$ and $\psi$: $R_{0(1)}$, $R_{0(2)}$,
$f_{2(2)}$, $\psi_{0(0)}$, $\psi_{0(1)}$, $\psi_{0(2)}$, $\psi_{1(1)}$
and $\psi_{2(2)}$. $\psi_{0(0)}$ must be fitted for consistency but is
not used.

We have implemented both a least-squares fit of, say, $\psi_l$ to
$ax^l+bx^{l+1}$ (direct fit), and a least-squares fit of $\psi_l/x^l$
to $a+bx$ (linear fit), and similarly for $f_l$ and $R_l$. The direct
fit weights grid points by $x^l$ relative to the linear fit. In each
case we fit to the first $n_\text{fit}$ points. We obtain $R_{0(1)}$
as the value of the left difference at $x=0$, as this term will have
to cancel the equivalent transport term, and we then fit to
$R-R_{0(1)}(u)x$ with the general method. We choose to also fit
$R_{0(3)}$, $f_{2(3)}$, $\psi_{1(2)}$ and $\psi_{2(3)}$, which are not
used, in order to fit two powers of $x$ to each function. The exception
is that we fit three powers to $\psi_0$, which then requires $n_\text{fit}\ge 3$.

If we restrict to linear perturbations about Minkowski spacetime in
Bondi gauge, with $R=R_{0(1)}(u)\,x$ exactly in the background
solution, we are dropping all other $R_l(k)$ and all products of
expansion coefficients. In an early version of the code, we did that,
and also truncated the expansions at different orders from the above,
resulting in the following expansion:
\begin{eqnarray}
\label{oldexpansiongamma}
\gamma_{0,1,2}&=& O(x^2), \\
b_1&=& O(x^2), \\
{R^4b_{2,x}\over G}&=&-4R_{0(1)}^2\left(f_{2(2)}x^4+{6\over
    5}f_{2(3)}x^5\right), \nonumber \\ \\
b_2&=&-4R_{0(1)}^{-1}\left(f_{2(2)}x+{3\over
    5}f_{2(3)}x^2\right), \nonumber \\ \\
(R\Xi R)_0&=&-{1\over 2}R, \\
(R\Xi f)_2&=&-f_{2(2)}x^2-{3\over 10}f_{2(3)}x^3, \\
(R\Xi\psi)_0&=&{1\over 2}\psi_{0(1)}x, \\
(R\Xi\psi)_1&=&-{1\over 2}\psi_{1(1)}x, \\
\label{oldexpansionpsi}
(R\Xi\psi)_2&=&-\psi_{2(2)}x^2,
\end{eqnarray}
with all other components initialised to zero. The almost-linear
simulations presented here were carried out with this expansion, but
we have checked since that the fully nonlinear expansion makes only a
very small difference to the error we have measured in
convergence tests. In particular, the magnitude and qualitative
behaviour of the error is unchanged.


\subsubsection{Integration of the hierarchy equations}


We initialise the integrals for all hierarchy equations up to
$i=i_\text{expand}\ge 1$, and integrate from there. We truncate the
integrand to $l_\text{max,local}(i)$ when integrating from $x_{i-1}$
to $x_i$. We then use $\Xi R$ to find the $x$-shift $B$. This allows
us to calculate the upwinded $x$-derivatives of the evolved variables
$R$, $f$ and $\psi$, as these depend on the local sign of $B$.

In the integrals for $R\Xi f$ and $R\Xi\psi$ we truncate not the
integrand but the whole integral to $l\le l_\text{max,local}(i)$. We
also set
\begin{equation}
\label{RXipsistartup}
(R\Xi \psi)_l=-\left(R B\psi_{,x}\right)_{\text{upwind},l},
\qquad l>l_\text{max,local}(i).
\end{equation}
With (\ref{dphidu}), this gives $(R\psi_{,u})_l=0$ for
$l>l_\text{max,local}(i)$. As $R$ does not depend on $y$ in lsB and
gsB gauges, this then also sets $(\psi_{,u})_l=0$, consistent with the
boundary condition $\psi_l=0$ that we impose on
$l>l_\text{max,local}(i)$. We treat $R\Xi f$ similarly.


\subsection{Time evolution}


We set initial data for $f$ and $\psi$ on $u=0$,
$0\le x\le x_\text{max}$.  In gauges other than affine gauge we must
also initialise $R$, and in these gauges we think of this
initialisation as pure gauge. We choose
\begin{equation}
\label{myRinit}
R(0,x,y)={x/2},
\end{equation}
in analogy with $R=(v+u)/2$ in the standard double null coordinates
on Minkowski spacetime. 

After solving all hierarchy equations, and applying the shift terms in
$\Xi$, the resulting ``time'' derivatives $\phi_{{\rm evol},u}$ are
discretized using the second-order Runge-Kutta method, with all
hierarchy equations and gauge conditions evaluated at each Runge-Kutta
sub-step, so that our time update can be characterised as the ``method
of lines''.

In double-null or sdn gauges, $R(u,x,y)$ is genuinely evolved, but in
Bondi gauge $R=x$, and only $f$ and $\psi$ are evolved. In gsB gauge
we have $R(u,x,y)=s(u)x$. Numerically, we evolve only $s(u)$ (as an
auxiliary variable). In lsB gauge we have $R(u,x,y)=\bar
R(u,x)$. Numerically, we evolve $R(u,x,y)$ but filter out the $l>0$
components that are created by numerical error after each full time
step.


\section{Tests in the almost-linear regime}
\label{sec:almostlin}


We test convergence of the full nonlinear code in a regime of small
deviations from Minkowski spacetime (with $\psi=0$). We evolve
in several of the nonlinear gauge choices we have discussed
above. Specifically, we compare sdn gauge (\ref{sdnshift2}), gsB gauge
(\ref{gsBshift}) with $s(u)$ given by (\ref{gsBx0min}), lsB2 gauge
(\ref{lsb-XiRmeanminx0}), and lsBtosdn gauge (\ref{lsBtosdn}), each
using the $(G,x)$ formulation and the $(\gamma,R)$ formulation, the
latter with and without the integration by parts (\ref{barSbbyparts}).
All tests use direct fits near the origin with $n_\text{fit}=2$ and
$n_\text{expand}=1$ and the expansion
(\ref{oldexpansiongamma}-\ref{oldexpansionpsi}). 


\subsection{Linearised solutions as testbeds}


In the small-data regime we can use exact solutions of the {\em
  linearised} field equations as testbeds. Small data here means in
practice that the difference between the solutions of the linearised
and nonlinear field equations can be neglected in comparison with the
numerical error in the nonlinear evolution.

For clarity, in this subsection we denote the exact solutions of the
linearised equations by $\delta\psi$, $\delta f$, $\delta b$,
$\delta R$. These will then be good approximations to nonlinear but
small $\psi$, $f$ and $b$, and a small nonspherical part of $R$.

In the linearised equations, $\delta\psi$ on the one hand, and
$\delta f$, $\delta b $ and $\delta R$ on the other, evolve
independently, while different spherical harmonic components $l$ also
decouple from each other.

We write the Minkowski background $f=b=\psi=0$ in a gauge which agrees
with all of our gauge choices. The spherical background metric
coefficients $G$, $R$ and $H$ in this gauge are given by
Eqs.~(\ref{G0exact}-\ref{H0exact}) in Appendix~\ref{appendix:Minkowski}.

The quantities $\delta b$ and $\delta R$ for the same physical
solution disagree in different gauges. By contrast $\delta\psi$ is
linearly gauge-invariant, and $\delta f$ is linearly gauge-invariant
within the class of lsB and gsB gauges; see also
Appendix~\ref{appendix:otherlineargauges}. This means that the
linearised solution in Bondi gauge also gives us $\delta\psi$ in any
other gauge, and $\delta f$ in any lsB or gsB gauge. In these gauges,
$\delta R=0$ (in the linearised equations). By contrast, in sdn gauge
$\delta R$ develops dynamically even if it is set to zero in the
initial data.


\subsection{Convergence test method}


To look for second-order self-convergence of a variable
  $\phi$ with respect to $\Delta x$, we
assume that the Richardson expansion
\begin{equation}
\phi_{\Delta x}(x)=\phi_0(x)+\phi_2(x)\, \Delta x^2+O(\Delta x^3)
\end{equation}
holds, where we have suppressed the other arguments of
$\phi$. $\phi_0(x)$ is the solution in the continuum limit in $x$,
and $\phi_2(x)$ is the second-order error, assumed to be leading.

We now distinguish two cases. If $\phi(u,x,y)$ obeys a hypersurface
equation (PDE in $x$ and $y$ only), then $u$ is just a parameter for
the purposes of the convergence test. At fixed finite resolution in
$y$ we can then think of the PDE as a (large) system of ODEs in $x$.

If, on the other hand, $\phi$ obeys an evolution equation (PDE in $u$,
$x$ and $y$), then at fixed finite resolution in $y$ we can consider
it as a (large) system of PDEs in $u$ and $x$. Our time step criterion
(\ref{timestep}) scales $\Delta u$ in proportion to $\Delta x$. If the
discretisation in $u$ is also at least second-order accurate, then we
expect that the discretisation errors in both $u$ and $x$ are
proportional to $\Delta x^2$, so we are testing convergence in $u$ and
$x$ together.

In halving $\Delta x$ exactly, $\Delta u$ is halved approximately, but
not exactly, by the application of the time step condition
(\ref{timestep}). To compensate for this, we align the output times at
both resolutions exactly by adjusting the last time step coming up to
the scheduled output time.

From pairs of numerical evolutions, we calculate the self-convergence
error estimate
\begin{eqnarray}
\label{calEdef}
{\cal E}_{\phi,\Delta x}&:=&{\phi_{\Delta x}-\phi_{\Delta x/2}
\over \Delta x^2-(\Delta x/2)^2}\, \Delta x_\text{ref}^2 \\
&=&{3\over 4}\left({\Delta x_\text{ref}\over \Delta x}\right)^2
\left[\phi_{\Delta x}-\phi_{\Delta x/2}\right] \\
&\simeq& \Delta x_\text{ref}^2\,\phi_2.
\end{eqnarray}
Any pair of resolutions could be used to estimate the error, but for
simplicity we use $\Delta x$ and $\Delta x/2$, as the coarse grid is
then aligned with the fine grid, so we can evaluate the difference on
the coarse grid without interpolation.
If we know the continuum solution $\phi_0$, we can also compute
the alternative error estimate
\begin{eqnarray}
\label{calEdefbis}
{\cal E}_{\phi,\Delta x}&:=&\left({\Delta x_\text{ref}\over \Delta x}\right)^2
\left[\phi_{\Delta x}-\phi_0\right] \\
&\simeq& \Delta x_\text{ref}^2\,\phi_2.
\end{eqnarray}

Note that ${\cal E}_{\phi,\Delta x}$ depends on the reference
resolution $\Delta x_\text{ref}$, the fixed resolution $\Delta y$ (or
$l_\text{max}$), and $(u,x,y)$, but for brevity we do not write these
arguments. If ${\cal E}_{\phi,\Delta x}$ calculated for two or more
pairs of resolutions is similar at all $\Delta x$ below some
threshold, we have pointwise second-order convergence in $\Delta x$,
and ${\cal E}_{\phi,\Delta x}$ itself is approximately equal,
pointwise, to the discretisation error in $x$, or in $u$ and $x$, at
the reference resolution $\Delta x_\text{ref}$ and the fixed
resolution $\Delta y$. The error at any other (smaller) $\Delta x$ can
be estimated by scaling with $(\Delta x/\Delta
x_\text{ref})^2$.

In this paper we do not yet carry out systematic convergence testing
in $y$, as we expect different spherical harmonics to decouple in
almost-linear evolutions. Hence the error in $\phi_l(u,x)$ becomes
negligible once $l_\text{max}>l$, and otherwise $\phi_l(u,x)$ cannot
be represented at all. However, for completeness, looking ahead to
Paper~II, we discuss testing convergence in $y$ already here.

On a smooth nonlinear solution, a spectral method should converge
exponentially, but we shall see in Paper~II that our code converges
only to second order in $\Delta y$. Hence, to look for second-order
convergence with respect to $\Delta y\propto 1/l_\text{max}\propto
1/N_y$, we assume that the Richardson expansion
\begin{equation}
\phi_{l_\text{max}}(y)=\phi_0(y)+\phi_2(y)\,l_\text{max}^{-2} 
+O(l_\text{max}^{-3})
\end{equation}
holds, where we have again suppressed the other arguments of $\phi$
and are keeping the resolution in them fixed. We can then consider the
PDE as a large system of ODEs in $y$. $\phi_0(y)$ is the solution
in the continuum limit in $y$ (but at fixed finite resolution in $x$
and $u$), and $\phi_2(y)$ is the second-order error, assumed to be
leading. From pairs of numerical evolutions, we then calculate the
quantity
\begin{eqnarray}
\label{calEdef}
{\cal E}_{\phi,l_\text{max1}}(x,y)&:=&
{\phi_{l_\text{max1}}-\phi_{l_\text{max2}}
\over l_\text{max1}^{-2}-l_\text{max2}^{-2}}\,l_\text{maxref}^{-2} \\
&\simeq& l_\text{maxref}^{-2}\, \phi_2.
\end{eqnarray}
We can compare $\phi_l(u,x)$ at different $l_\text{max}$, and so do
not need to align $y$-grids at different resolutions.

For code checks, it is often useful to plot single Fourier components
${\cal E}_{\phi_l,\Delta x}(x,u)$ of the error against $x$, and
animate these plots with time $u$. In
Figs.~\ref{fig:psi02convergence}-\ref{fig:b02convergence}, for data
dominated by a single spherical harmonic, we show such single-$l$
errors against $x$, at a representative moment of time $u$. In
Fig.~\ref{fig:planewave_convergence}, for data containing all
spherical harmonics, we instead take the root-mean-square (from now,
rms) norm of ${\cal E}_{\phi,\Delta x}(x,u,y)$ over
$0\le x\le x_\text{max}$ and $-1\le y\le 1$, and plot this against
$u$. We also evaluate the maximum norm of
${\cal E}_{\phi_l,\Delta x}(x,u)$ over $x$, and the maximum norm of
${\cal E}_{\phi,\Delta x}(x,u,y)$ over $x$ and $y$, but we do not
present plots here.


\subsection{Single-$l$ tests}
\label{sec:convergence_dAlembert}



\subsubsection{Generalised d'Alembert exact solutions}


We have tested two kinds of exact solutions. The first are the
generalised d'Alembert solutions for a single spherical harmonic
derived in Appendix~\ref{appendix:linperts} and given in
(\ref{psilexplicit}) for the scalar field $\delta\psi_l$, and in
(\ref{bl}-\ref{fl}) for the coupled metric perturbations
$(\delta b_l,\delta f_l)$, both in terms of an arbitrary function
$\chi$ of one variable.  These were derived, and used as testbeds,
previously in \cite{GomezPapadopoulosWinicour1994}.

We choose the free function $\chi$ to be a Gaussian with centre at
$0.8$ and width $0.2$. As the amplitude for $\chi$ we choose
$10^{-11-l}$, which results in a maximum value $\sim 10^{-11}$ for
$\psi$ and $f$ in the initial data. The numerical domain is
$0\le x\le 3$ with $x_0=2$ and $0\le u\le 1.9$. We set $N_y=5$, and
$N_x=64...8192$, increasing by fSactors of two. For these tests, and in
contrast to critical collapse applications, the fact that the grid
shrinks to a point is irrelevant, and this is why we stop at $u=1.9$,
when the grid has shrunk to $0.05$ of its original size, and the
waves have essentially left the grid.

At high resolution, our numerical evaluation of the d'Alembert exact
solution at the innermost grid points suffers from large round-off
error, resulting from the division by powers of $r$ up to
$r^{l+1}$. To get round this, at small $r$ we implement a truncated
power-series expansion of the d'Alembert solution.

In initial data in lsB and gsB gauge, $R$ is set to the spherical
background value given in Eq.~(\ref{R0exact}), that is
$R_0(0,x)=x/2$. In sdn gauge, we add to this a Gaussian $\delta R$, so
that we do not have to wait for $\delta R$ to develop dynamically.

To start with a summary of the numerical results that follow, sdn and
lsBtosdn gauge are unstable, our flavour of gsb is unstable in the
$(G,x)$ formulation but stable in the $(\gamma,R)$ formulation (with
and without integration by parts), and the lsB2 flavour of lsB gauge
is stable in all three formulations. We then find pointwise
second-order convergence (both self-convergence and convergence
against the exact solution) for all $l$-components of all variables.


\subsubsection{Results in sdn gauge} 


In sdn gauge, for $l=3$ d'Alembert data and $l=3$ d'Alembert plus
Gaussian-in-$R$ data, we have only gone to 1024 grid points to see
that in all three formulations the error (against the exact solution)
has a constant in $x$, oscillating in $u$, part that does not decrease
with resolution. In other words, in sdn gauge there is an instability
at the origin.


\subsubsection{Results in lsB tosdn gauge}


In lsBtosdn gauge, $f$ is unstable at the centre and not
converging from the start for $l=2,3$. ($\psi$ still converges,
presumably because it sees essentially flat spacetime). It is
sufficient to go to 1024 grid points to see this. We have only tried
the $(\gamma,R)$ formulation with integration by parts.

A mild instability at the centre is still present when the blending is
between $0.3\,x_0$ and $x_0$, with pure lsB gauge for $x\le
0.3\,x_0$. This could be because we are then not projecting $R$
down to $l=0$ even at those $x$. We also see something like a linear
gauge shock at $x\sim 1.45$, in the middle of the blending zone.


\subsubsection{Results in gsB gauge}


In the $(G,x)$ formulation, $l=2$ is unstable in our flavour of gsb
gauge, with again an oscillating instability at the centre giving rise
to an error constant in $x$ that increases with better resolution. We
needed to go to 4096 points to see this. In the $(\gamma,R)$
formulation with and without integration by parts, it is stable and
perfectly second order convergent to 8192 points.

By contrast, $l=3$ in gsb gauge is stable up to 8192 points
already in the $(G,x)$ formulation.


\subsubsection{Results in lsB2 gauge}


We have tested lsB2 gauge in the $(G,x)$ formulation only for $l=3$ and
only up to 1024 gridpoints. The error at 1024 grid points is visually
identical with that in the $(\gamma,R)$ formulation with integration
by parts. We have tested lsB2 gauge in the $(\gamma,R)$ formulation
with integration by parts in more detail. All error plots in the
following are in this gauge and formulation.

In particular, we have tested the d'Alembert solution for $l=0,1$
($\psi$ only), $l=2,3,4$ ($\psi$, $f$ and $b$) and $l=5$ ($f$ and $b$
only, not suppressing this highest frequency for once).
We find second-order pointwise convergence of $\psi_0$, $\psi_1$
and, for $l\ge 2$, of $\psi_l$, $f_l$ and $b_f$, from $N_x=64$ to
$N_x=8092$ radial grid points.

\begin{figure}
\includegraphics[width=0.4\textwidth, angle=0]{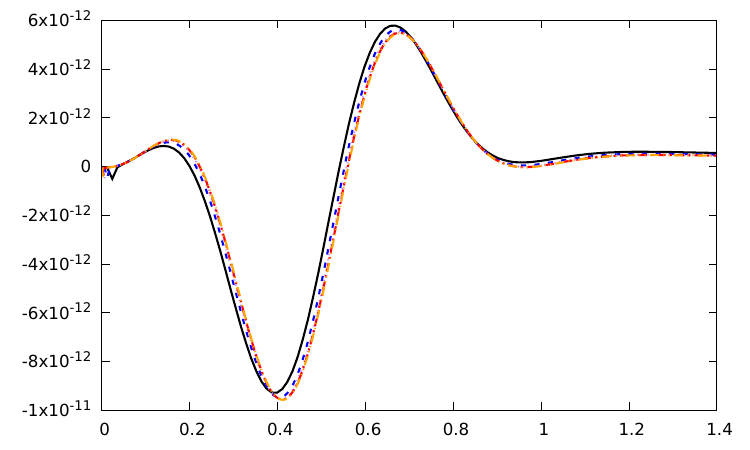} 
\includegraphics[width=0.4\textwidth, angle=0]{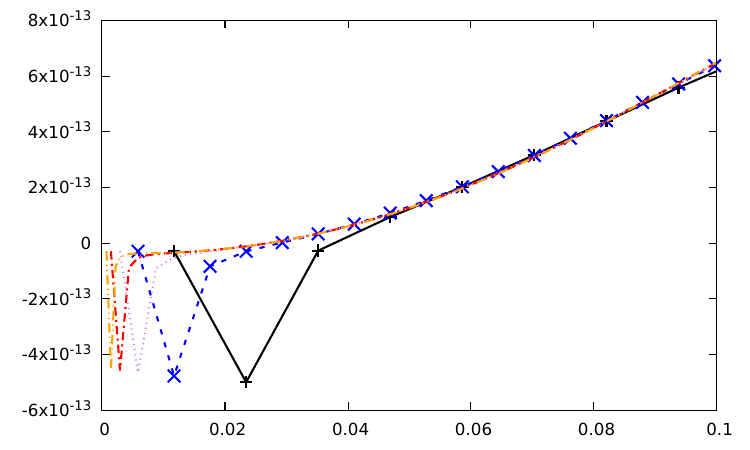} 
\caption{D'Alembert test in lsB gauge: A snapshot of the scaled errors
  ${\cal E}_{\psi_2,\Delta x}(x)$ for the quadrupole component
  $\psi_2$ of the scalar field $\psi$, at five resolutions from
  $N_x=256$ to $N_x=4096$. We show a moment of time $u=0.48$ where the
  wave passes through the centre. In the upper plot we truncate the
  numerical domain $0\le x\le 3$ to $0\le x\le 1.4$, as nothing
  interesting happens at larger $x$. The lower plot, restricted to
  $0\le x\le 0.1$, focuses on the centre, and we show grid points for
  the two coarsest grids.}
\label{fig:psi02convergence}
\end{figure}

For $l=2$, the above statement needs to be qualified. The error in
$\psi_2$ is not smooth near the origin, but has a blip at the second
grid point (rather than a specific value of $x$). This is demonstrated
in Fig.~\ref{fig:psi02convergence}. However, we get pointwise
second-order convergence at fixed $x$ for all resolutions $\Delta x<x/2$.

\begin{figure}
\includegraphics[width=0.45\textwidth, angle=0]{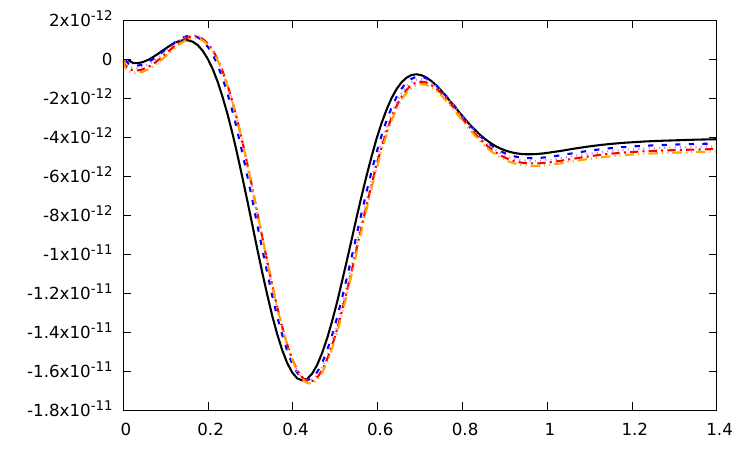} 
\includegraphics[width=0.45\textwidth, angle=0]{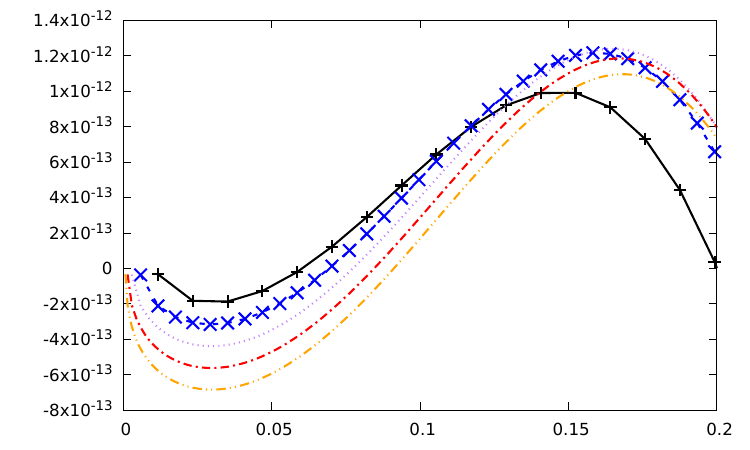} 
\caption{D'Alembert test: As in Fig.~\ref{fig:psi02convergence}, but
  now for the scaled error ${\cal E}_{f_2,\Delta x}(x)$. Again the
  lower plot is a detail of the upper one near the centre, restricted
  here to $0\le x\le 0.2$, and with grid points for the two lowest
  resolutions.}
\label{fig:f02convergence}
\end{figure}

Fig.~\ref{fig:f02convergence} shows the scaled error in $f_2$. Here,
it is not clear if we have reached convergence even at $N_x=4096$, and
it appears again that the error is not smooth at the origin, although
in a different manner to that in $\psi_2$.

\begin{figure}
\includegraphics[width=0.45\textwidth, angle=0]{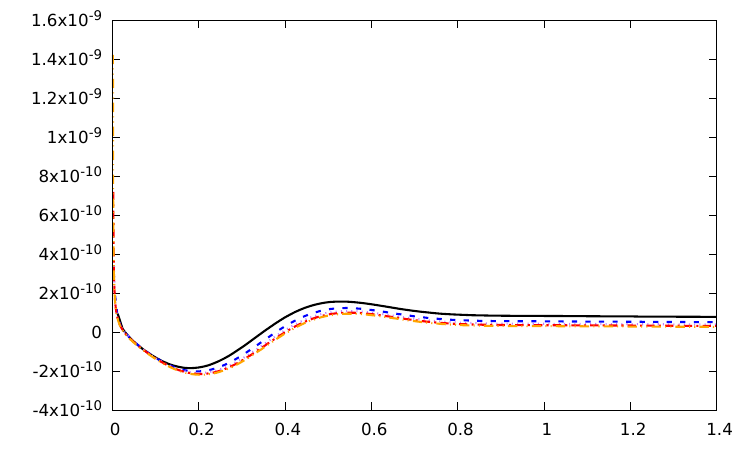} 
\includegraphics[width=0.45\textwidth, angle=0]{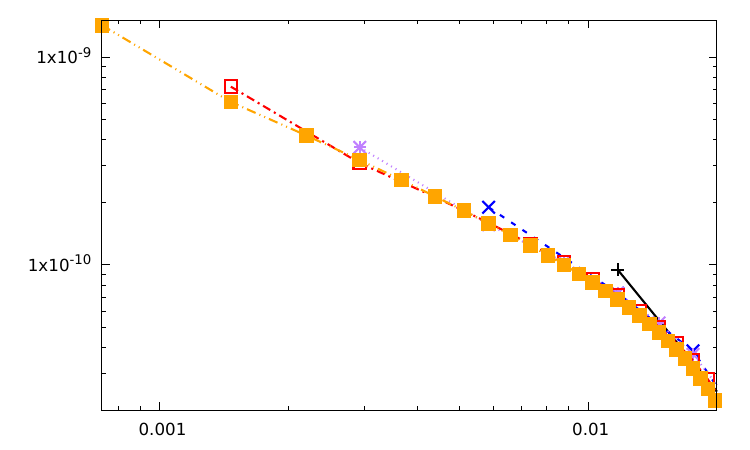} 
\caption{D'Alembert test: As in Fig.~\ref{fig:psi02convergence}, but
  now for the scaled error ${\cal E}_{b_2,\Delta x}(x)$. Again the
  lower plot is a detail of the upper one near the centre, but now
  restricted to $0\le x\le 0.02$, on a log-log scale, and showing grid
  points at all resolutions.}
\label{fig:b02convergence}
\end{figure}

Fig.~\ref{fig:b02convergence} shows the scaled error in $b_2$. Near
the origin, the scaled error is ${\cal E}_{b_2,\Delta x}(x)\sim
(\Delta x)^2/x$. At the innermost grid point this evaluates to $\sim
\Delta x$. Hence $b_2$ converges pointwise to second order at all
points, including near the origin, but converges only to first order
in the maximum norm, which at high resolution is dominated by that
innermost grid point. In other norms, such as the root-mean-square
norms, the convergence would be at an intermediate order.

For $l>2$, we suspect that the error is still technically unsmooth
near the origin, but this is hard to see as the errors, like the
variables themselves, are suppressed at the origin by powers of $x^l$
for $f$, $\psi$ and the other scalars, and $x^{l-1}$ for $b$. Hence
the second-order pointwise convergence looks fine. It is possible that
our methods can be improved to make the error better behaved at the
origin, but we leave this to future work.

We note finally that, where we have the exact solution, the
self-convergence estimate of the error is pointwise approximately
equal to the true error (difference from the exact solution).

To test convergence at $l>5$, we change from the d'Alembert solution (which
becomes increasingly complicated) to simple Gaussian data
for a single $l$. As we need larger $N_y$ to represent larger $l$, we
go to the half-range, with $\bar N_y=17$. We now set initial data
directly for $f_l$ and $\psi_l$, namely a Gaussian again with centre
at $x=0.8$, width $0.2$, and amplitude $10^{-11}$. The numerical
domain is again $0\le x\le 3$ with $x_0=2$ and $0\le u\le 1.9$, with
$\bar N_y=17$, and $N_x=64...8192$, increasing by factors of two. At
$l=2$, we find the same as with the d'Alembert $l=2$ data. At $l=4,8,16,32$
we find apparent perfect second-order pointwise convergence (as for
the $l=4$ d'Alembert data). 


\subsection{Plane-wave tests}



\subsubsection{Plane-wave exact solutions}


We show in Appendix~\ref{appendix:linpertsnodecomposition} that any
solution of the scalar wave equation gives rise to a solution of the
linearised Einstein equations, without the need for a spherical
harmonic decomposition. In particular, the scalar wave equation admits
plane-wave exact solutions, and in Appendix~\ref{appendix:planewave}
we give these in Eq.~(\ref{psiplanewave}) for $\delta\psi$ and in
(\ref{fplanewave}-\ref{bplanewave}) for the related $(\delta
b_l,\delta f_l)$. Of course, these are just the translation into our
coordinates of the linear limit of the well-known plane-symmetric
gravitational waves. This is our second exact solution testbed.

A scalar field plane wave moving in the negative $z$-direction on flat
spacetime takes the form
\begin{equation}
\psi=\chi(t+z)=\chi(u+r+z)=\chi(u+r(1+y)).
\end{equation}
Its contours at constant $z$ are parabolas in the $(\rho,z)$ plane
(where $\rho^2:=r^2+z^2)$, as
one expects of the intersection of a null plane with a null cone. When
this intersection reaches the vertex of the cone it closes up and
disappears.

We choose $\chi$ to be a Gaussian with centre $1.0$ and
width $0.1$, and with amplitude $10^{-11}$ for $\psi$, which means that
the maximum of $\psi$ is exactly $10^{-11}$, and amplitude
$10^{-13}$ for $f$, which means that the maximum of $f$ is
approximately $4\cdot 10^{-11}$. The numerical domain is again $0\le
x\le 3$ with $x_0=2$, now with $0\le u\le 1.5$. This means that the
plane wave crosses the origin and then disappears out of the numerical
domain during the evolution.

The (single) plane wave solutions contain both even and odd spherical
harmonics. In order to test plane waves with our half-range
formulation, we add a copy of the same wave moving in the opposite
direction to make $\psi$ and $f$ even functions of $y$ (at constant
$u$ and $x$), and $b$ an odd function.


\subsubsection{Results in lsB2 gauge}


We evolve with all combinations of $N_y = 17,33,65,129$, $\bar N_y=9,17,33,65$
(for the double plane wave only), and $N_x = 64...1024$. 

This double plane wave solution contains only even spherical
harmonics, and so can be run on the full or half range in $y$. We have
verified that the errors in the double plane wave test are identical
in the equivalent resolution pairs.  We remove the top two frequencies
in $f$ and top frequency in $b$ in the exact solution before the
comparison with the numerical solution, as the numerical solution is
similarly truncated.

\begin{figure*}
\centering
\includegraphics[width=0.325\textwidth,
  angle=0]{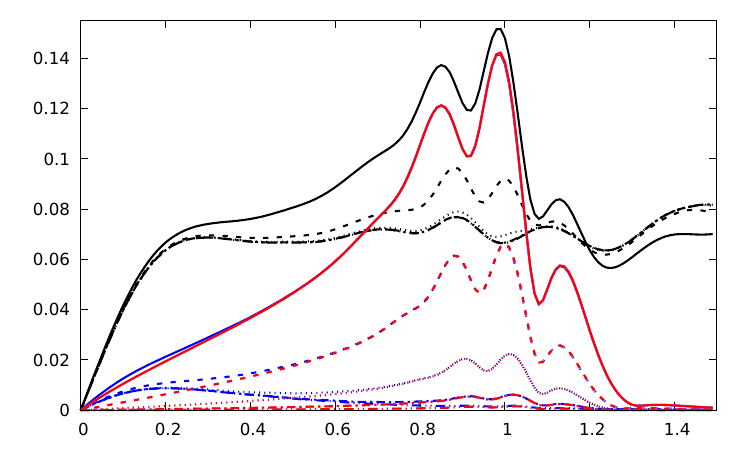} 
\includegraphics[width=0.325\textwidth,
  angle=0]{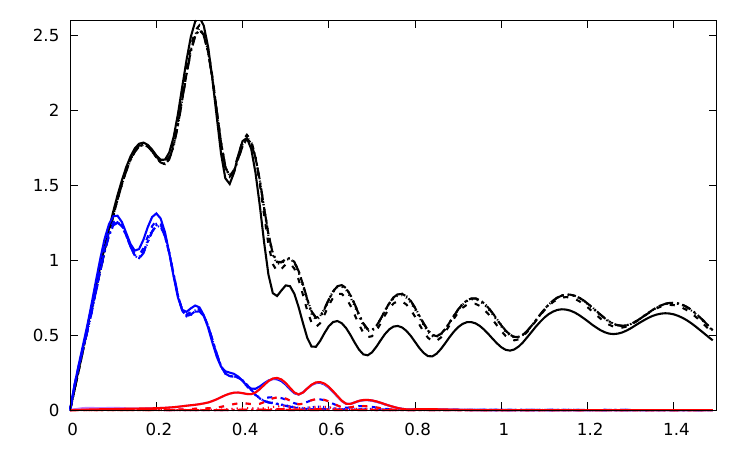} 
\includegraphics[width=0.325\textwidth,
  angle=0]{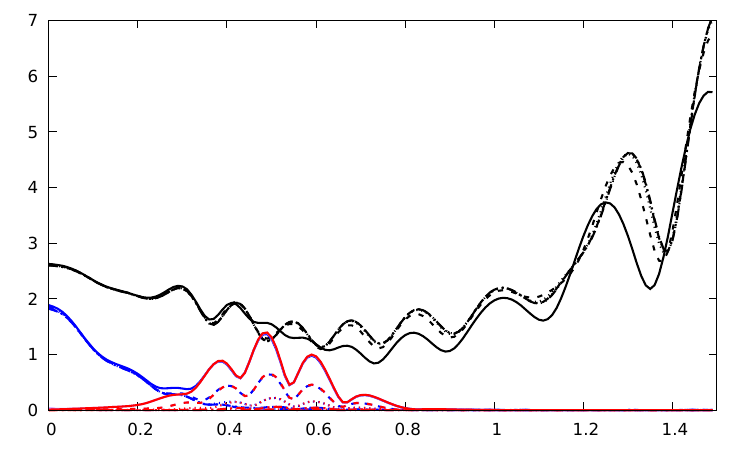} 
\includegraphics[width=0.325 \textwidth,
  angle=0]{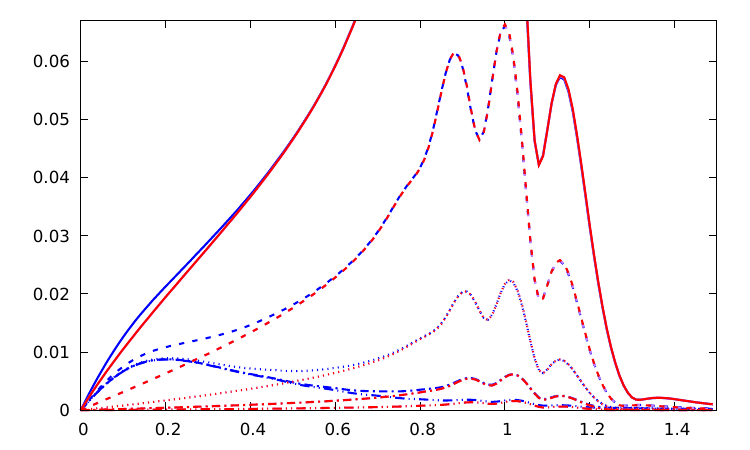} 
\includegraphics[width=0.325 \textwidth,
  angle=0]{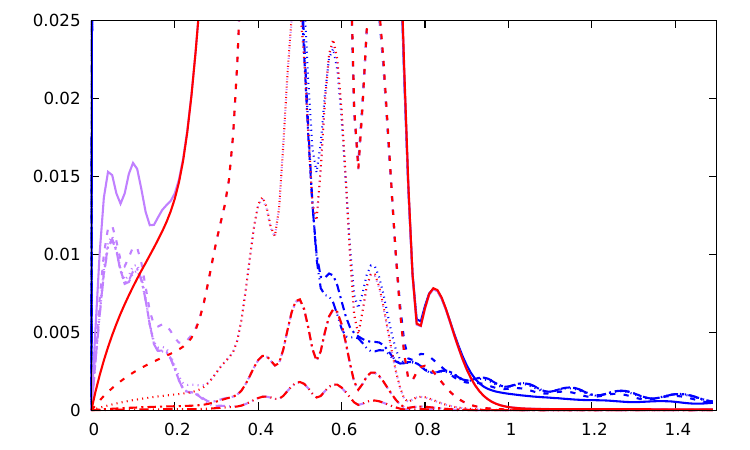} 
\includegraphics[width=0.325 \textwidth,
  angle=0]{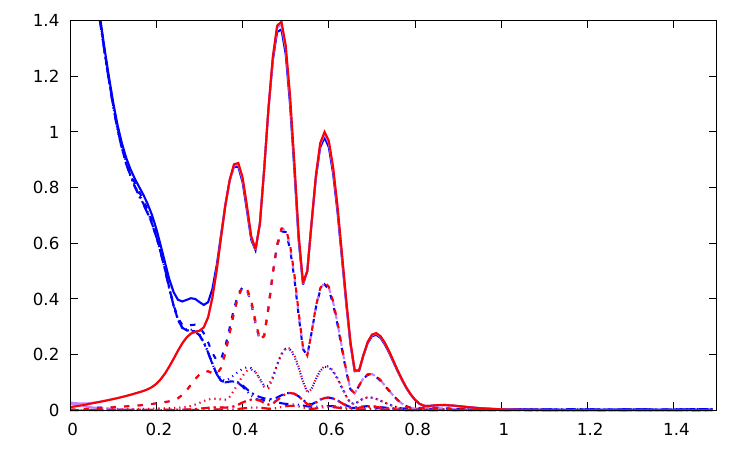} 
\includegraphics[width=0.325 \textwidth,
  angle=0]{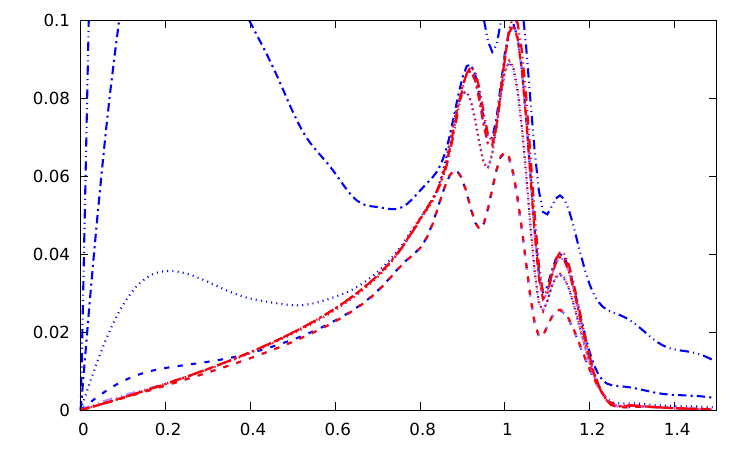}  
\includegraphics[width=0.325 \textwidth,
  angle=0]{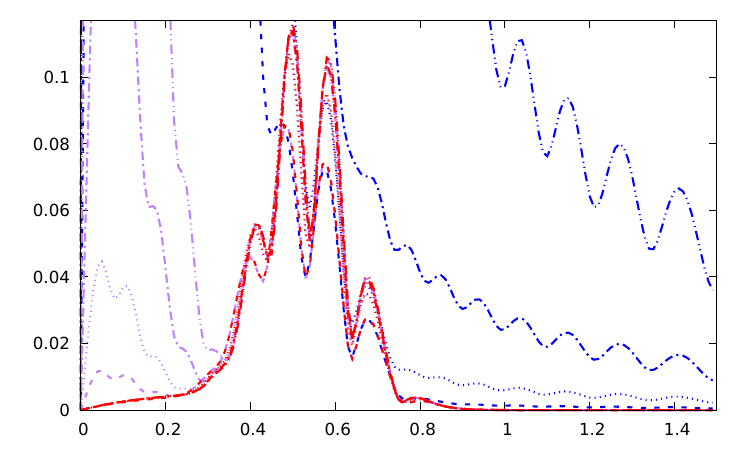}  
\includegraphics[width=0.325 \textwidth,
  angle=0]{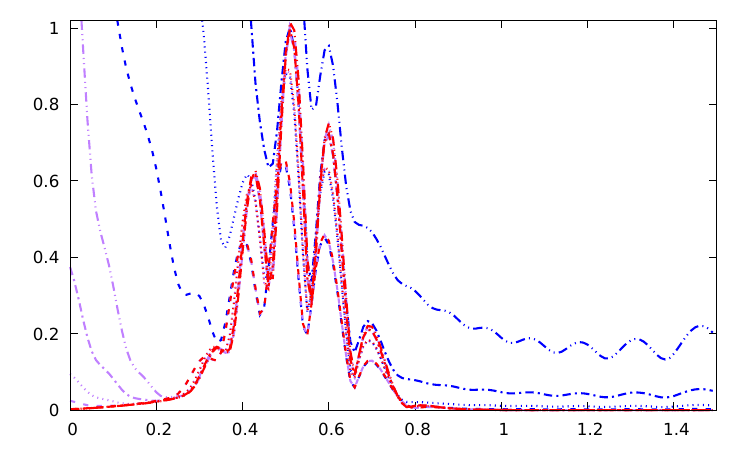}  
  \caption{(Single) plane-wave tests: the rms (over $x$ and $y$) norm
    of the error in $\psi$, $f$ and $b$ (scaled by $10^{-11}$ to give
    an indication of relative erorr), against $u$. $\psi$ is in the
    left column, $f$ in the middle columng, and $b$ in the right
    column. Angular resolution is indicated by line colour: $N_y=17$
    black, $N_y=33$ blue, $N_y=65$ purple and $N_y=129$ red. Radial
    resolution is indicated by line type: $N_x=64$ solid, $N_x=128$
    dashed, $N_x=256$ dotted, $N_x = 512$ dot-dashed, and $N_x=1024$
    dot-dot-dashed. The middle plot in each column enlarges the lower
    part of the upper plot, and for clarity omits $N_y=17$. The lower
    plot additionally omits $N_x=64$, and is scaled with
    $(N_x/128)^2$, to test for (local-in-$u$) second-order convergence
    with $N_x$. The horizontal range is always $0\le u\le 1.5$, but
    the vertical range (rms error) has been chosen differently in each
    of the nine plots.}
\label{fig:planewave_convergence}
\end{figure*}

Beginning with the single plane wave, we evaluate the maximum and rms
norms of the error against $u$. The norms are taken over all grid
values of $x$ and $y$, with $y=\pm-1$ weighted half in the rms norm in
order to make it the same on the full and half-grid. The rms error in
$\psi$, $f$ and $b$ for the single plane wave at the different
resolutions is shown in Fig.~\ref{fig:planewave_convergence}.

The figure consists of nine plots laid out in a square. The three
columns from left to right show the variables $\psi$, $f$ and
$b$. Focus initially on the middle column, showing $f$. Each plot shows
four differences of angular resolution and five differences of radial
resolution. Different radial resolutions are distinguished by line
type, and different angular resolution by line colour.

In the top plots, the rms error is {\em not} scaled. We see that the
error decreases quickly with angular resolution $N_y$, but is almost
independent of radial resolution $N_x$. This indicates that for
$N_y=17,33$ and $N_x=64...1024$, the total error budget is
dominated by angular discretisation error.

The middle plots zoom in on the smallest errors, which arise from
the highest angular resolutions. They demonstrate that at $N_y=65,129$
the error does decrease with increasing radial resolution $N_x$. In
the lowest plot these errors are scaled with $(N_x/128)^2$. The
scaled curves with $N_y=129$ (omitting $N_x=64$) lie on top of each
other, indicating second-order convergence with $N_x$. This indicates
that at $N_y=129$, for $N_x=128...1024$ the error budget is dominated
by the radial discretisation error.

Comparing now the three variables $\psi$, $f$ and $b$, we note that as
$b$ is computed from $f$ at each time step, it shows a numerical error
already in the initial data, whereas $\psi$ and $f$ are evolved from
analytic initial data, and so their error is zero at $u=0$. For
$N_y=129$ only, the initial error in $b$ is neglible, and the
subsequent error in $b$ is similar to the error in $f$ or $\psi$.

We see perfect second-order convergence with $N_x$ in $\psi$ (but not
$f$ and $b$) already at $N_y=65$. Moreover, the errors in $\psi$ with
$N_y=65$ and $129$ (purple and red curves) are indistinguishable in
our plots.  We conclude that a plane scalar wave requires less angular
resolution than a plane gravitational wave of the same shape $\chi$
(where $\chi$ is a solution of the scalar wave equation). This may be
because the exact solution for $f$ and $b$ involves derivatives of $\chi$.

A closer look at the middle and bottom plots shows a transition from
an error dominated by $y$-discretisation at early times and small
$N_y$ and by $x$-discretisation at late times and large
$N_y$. Plotting single-$l$ components of the error against $u$ and $x$
further shows that the early discretisation error in $y$ arises mostly
at $x\gtrsim 1.5$ and the late error discretisation error in $x$
mostly at $x\lesssim 1.5$.

The rms errors in the double plane wave test are very similar as
functions of $u$, but larger by roughly $\sqrt{2}$, as there are now
two identical waves of error instead of one in the same numerical
domain. We therefore do not present plots here.

\subsection{Tests of the computation of the Hawking mass}


As a different indication of numerical error, we have implemented the
expression for $M$ given by (\ref{MHaw}) with (\ref{CHawux}), a
centred finite differencing of this to compute the resulting $M_{,x}$,
and, independently, the direct expression (\ref{DMlsB}) for
$M_{,x}$. The two expressions for $M_{,x}$ are affected in different
ways by finite differencing error in $x$ and spectral error in $y$,
and so their agreement in the continuum limit is a non-trivial test of
the correctness of (\ref{MHaw}) with (\ref{CHawux}), (\ref{DMlsB}),
the hierarchy equations, and their discretisations. The cleanest test
is one where the solution is close to Minkowski, so that the hierarchy
equations are approximately linear in $\psi$, $b$ and $f$, and the
expressions for $\gamma$ and $M$ are approximately quadratic.

We have tested single-$l$ Gaussian initial data in $f$ or $\psi$, and
have varied the amplitude, $N_x$, $N_y$, and the numerical method. We
have not evolved these data.

We first take a Gaussian in $f$ only with centre at $x=1$, width
$0.25$, $x_{\rm max}=5$, $l=2$, and amplitude $10^{-4}$. Then
$M\sim 9\cdot 10^{-8}$ at the outer boundary, and $M_{,x}$ has a
maximum of $\sim2.5\cdot 10^{-7}$. Our baseline resolution is
$N_x=1000$ and $N_y=5$. At this baseline, the difference between the
two versions of $M_{,x}$, an estimate of the error in either, has
maximum $\sim 1.1\cdot 10^{-10}$, a relative error of
$\sim 4\cdot 10^{-4}$.

Decreasing the amplitude by a factor of 10 reduces $M$ and the error
in $M_{,x}$ by a factor of 100, as expected. If we decrease the
amplitude much more, the error is dominated by noise. This would be
expected as round-off-error in the cancellation between $1$ and
$2R^2\rho_+\rho_-$ in $C$.

The error with $N_y=5,9,17,33,65,129$ is the same, and so for $\bar
N_y=3,5,9,17,33,65$ on the half-range. Doubling $N_x$ from the
baseline reduces the error by a factor of 4. Hence the error in $M$ in
almost-linear evolutions is dominated by finite differencing in $x$,
while spectral error in $y$ is negligible.

Initial data in which only $\psi$ is non-vanishing test other parts of
the two expressions for $M_{,x}$. Note that $b\ne 0$ when hierarchy
equations are solved for these data even with $f=0$. We use a Gaussian
with the same centre and width as for $f$ above. An amplitude of
$10^{-4}$ gives $M\sim 5.5\cdot 10^{-8}$. $M_{,x}$ has a maximum of
$\sim 1.5\cdot 10^{-7}$, and the error at baseline has a maximum of
$\sim 4.5\cdot 10^{-11}$, a relative error of $\sim 3\cdot
10^{-4}$. All other comments for pure $f$ data just above also apply
here.


\section{Conclusions}
\label{sec:conclusions}


We have begun an investigation of the use of null coordinates in
numerical relativity, applied to gravitational collapse. In this first
paper we have both made progress and identified new difficulties.


\subsection{Progress}


On the purely mathematical side, we have shown that the ingoing null
derivative $\Xi$ normal to the surfaces of constant $(u,x)$ plays a
role similar to the Lie derivative ${\cal L}_n$ normal to spacelike
time slices. Specifically, and oversimplifying a bit, the Einstein
equations give us the geometric time derivative ${\cal L}_n{\cal
  L}_n\,g_{ij}$ on the usual spacelike time slices, but $\Xi g_{ij}$
on null slices. Writing the hypersurface equations in terms of $\Xi$
removes any explicit appearance of the radial shift $B$, just as
writing the 3+1 equations in terms of ${\cal L}_n$ removes any
explicit appearance of the lapse and shift.

On the numerical side, we have removed a practical obstacle to using
null cones with a regular centre, already identified in
\cite{GomezPapadopoulosWinicour1994}, namely that the time step
$\Delta u\sim \Delta x(\Delta\theta)^2$ is unreasonably small at high
angular resolution. We have been able to replace this by
$\Delta u\sim \Delta x$, similar to Cartesian coordinates, by reducing
the angular resolution at small radius.

For applications to critical collapse in spherical polar coordinates,
we have found an equivalent of the method of repeated radial
regridding of \cite{Garfinkle1995} that works beyond spherical
symmetry. Essentially this is done by adding to a standard gauge
choice, such as Bondi gauge, an ingoing radial shift that shrinks the
grid continuously, with the outer boundary becoming future spacelike
\cite{Rinne2020}.

In the present paper, we have demonstrated convergence of these
numerical methods in the evolution of weak data, and in Paper~II we
successfully apply them axisymmmetric scalar field critical
collapse.


\subsection{Problems}


As we have discussed, one cannot find marginally outer-trapped
surfaces embedded in null cones with a regular centre, and in vacuum
axisymmetry not even any (non-marginally) outer-trapped surfaces. We
propose using closed 2-surfaces of large Hawking compactness (Hawking
mass/area) as an alternative diagnostic of black hole formation. In
Paper~II, this allows us to find the threshold of black-hole formation
by bisection, and demonstrate critical scaling of the black hole mass.

It becomes clear in Paper~II that in sufficiently non-spherical
spacetimes the divergence of the congruence of the generators of our
null cones becomes negative in some directions even when no black hole
is formed. Bondi coordinates and double-null coordinates, and the
generalised Bondi coordinatees of Paper~II, break down when this happens. 


\subsection{Outlook}


Two key questions remain open. The first is wheter outgoing null cones
emanating from a regular centre remain regular in strong gravity
further away from spherical symmetry, and in particular in
electromagnetic or vacuum critical collapse. This is a purely
geometric question, independent of gauge or numerical methods. Note
also that the potential problem is non-sphericity, rather than black
hole formation. One reason to be optimistic is that in
\cite{Vacuum_Collapse} we have constructed outgoing null cones
emanating from a regular vertex in post-processing of 3+1
near-critical vacuum evolutions, as a way of comparing evolutions in
different coordinate systems, without finding caustics.

A second question is how much useful information about any newly
formed black hole we can find, given that outgoing null coordinates
cannot penetrate into the horizon, or at least not very deeply, and
that we cannot find MOTS on a single coordinate null cone.

As already discussed, going further will probably require a change to
a generalised affine parameter gauge. We leave this to future work.


\acknowledgments


We would like to thank Luis Lehner, Nigel Bishop, David Garfinkle,
Bernd Br\"ugmann, Tom\'a\v{s} Ledvinka, Anton Khirnov and Jonathan Luk
for helpful discussions, and the Mathematical Research Institute
Oberwolfach for supporting this work through its ``Oberwolfach
Research Fellows'' scheme. Bastien Roy and Bernardo Porto-Veronese
contributed through M1 internships in 2021 funded by \'Ecole
Polytechnique. DH was supported in part by FCT (Portugal) Project
No. UIDB/00099/2020. TWB was supported in part by National Science
Foundation (NSF) grants PHY-2010394 and PHY-2341984 to Bowdoin
College.


\appendix



\section{The time step problem in null coordinates}
\label{appendix:timestepproblem}


We show here that the time step for explicit finite differencing
schemes is even more severely limited in the combination of polar
spatial coordinates with a regular centre and a null ``time''
coordinate $u$ than for polar coordinates and the usual (spacelike)
time coordinate $t$. For simplicity we restrict to axisymmetry in 3+1
dimensions. We focus on the causal geometry, rather than giving a
rigorous argument. Our presentation closely follows
\cite{GomezPapadopoulosWinicour1994}.

Consider at first the Minkowski metric in standard spherical polar
coordinates,
\begin{equation}
ds^2 = -dt^2+dr^2+r^2\left(d\theta^2 +
\sin^2\theta\,d\varphi^2\right).
\end{equation}
In an abuse of notation, we now let $dx^\mu$ stand for the small
finite coordinate distance from the grid point that is being updated
to any one of the other grid points from which the update is
calculated. These will be integer multiples of the grid spacings
$\Delta x^\mu$.  For an explicit numerical method, the update is found
from only a few neigbouring grid points to the past. The set of these
points is called the ``stencil'' of the point being updated. Causality
requires that the numerical stencil be wider than the physical light
cone, or, in our notation, $ds^2>0$ for the outermost grid points in
the stencil. Empirically, this is also a necessary criterion for the
stability of any explicit finite difference scheme, often referred to
as the Courant-Friedrichs-Lewy (CFL) condition.

For simplicity we assume that the stencil involves, besides the point
being updated, only grid points on the previous time level. We
parameterise such a stencil by $dt=-\Delta t$, $dr=s\Delta r$,
$d\theta=q\Delta\theta$, with $\Delta t,\Delta r,\Delta\theta>0$ by
definition, and $s,q=-1,0,1$ as a simple example of a $3\times 3$
point stencil in $r$ and $\theta$. We assume axisymmetry, so nothing
depends on $\varphi$. We also have $r=p\Delta r$, where,
for example, $p=1,2,3,\dots$ on a centred equally spaced grid. What
matters here is only that $s$ and $q$ are of order unity and so is the
smallest possible value of $p$, which occurs next to the centre.

The CFL condition $ds^2>0$ translates into
\begin{equation}
\label{Deltatcondition}
\Delta t<\Delta r\sqrt{s^2+p^2q^2\Delta\theta^2}.
\end{equation}
The choice $s=0$, $q=\pm 1$ of points in the stencil, with the choice
$p=1$ to locate the stencil next to the centre, gives
\begin{equation}
\label{Deltatconditionsharp}
\Delta t\lesssim \Delta r\,\Delta \theta,
\end{equation}
where we have replaced $<$ by $\lesssim$ to allow for more general
stencils and for other $O(1)$ factors in the argument. As is
well-known, this is worse than the time step restriction
$\Delta t\lesssim \min(\Delta x^i)$ in Cartesian coordinates by the
factor $\Delta\theta\ll 1$. Put differently, the right-hand side of
(\ref{Deltatconditionsharp}) is quadratic in small quantities.
Halving the grid spacing in all spatial directions halves $\Delta t$
in Cartesian coordinates, but reduces it to a quarter in polar
coordinates.

Consider now the Minkowski metric in the null coordinate form
\begin{equation}
ds^2 = -2G\,du\,dx-Hdu^2+R^2\left(d\theta^2 +
\sin^2\theta\,d\varphi^2\right).
\end{equation}
We let $du=-\Delta u$, $dx=-s\Delta x$, $d\theta=q\Delta\theta$, and
$R=p\Delta x$. At the centre $R=0$, any radial gauge must approach
Bondi gauge to keep it at $x=0$, and setting $H=1$ and $R=x$ there without loss
of generality, we also have $G=1$.  The CFL
condition(\ref{Deltatcondition}) is now
\begin{equation}
\Delta u\lesssim \Delta x\left(\sqrt{s^2+p^2q^2(\Delta \theta)^2}-s\right).
\end{equation}
With $s=1$, $p=1$, $q=\pm 1$, and assuming $\Delta\theta\ll 1$,
the equivalent of (\ref{Deltatconditionsharp}) is now
\begin{equation}
\label{Deltauconditionsharp}
\Delta u\lesssim {1\over 2}\Delta x\,(\Delta \theta)^2,
\end{equation}
worse by a second factor of $\Delta\theta\ll 1$ than for polar
coordinates on spacelike time slices. Except in spherical
symmetry, this is a problem of any explicit numerical time evolution
scheme on null cones with a regular vertex.

At large $R$, the sharpest CFL condition arises from stencil points with
$q=0$ (and $p$ then drops out). We now consider arbitrary $G$ and $H$.
The CFL condition becomes
\begin{equation} 
\Delta u<-{2sG\over H}\,\Delta x.
\end{equation}
Recall that $G>0$.  Choosing $s=\pm 1$ with sign opposite to that of
$H$ we obtain
\begin{equation} 
\Delta u<{\Delta x\over B},
\end{equation}
where $B:=H/(2G)$ as defined previously. This is just the CFL
condition for the $x$-advection term in
$\partial_u=\Xi+B\partial_x+Sb\partial_y$.


\section{Minkowski spacetime}
\label{appendix:Minkowski}


In Minkowski spacetime, we can choose coordinates where $f=b=0$ and
the remaining metric coefficients $R$, $G$ and $H$ depend only on
$u$ and $x$. We denote them by $R_0$, $G_0$ and $H_0$.  With these
assumptions, there are only two nontrivial hierarchy equations, namely
\begin{eqnarray}
\label{G0equation}
\left(\ln{G_0\over R_{0,x}}\right)_{,x}&=&0, \\
\label{Xi0R0equation}
\left(R_0\Xi_0R_0\right)&_{,x}=&-{1\over 2}G_0.
\end{eqnarray}
Clearly, these can be integrated in terms of two arbitrary functions of $u$.

To clarify what these two functions are in a regular spacetime, we
note that in the standard double null coordinates $(U,V)$, the
Minkowski metric is
\begin{equation}
\label{flatmetricMinkowski}
ds_0^2=-dU\,dV+\left({V-U\over 2}\right)^2\,d\Omega^2.
\end{equation}
We now change to the most general null coordinates $u$ and $x$ adapted
to the spherical symmetry, defined by
\begin{eqnarray}
U&=&U(u), \\
V&=&2R_0(u,x)+U(u),
\end{eqnarray}
and obtain the metric
\begin{equation}
\label{myflatmetric}
ds^2=-2G_0\,du\,dx-H_0\,du^2+R_0^2\,d\Omega^2,
\end{equation}
where
\begin{eqnarray}
\label{G0Mink}
G_0&=&U'R_{0,x}, \\
H_0&=&U'(U'+2R_{0,u}),
\end{eqnarray}
and hence 
\begin{equation}
\label{Xi0R0Mink}
\Xi_0R_0=-{1\over 2}U'.
\end{equation}
Therefore the general solution of (\ref{G0equation},
\ref{Xi0R0equation}) with a regular centre is (\ref{G0Mink}) and
(\ref{Xi0R0Mink}): note this has only one free function
$U(u)$. (\ref{G0Mink}) can be written as
\begin{equation}
g_0=U'(u).
\end{equation}
In Bondi coordinates, where $R_{,u}=0$, we have
\begin{equation}
H_{0,\text{Bondi}}=U'(u)^2.
\end{equation}

If we choose $x=0$ to be a geodesic and the coordinate basis vectors
to be parallely transported along it, the metric at $x=0$ is given by the
Minkowski metric (\ref{myflatmetric}) at $x=0$ for all times. 

In flat spacetime, shifted double null coordinates, shifted global
Bondi coordinates and a natural choice of local shifted Bondi
coordinates are all identical. To find the metric in these coordinates, we
start again from the metric (\ref{flatmetricMinkowski}) and make the
specific coordinate transformation
\begin{eqnarray}
U&=&u-x_0, \\
V&=&(u-x_0)\left(1-{x\over x_0}\right),
\end{eqnarray}
where $x_0>0$ is a parameter,
to obtain (\ref{myflatmetric}) with 
\begin{eqnarray}
\label{R0exact}
R_0&=&{1\over 2}\left(1-{u\over x_0}\right)x, \\
\label{G0exact}
G_0&=&{1\over 2}\left(1-{u\over x_0}\right), \\
\label{H0exact}
H_0&=& 1-{x\over x_0}.
\end{eqnarray}
It follows that
\begin{eqnarray}
g_0&=&1, \\
\label{H0o2G0}
B_0:={H_0\over 2G_0}&=& {x_0-x\over x_0-u}, \\
\label{Xi0R0exact}
\Xi_0R_0&=&-{1\over2}.
\end{eqnarray}
We call this the shifted Minkowski (from now on, sM) background gauge.


\section{Spherical symmetry}
\label{appendix:sphericalsymmetry}


We restrict to spherical symmetry, but bring in a spherical scalar
field as matter, by setting $f=b=0$ and making $R$, $G$, $H$ and
$\psi$ functions of $u$ and $x$ only. The metric is
\begin{equation}
\label{sphericalnull}
 ds^2=-2G\,du\,dx-H\,du^2
+R^2\,d\Omega^2.
\end{equation}
The hierarchy equations become
\begin{eqnarray}
\label{dmygdxspherbis}
{\cal D}\ln g&=&4\pi
R\left({\cal D}\psi\right)^2 \\
\label{myximyRspherbis}
{\cal D}(R\Xi R)&=&-{g\over2}, \\
\label{myximypsispherbis}
{\cal D}(R\Xi\psi)&=&-(\Xi R){\cal D}\psi,
\end{eqnarray} 
where
\begin{equation}
\label{Xispherbis}
\Xi=\partial_u-{H\over g}{\cal D}.
\end{equation}

In spherical symmetry, diagnosing collapse and estimating the horizon
mass is straightforward. The Hawking compactness $C$ and mass $M$
are given by
\begin{equation}
M={R\over 2}\,C, \qquad C=1+{2\Xi R\over g},
\end{equation}
and, with $|\nabla R|^2=-2\Xi R/g$, the Hawking mass in spherical
symmetry is equal to the well-known Misner-Sharp mass
\begin{equation}
M={R\over 2}\left(1-|\nabla R|^2\right).
\end{equation}
(Unlike the Hawking mass in general, the Misner-Sharp mass in
spherical symmetry derives from a conserved stress-energy current
\cite{Kinoshita}.) The mass aspect (\ref{DMlsB}) is
\begin{equation}
M_{,x}=-4\pi R^2R_{,x}{\Xi R\over g}({\cal D}\psi)^2,
\end{equation}
and the redshift defined in (\ref{Zdef}) is
\begin{equation}
Z=\sqrt{-2g\Xi R}, 
\end{equation}
We see that, as long as $\Xi R$ remains finite, $g\to\infty$ gives
both $Z\to\infty$ and $C\to 1$, so this is the obvious criterion for
black hole formation, both from the compactness reaching one and the
red shift from the centre to infinity diverging.

Assuming that the time step is limited by the Bondi shift
(\ref{Bondishift}), as in Eq.~(\ref{timestep}), we have
\begin{equation}
{\Delta u\over \Delta x}\lesssim \left|{R_{,x}\over \Xi R}\right|
= {G\over |g||\Xi R|}.
\end{equation}
At finite $G$ and $\Xi R$, this goes to zero again as $g\to\infty$, or
equivalently $R_{,x}\to 0$.

A commonly used non-null coordinate system in spherical symmetry is
the polar-radial one, defined by
\begin{equation}
\label{sphericalpolarradial}
ds^2=-\alpha^2\,dt^2+a^2\,dR^2+R^2\,d\Omega^2,
\end{equation}
where $R$ is now a coordinate. It is instructive to relate this to our
null coordinates.

Expressing $(t,R)$ in terms of $(u,x)$, and comparing coefficients of
(\ref{sphericalpolarradial}) and (\ref{sphericalnull}), we can write
the resulting three equations as
\begin{eqnarray}
\label{slicingcondition}
t_{,x}&=&{a\over\alpha}R_{,x}, \\
g&=&-a^2\left(R_{,u}-{\alpha\over a}t_{,u}\right), \\
H&=&g\left(R_{,u}+{\alpha\over a}t_{,u}\right),
\end{eqnarray}
and from this we can derive
\begin{eqnarray}
\Xi R&=&-{g\over 2a^2}
={1\over 2}\left(R_{,u}-{\alpha\over a}t_{,u}\right), \\
C&=&1-{1\over a^2}=1-|\nabla R^2|, \\
Z&=&{g\over a}=\alpha t_{,u}-aR_{,u}. \label{Zexpr}
\end{eqnarray}
From (\ref{slicingcondition}) and (\ref{Zexpr}) we find
\begin{equation}
Z\,du=\alpha\,dt-a\,dR,
\end{equation}
and so the redshift of the centre with respect to other observers at
constant $R$ is
\begin{equation}
Z=\alpha\left.{dt\over du}\right|_R
=\left.{dt_0\over du}\right|_ R,
\end{equation}
where $t_0$ is proper time along worldlines of constant $R$. In a
static spacetime only, we can set $t_{,u}=1$ without loss of
generality and $R_{,u}=0$, and we obtain $Z=\alpha$.


\section{Axisymmetric linear perturbations of Minkowski spacetime}
\label{appendix:linperts}



\subsection{Linearised field equations in any radial gauge}


As a test of our formulation and numerical methods beyond spherical
symmetry, we linearise about flat spacetime. We denote the background
values of all fields by a subscript $0$, and so $\psi_0=0$. We adapt
the background coordinates to spherical symmetry by making $G_0$,
$H_0$ and $R_0$ functions of $u$ and $x$ only. Although $G_0=R_{0,x}$
holds in the background, for clarity we write either $G_0$ or
$R_{0,x}$ in the linearised equations, as they appear in the
linearisation process.

The linearised hierarchy equations are
\begin{equation}
\label{linGeqn}
\left({\delta G\over G_0}-{\delta R_{,x}\over
  R_{0,x}}\right)_{,x}=0
\end{equation}
for $\delta G$, then
\begin{eqnarray}
\left({R_0^4 \delta b_{,x}\over G_0}\right)_{,x}&=& 
R_0^2
\left(2S \delta f_{,xy}-8y\delta f_{,x}\right) \nonumber \\
&&-{R_0^2\over G_0}\delta G_{,xy} -2R_0\delta R_{,xy}+2R_{0,x}\delta
R_{,y} \nonumber \\
&&+\left({R_0^2G_{0,x}\over G_0^2}+2{R_0R_{0,x}\over G_0}\right)
\delta G_{,y},  \label{linb} \\
\left(R_0\Xi_0 \delta f\right)_{,x}&=& 
{R_0\over 4x}\delta b_{,xy}+{R_{0,x}\over 2}\delta b_{,y}
-(\Xi_0R_0)\delta f_{,x} \nonumber \\
&&+{1\over 4R_0}\delta G_{,yy} \label{linf}
\end{eqnarray}
for $\delta b$ and $\delta f$,
\begin{equation}
\left(R_0\Xi_0\delta R+(\Xi_0R_0)\delta R
-R_0R_{0,x}\delta B 
\right)_{,x}=\delta S_{R}
\label{linR}
\end{equation}
for either $\delta R$ or $\delta H$ (we have not written out $\delta
S_R$ as it is long), and 
\begin{equation}
\left(R_0\Xi_0\delta\psi\right)_{,x}=
{G_0S\over 2R_0}\delta\psi_{,yy}-{G_0y\over R_0}\delta\psi_{,y}
-(\Xi_0R_0)\delta\psi_{,x}
\label{linpsi}
\end{equation}
for $\delta\psi$. We see that $\delta\psi$ decouples from the metric equations in
any radial gauge, and is independent of the choice of linearised
  radial gauge.

With the background solution $G_0=R_{0,x}$, and the boundary
condition $\delta G=\delta R_{,x}$ at the origin, which follows
from linearising the gauge condition $G=R_{,x}$ that $u$ is proper
time at the origin, the solution of
(\ref{linGeqn}) is
\begin{equation}
\label{deltaGeqnbis}
\delta G=\delta R_{,x}
\end{equation}
everywhere. This is far as we can go without choosing a (linearised)
radial gauge. 


\subsection{Linearised field equations in linearised Bondi gauge}


The perturbation equations take their simplest form in Bondi
gauge. This is defined by $R=x=:r$ in the full equations, and hence by
$R_0=x$, $G_0=H_0=1$ in the background, and $\delta R=\delta G=0$ for
the perturbations. 

However, the choices of background gauge and linear perturbation gauge
are in principle independent.  In particular, we can choose to use sM
gauge in the background, but linearised Bondi gauge $\delta R=\delta
G=0$ for the perturbations.  Under a change of background gauge, for
example from Bondi to sM, the linear perturbations $\delta b$, $\delta
f$ and $\delta\psi$ change only their argument, as if they were
scalars.

In linearised Bondi gauge, defined by $\delta R=\delta G=0$, $\delta
f$ and $\delta b$ obey the coupled equations
\begin{eqnarray}
\label{linbBondi}
\left({R_0^4 b_{,x}\over G_0}\right)_{,x}&=& 
R_0^2
\left(2S \delta f_{,xy}-8y\delta f_{,x}\right), \\
\left(R_0\Xi_0 \delta f\right)_{,x}&=& 
{R_0\over 4}\delta b_{,xy}+{R_{0,x}\over 2}\delta b_{,y}
-(\Xi_0R_0)\delta f_{,x}. \nonumber \\
\label{linfBondi}
\end{eqnarray}
These are just the first lines of (\ref{linb}), (\ref{linf}) above.
The linearised wave equation (\ref{linpsi}) does not simplify further.
Using $G_0=R_{0,x}$, as well as $\delta G=\delta R=0$, the linearised
hierarchy equation (\ref{linR}) becomes
\begin{eqnarray}
\label{linHBondi}
\left(-{R_0\over 2}\delta H\right)_{,x}&=&{R_0^2\over 4}
\left(S\delta b_{,xy}-2y\delta b_{,x}\right) \nonumber \\
&&+R_{0,x}\Bigl[(6S-4)\delta f
-2yR_0\delta b \nonumber \\
&&+{S\over 2}\left(2R_0\delta b_{,y}+8y\delta f_{,y}
-S\delta f_{,yy}\right)\Bigr]. \nonumber \\
\end{eqnarray}
Given a solution $(\delta f,\delta b)$ of (\ref{linbBondi},\ref{linfBondi}),
(\ref{linHBondi}) can be solved for $\delta H$ by
integration, but as $\delta H$ does not couple back
into the equations for $\delta b$ and $\delta f$, we can ignore it. 

The perturbation equations in linearised Bondi gauge
(\ref{linfBondi},\ref{linbBondi}) are given in Bondi background
coordinates as Eqs.~(\ref{deltabBondi}), (\ref{deltafBondi}) below.


\subsection{Other linearised radial gauges}
\label{appendix:otherlineargauges}


Consider now the equations linearised about Minkowski in any linear
gauge obtained by linearising a nonlinear gauge. The different
spherical harmonics $l$ still evolve independently.

If $R_{,y}=0$ in the nonlinear gauge, such as in gsB and lsB gauge,
the $l\ne 0$ components of $\delta R$ are absent, and hence, from
(\ref{deltaGeqnbis}), also the $l\ne 0$ components of $\delta
G$. (\ref{linf},\ref{linb}) then still reduce to
(\ref{linfBondi},\ref{linbBondi}), and exact solutions derived in
linear Bondi gauge are still relevant, after changing only their
argument.

By contrast, in any gauge where $R_{,y}\ne 0$, such as sdn gauge,
$\delta R$ and $\delta G$ couple back to $\delta f$ and $\delta b$.
However, $\delta f$ transforms as a scalar under changes of radial gauge,
even nonlinearly, and because $f$ is already a perturbation the change
due to the change of argument is quadratically small. By contrast,
$\delta b$ changes already to linear order.  $\delta R$ does transform
as a scalar, but $R_{,x}\ne 0$ in the background solution, so the
change of argument changes $\delta R$ to linear order. In summary, we
can still use the exact solution for $\delta f$ as a testbed, but
not $\delta b$.


\subsection{Solution of the scalar wave equation in Bondi gauge}


We now find explicit solutions of the scalar wave equation for
$\delta\psi$, and of the coupled equations for $\delta f$ and $\delta
b$ in linear Bondi gauge, using separation of variables. We use Bondi
gauge $R=x$ in the background, and to indicate this we write $r$ for
$x$. At the end we translate the results back into sM background
gauge. We begin in this subsection with the wave equation.

The linear wave equation on flat spacetime in Bondi coordinates is
\begin{equation}
-2\delta\psi_{,ur}+\delta\psi_{,rr}+{2\over r}(\delta\psi_{,r}-\delta\psi_{,u})
+{1\over r^2}\left[(1-y^2)\delta\psi_{,y}\right]_{,y}=0.
\end{equation}
Replacing the retarded time coordinate $u$ with the Minkowski time coordinate
\begin{equation}
t:=u+r,
\end{equation}
and writing
\begin{equation}
\label{psiphi}
\delta\psi(u,r,y)=:{\delta\phi}(t,r,y)={\delta\phi}(u+r,r,y),
\end{equation}
this takes the more familiar form 
\begin{equation}
-{\delta\phi}_{,tt}+{\delta\phi}_{,rr}+{2\over
  r}{\delta\phi}_{,r}+{1\over r^2}\left[(1-y^2){\delta\phi}_{,y}\right]_{,y}=0.
\end{equation}
We initially work in coordinates $(t,r,y)$ and switch back to
$(u,r,y)$ later.


\subsubsection{Separating off the $y$-dependence}


Spherical symmetry of the background allows us to separate the angle
$y$ with the ansatz
\begin{equation}
{\delta\phi}(t,r,y)=:\sum_{l=0}^\infty{\delta\phi}_l(t,r)P_l(y),
\end{equation}
where $P_l(y)$ is the Legendre polynomial of order $l$, obeying
\begin{equation}
\label{L2Y}
L^2P_l:=(1-y^2)P_l''-2yP_l'=-l(l+1)P_l.
\end{equation}
We then have the one-dimensional wave equation
\begin{equation}
\label{phileqn}
-{\delta\phi}_{l,tt}+{\delta\phi}_{l,rr}+{2\over r}{\delta\phi}_{l,r}-{l(l+1)\over r^2}{\delta\phi}_l=0,
\end{equation}
for the $l$-th partial wave, or equivalently
\begin{equation}
\label{psileqn}
-2\delta\psi_{l,ur}+\delta\psi_{l,rr}+{2\over r}(\delta\psi_{l,r}-\delta\psi_{l,u})
-{l(l+1)\over r^2}\delta\psi_l=0, 
\end{equation}
where
\begin{equation}
\label{psiurysum}
\delta\psi(u,r,y)=:\sum_{l=0}^\infty\delta\psi_l(u,r)P_l(y),
\end{equation}


\subsubsection{General solution of $l$-th partial wave equation}
\label{appendix:pertscalar}


We can find the general solution of (\ref{phileqn}) that is regular at
$r=0$ in terms of a single free function $\chi_l$ of
one variable, in the form of a generalised d'Alembert solution
\cite{Price}, as
\begin{eqnarray}
\label{philexplicit}
{\delta\phi}_l(t,r)&=&\sum_{p=0}^l A_l^p r^{-p-1}
\Bigl[\chi_l^{(l-p)}(t-r) \nonumber \\ 
&&\ \ \ \ -(-1)^{l-p}\chi_l^{(l-p)}(t+r)\Bigr],
\end{eqnarray}
where 
\begin{equation}
A_l^p:={(l+p)!\over 2^p\,p!\,(l-p)!},
\end{equation}
and $\chi^{(n)}$ denotes the $n$-th derivative of $\chi$.

Expanding $\chi(t\pm r)$ into its Taylor series about $r=0$, we obtain the
double sum
\begin{equation}
{\delta\phi}_l(t,r)=2\sum_{p=0}^l\sum_{\substack{q=0 \\ l-p+q \text{ odd}}}^\infty
{(-1)^q A_l^p\over q!} r^{q-p-1} \chi_l^{(l-p+q)}(t).
\end{equation}
Noting that $l-p+q\ge 0$ and odd in all terms, we parameterise $q$ as
$q=2k+1-l+p$. We then have
\begin{eqnarray}
\label{philinpowersofr}
{\delta\phi}_l(t,r)&=&2\sum_{k=0}^\infty\left(\sum_{\substack{p={\rm max}\\(0,l-2k-1)}}^{l}
{(-1)^{l-p}A_l^p\over (2k+1-l+p)!}\right)
\nonumber \\ && r^{2k-l} \chi_l^{(k)}(t) .
\end{eqnarray}
The inner sum (in round brackets) vanishes for $k<l$, and so the outer
sum starts up only at $k=l$. Hence, the expression
(\ref{philexplicit}) admits a formal expansion in even(odd) powers of
$r$ at constant $t$, for $l$ even(odd), starting at $r^l$. In
particular, $\delta\psi_l(t,-r)=(-1)^l\delta\psi_l(t,r)$, and
$\delta\psi_l\sim r^l$ at the origin.

Switching back from $(t,r,y)$ to $(u,r,y)$, we trivially find
(\ref{psiurysum}) with
\begin{eqnarray}
\label{psilexplicit}
\delta\psi_l(t,r)&=&\sum_{p=0}^l A_l^p r^{-p-1}
\Bigl[\chi_l^{(l-p)}(u) \nonumber \\ 
&&\ \ \ \ -(-1)^{l-p}\chi_l^{(l-p)}(u+2r)\Bigr],
\end{eqnarray}
Similarly, by setting $t=u+r$, in (\ref{philinpowersofr}), we
trivially obtain
\begin{eqnarray}
\label{psilinpowersofr}
\delta\psi_l(t,r)&=&2\sum_{k=0}^\infty\left(\sum_{\substack{p={\rm max}\\(0,l-2k-1)}}^{l}
{(-1)^{l-p}A_l^p\over (2k+1-l+p)!}\right)
\nonumber \\ && r^{2k-l} \chi_l^{(k)}(u+r) .
\end{eqnarray}


\subsection{Relating the perturbed Einstein equations in Bondi gauge
  to the scalar wave equation, in spherical harmonics}
\label{appendix:gwtoscalar}


In the vacuum Einstein equations in Bondi gauge, linearised about flat
spacetime, the coupled equations for $\delta b$ and $\delta f$
(\ref{linfBondi},\ref{linbBondi}) become
\begin{eqnarray}
\label{deltabBondi}
r^2\delta b_{,rr}+4r \delta b_{,r}-2(1-y^2)\delta f_{,r
  y}+8y\delta f_{,r}&=&0, \nonumber \\ \\
r(4\delta f_{,ur}-2\delta f_{,rr})+4(\delta f_{,u}-{\delta
  f}_{,r}) && \nonumber \\
\label{deltafBondi}
-r \delta b_{,r y}-2\delta b_{,y}&=&0, \nonumber \\
\end{eqnarray}
where we have used $x=R_0=:r$ and $\Xi_0 R_0=-1/2$. These equations
are equivalent to Eqs.~(6,7) of
\cite{GomezPapadopoulosWinicour1994}. In that paper solutions of these
coupled equations were related to solutions of the scalar wave
equation. Here we give a self-contained derivation of that relation.


\subsubsection{Separating off the $y$-dependence}


Spherical symmetry of the background means we can make the separation
of variables ansatz
\begin{eqnarray}
\label{bsep}
\delta b&=&\delta b_l(u,r)P^{(1)}_l(y), \\
\label{fsep}
\delta f&=&\delta f_l(u,r)P^{(2)}_l(y), 
\end{eqnarray}
with the $P^{(1,2)}_l(y)$ to be found below.
We obtain the four separated equations
\begin{eqnarray}
\label{Pp2Q}
{P_l^{(1)}}'&=&P_l^{(2)},\\
\label{Qp2P}
(1-y^2){P_l^{(2)}}'-4y{P_l^{(2)}}&=&\lambda P_l^{(1)}, \\
\label{AB1}
(r^4\delta b_{l,r})_{,r}&=&2\lambda r^2\delta f_{l,r}, \\
\label{AB2}
2r(r \delta f_{l,u})_{,r}-(r^2\delta f_{l,r})_{,r}&=&
  {1\over 2}(r^2\delta b_l)_{,r},
\end{eqnarray}
where, without loss of generality, we have fixed a separation
constant in (\ref{Pp2Q}) to one. Clearly (\ref{Pp2Q}) and (\ref{Qp2P})
together give
\begin{equation}
\label{Peqn}
(1-y^2){P_l^{(1)}}''-4y{P_l^{(1)}}'=\lambda P_l^{(1)}.
\end{equation}


\subsubsection{Tensor spherical harmonics}


We would like to relate the unfamiliar spherical harmonic-like
functions $P^{(1)}_l$ and $P^{(2)}_l$ to the scalar spherical harmonics
$P_l$. To motivate this, we note, following \cite{GerlachSengupta},
that a geometrically natural ansatz for a perturbation $\delta
g_{ij}$ of the round unit metric $g_{ij}^0$ on $S^2$ with covariant
derivative $\bar\nabla_i$ is
\begin{equation}
\label{GS1}
\delta g_{ij}=g_1Y_{lm}\,g_{ij}^0+g_2 \bar\nabla_i\bar\nabla_jY_{lm}
\end{equation}
where we have gone beyond axisymmetry but have restricted to polar
perturbations, and where $Y_{lm}$ is a scalar spherical
harmonic. Similarly, the perturbation of any vector on $S^2$ such as
the shift must take the form
\begin{equation}
\label{GS2}
\delta \beta^i=g_3\,g^{ij}_0\bar\nabla_jY_{lm}.
\end{equation}
Here $g_{1,2,3}$ are functions of $u$ and $r$.  In axisymmetry, the
spherical harmonic $Y_{lm}$ reduces to the Legendre polynomial
$Y_{l0}=P_l$. Comparing (\ref{GS1},\ref{GS2}) with our metric ansatz
(\ref{ymetric}) and separation of variables ansatz
(\ref{bsep},\ref{fsep}) suggests
\begin{equation}
\label{PQY}
P^{(1)}_l= P_l', \quad P^{(2)}_l=P_l'',
\end{equation}
consistent with (\ref{Pp2Q}). Substituting (\ref{PQY}) into
(\ref{Peqn}), and using (\ref{L2Y}) and its $y$-derivative, we find
that (\ref{Peqn}) is indeed obeyed with the separation constant given
by
\begin{equation}
\label{lambdaval}
\lambda=-l(l+1)+2 = -(l+2)(l-1).
\end{equation}
We note in passing that $P_l^{(2)}$ obeys the ODE
\begin{equation}
\label{Qeqn}
 (1-y^2){P_l^{(2)}}''-6{P_l^{(2)}}'=\tilde \lambda P_l^{(2)},
\end{equation}
where
\begin{equation}
\tilde\lambda:=-l(l+1)+6=-(l+3)(l-2).
\end{equation}
We are now left with the two coupled PDEs (\ref{AB1},\ref{AB2}) for
$\delta b_l$ and $\delta f_l$, with $\lambda$ given by (\ref{lambdaval}).

\subsubsection{Potential ansatz}


We can solve (\ref{AB1}) identically by introducing the potential
$\Psi_l$, in terms of which
\begin{eqnarray}
\label{bpsi}
\delta b_l(u,r)&=& 2\lambda\int_{r_l(u)}^r {\Psi_l(u,\bar r)\over
  \bar r^4}d\bar r, \\
\label{fpsi}
\delta f_l(u,r)&=&\int_{r_l(u)}^r{\Psi_{l,r}(u,\bar r)\over
  \bar r^2}d\bar r.
\end{eqnarray}
$r_l(u)$ is arbitrary, but will later be set to zero.
Substituting this into (\ref{AB2}), dividing by $r$ and
differentiating with respect to $r$ in order to eliminate the
integrals, we obtain the third-order PDE $E_3=0$, where
\begin{equation}
\label{E3PDE}
E_3:=-2r^3\Psi_{l,urr}+r^3\Psi_{l,rrr}-r^2\Psi_{l,rr}+\lambda(r\Psi_{l,r}-\Psi).
\end{equation}
We can write
\begin{equation}
E_3=r^2\left({E_2\over r}\right)_{,r},
\end{equation}
where we have defined
\begin{equation}
E_2:=-2r^2\Psi_{l,ur}+r^2\Psi_{l,rr}+2r\Psi_{l,u}-2r\Psi_{l,r}+\lambda\Psi_l.
\end{equation}
Hence the general solution of $E_3=0$ is 
\begin{equation}
E_2=Cr,
\end{equation}
where $C$ is an arbitrary constant, and the general solution of this
is in turn
\begin{equation}
\Psi_l^{(3)}(u,r)=\Psi_l^{(2)}(u,r)+{Cr \over l(l+1)},
\end{equation}
where $\Psi_l^{(2)}$ is the general solution of the second-order ODE
$E_2=0$. With the substitution
\begin{equation}
\label{Psipsi}
\Psi_l=:r^2\delta\psi_l,
\end{equation}
$E_2=0$ for $\Psi_l$ becomes the scalar wave equation (\ref{psileqn})
for $\delta\psi_l$.

Substituting (\ref{Psipsi}) into (\ref{bpsi},\ref{fpsi}), we have
\begin{eqnarray}
\label{bl}
\delta b_l(u,r)&=&2\lambda\int_{r_l(u)}^r {\delta\psi_l(u,\bar r)\over
  \bar r^2}d\bar r, \\ 
\label{fl}
\delta f_l(u,r)&=&\delta\psi_l(u,r)-\delta\psi_l(u,r_l(u))
\nonumber \\ &&
+2\int_{r_l(u)}^r{\delta\psi_l(u,\bar r)\over \bar r}
\,d\bar r. 
\end{eqnarray}

Note that, in addition to an arbitrary solution $\delta\psi_l$ of the scalar
wave equation, the gravitational wave $(\delta f_l,\delta b_l)$ is parameterised
also by the constant $C$ and the function $r_l(u)$. The particular
integral parameterised by the constant $C$ is both non-dynamical and
gives rise to $\delta b$ and $\delta f$ that are singular at $r=0$. We therefore set
$C=0$, and so obtain $E_2=0$ as the master equation governing
(axisymmetric, polar) linear gravitational waves. We believe that
$r_l(u)$ parameterises some kind of gauge freedom, but have not tried
to show this. We now set it to zero. 

Using the fact that $\delta\psi_l\sim r^l$ at constant $t$, we see from
(\ref{bl},\ref{fl}) that near the centre, each spherical harmonic
component of the linear   gravitational wave has the spatial dependence
$\delta b\sim P_l'r^{l-1}$ and $\delta f\sim P_l'' r^l$, for $l\ge 2$. In
particular, near the origin $r=0$ we therefore have $\delta b\sim ry$ and
$\delta f\sim r^2$ for generic regular solutions with non-vanishing $l=2$
components.

Substituting the general regular solution (\ref{psilexplicit}) for
$\delta\psi_l(u,r)$ with a sufficiently simple function $\chi_l(u)$ into
(\ref{bl},\ref{fl}), and with $r_l(u)=0$, we can carry out the
integrations and so obtain $\delta b_l(u,r)$ and $\delta f_l(u,r)$ in closed
form. In particular, this works when $\chi_l(u)$ is a Gaussian, and
gives us a class of exact solutions for testing our code.

The general solution of (\ref{deltabBondi},\ref{deltafBondi}) is
obtained by summing
\begin{eqnarray}
\label{deltafsum}
\delta f(u,r,y)&=&\sum_{l=2}^\infty \delta f_l(u,r)P_l''(y), \\
\label{deltabsum}
\delta b(u,r,y)&=&\sum_{l=2}^\infty \delta b_l(u,r)P_l'(y).
\end{eqnarray}
If we now substitute (\ref{bl}) and (\ref{fl}), and then
(\ref{psilinpowersofr}), we see that at constant $u$, $\delta f_l$ admits a
formal expansion in even (odd) powers of $r$ when $l$ is is even
(odd), starting at $r^l$. This is just as for $\delta\psi_l$. Similarly,
$\delta b_l$ admits a formal expansion in odd (even) powers of $r$ when $l$
is is even (odd), starting at $r^{l-1}$.

Obviously, the scalar field $\delta\psi$ governing the linearised
gravitational waves is completely independent from the matter scalar
field $\delta\psi$. We have used the same notation merely to stress that
they obey identical wave equations. 


\subsubsection{The cases $l=0$ and $l=1$}


Our separation of variables ansatz gives $b_0=f_0=f_1=0$ because
$P_0'=P_0''=P_1''=0$. In addition the potential ansatz (\ref{bl})
gives $\delta b_1=0$ because $\lambda=0$. However, as $P_1=y$ and
$P_1'=1$, there can be nonvanishing perturbations $\delta b_1$, $\delta
H_1$, even though there is no $f_1$.

We therefore look at the case $l=1$ of the linearised equations in
Bondi gauge without making the potential ansatz
(\ref{bl},\ref{fl}). With $\delta b=\delta b_1(u,r)$ and $\delta f=0$,
(\ref{deltabBondi}) reduces to
\begin{equation}
\label{b1ODE}
(r^4\delta b_{1,r})_{,r}=0,
\end{equation}
while (\ref{deltafBondi}) is obeyed trivially. The solution of
(\ref{b1ODE}) that is finite at the centre is constant in $r$. 
The resulting regular perturbation of $H$ is given by
(\ref{linHBondi}). Putting both together, we have 
\begin{equation}
\label{leq1perturbation}
\delta b_1(u,r)=\delta b_1(u), \qquad \delta H_1=2r\delta b_1(u).
\end{equation}
We show in Appendix~\ref{appendix:residualgaugefreedom}
that this represents a linear gauge transformation, which
physically corresponds to an acceleration $-\delta b_1(u)$ of the
origin of our coordinate system along the symmetry axis.


\subsubsection{Transformation to shifted Minkowski background gauge}


Having found $\delta b(u,r,y)$, $\delta f(u,r,y)$ and
$\delta\psi(u,r,y)$ in Bondi background gauge, we transform them to
shifted Minkowski background gauge, with the coordinate change from
$(u,r,y)$ to $(u,x,y)$ given by
\begin{equation}
\label{Minkowskixtor}
r={x\over 2}\left( 1-{u\over x_0}\right),
\end{equation}
compare (\ref{R0exact}). Under this coordinate transformation, only
$\delta H$ changes nontrivially. The metric coefficients $\delta b$,
$\delta f$ and $R$ and the scalar field $\delta\psi$ change only as if
they were scalar fields, and $\delta G$ remains zero. We are only
interested in $\delta b$, $\delta f$ and $\delta\psi$, and we only
need to transform their argument $r$ to $x$ using
(\ref{Minkowskixtor}).


\subsubsection{Consistent boundary conditions 
  for the linearised Einstein equations decomposed into spherical
  harmonics}


We now return to the case of a non-trivial $r_l(u)>0$.
From (\ref{bl}) and (\ref{fl}) we read off that 
\begin{eqnarray}
\delta b_l(u,r_l(u))&=&0, \\ 
\delta f_l(u,r_l(u))&=&0, \\ 
\delta b_{l,r}(u,r_l(u))&=&{2\lambda\over r_l(u)^2}\delta\psi_l(u,r_l(u)), \\  
\delta f_{l,r}(u,r_l(u))&=&\delta\psi_{l,r}(u,r_l(u))+{2\over
  r_l(u)}\delta\psi_l(u,r_l(u)).  \nonumber \\
\end{eqnarray}
The first three equations gives us a class of startup conditions for
integrating the linearised hierarchy equations for $\delta b_l$ and
$\delta f_l$ that have a consistent continuum limit, in terms of a
boundary condition at $r=r_l(u)$ for $\delta b_{l,r}$, which is
equivalent to a boundary condition on the underlying scalar wave
$\delta\psi_l$.

The simplest choice is to set $\delta b_{l,r}=0$ at $r=r_l(u)$. As
this corresponds to the homogeneous Dirichlet boundary condition
$\delta\psi_l=0$ at $r=r_l(u)$ on the underlying scalar field, this
boundary condition gives rise to a well-posed PDE problem for the
spherical harmonic component $\psi_l(u,r)$. We have therefore shown
that, for the equations linearised about flat spacetime in linear
Bondi gauge, it is consistent to impose the three boundary conditions
$\delta f_l=\delta b_l=\delta b_{l,r}=0$ at $r=r_l(u)$. 

This is of course what we do, as a numerical trick, for the full
nonlinear Einstein equations, with $r_l\sim l\Delta x$. What we have
shown here is that this has a continuum limit for the Einstein
equations linearised about Minkowski and split into spherical harmonics.


\subsection{Solution without spherical harmonic decomposition}
\label{appendix:linpertsnodecomposition}



\subsubsection{Relation between solutions for $\delta\psi$ and for $(\delta f,\delta b)$}


We can remove the spherical harmonic decomposition, and obtain
expressions for $\delta b(u,r,y)$ and $\delta f(u,r,y)$ in terms of a
solution $\delta\psi(u,r,y)$ of the scalar wave equation.  Combining
(\ref{psiurysum}), (\ref{fl}) and (\ref{deltafsum}), we obtain the
unseparated expression
\begin{eqnarray}
\label{fpsiunseparated}
\delta f(u,r,y)&=&\delta\psi_{,yy}(u,r,y)-\delta\psi_{,yy}(u,0,y) \nonumber \\ &&
+2\int_0^r{\delta\psi_{,yy}(u,\bar r,y)\over \bar r}\,d\bar r
\end{eqnarray}
for $\delta f$ in terms of $\delta\psi$.  Substituting
(\ref{deltabsum}) into (\ref{deltabBondi}) and integrating twice, we
obtain the expression
\begin{eqnarray}
\label{bpsiunseparated}
\delta b(u,r,y)&=&\int_0^r{1\over \bar r^4}\int_0^{\bar r} \tilde r^2\Bigl(
2(1-y^2)\delta f_{,ry}(u,\tilde r,y)\nonumber \\ &&-8y\delta f_{,r}(u,\tilde r,y)\Bigr)\,d\tilde
r
\end{eqnarray}
for $\delta b$ in terms of $\delta f$. 


\subsubsection{Plane-wave solutions}
\label{appendix:planewave}


The scalar wave equation admits the plane-wave solution
$\delta\psi=\chi(t\pm z)$, that is 
\begin{equation}
\label{psiplanewave}
\delta\psi(u,r,y)=\chi(u_\pm), \qquad u_\pm:=u+r(1\pm y),
\end{equation}
for arbitrary functions $\chi$. Substituting this into
(\ref{fpsiunseparated}), we can carry out the integration explicitly
to obtain the plane wave solution 
\begin{equation}
\label{fplanewave}
\delta f(u,r,y)=r^2\chi''(u_\pm)+2{r\chi'(u_\pm)\over 1\pm y}
-2{\chi(u_\pm)-\chi(u)\over (1\pm y)^2},
\end{equation}
and substituting this into (\ref{bpsiunseparated}), we can again carry
out the integrations in closed form to obtain 
\begin{eqnarray}
\delta b(u,r,y)&=&\pm 2\Bigl[r(1\mp y)\chi''(u_\pm) \nonumber \\
&& -{1\pm 3y\over
  1\pm y}\left(\chi'(u_\pm)-\chi'(u)\right)\Bigr].
\label{bplanewave}
\end{eqnarray}
Note that $\psi(u,r,-y)=\psi(u,r,y)$, $f(u,r,-y)=f(u,r,y)$ and $b(u,r,-y)=-b(u,r,y)$.

To see that these expressions are regular as $y\to \mp 1$, we define
the auxiliary variable $\epsilon:=r(1\pm y)$. We can then write
\begin{eqnarray}
\delta\psi(u,r,y)&=&\chi(u+\epsilon), \\
\delta f(u,r,y)&=&r^2\chi''(u+\epsilon)+2r^2{d\over
  d\epsilon}\left({\chi(u+\epsilon)-\chi(u)\over\epsilon}\right), \nonumber \\ 
\delta b(u,r,y)&=&\pm 2r\Bigl[(1\mp y)\chi''(u+\epsilon) \nonumber \\
&&-(1\pm 3y)\left({\chi(u+\epsilon)-\chi(u)\over\epsilon}\right)\Bigr],
\end{eqnarray}
where we can now see that the fraction in large round brackets is
regular as $\epsilon\to 0$ if $\chi(u)$ is regular.


\section{Residual gauge freedom}
\label{appendix:residualgaugefreedom}


A redefinition of the coordinates $(u,x,y)$ leaves the form
(\ref{mymetric}) of the metric invariant if $g_{xx}=g_{xy}=0$ in the
new, as in the old, coordinates. A nonlinear gauge transformation that
does this and that can be written explicitly in terms of free
functions is
\begin{eqnarray}
\label{residualgaugeu}
u&=&u(\tilde u), \\
\label{residualgaugey}
y&=&y(\tilde u,\tilde y), \\
\label{residualgaugex}
x&=&x(\tilde u,\tilde x,\tilde y).
\end{eqnarray}
This means we separately relabel the outgoing null cones ${\cal
  N}^+_u$, the outgoing null rays ${\cal L}^+_{u,y,\varphi}$ on each
outgoing cone, and the coordinate $x$ along each outgoing
ray. However, we cannot express the most general gauge freedom, which
also moves the central worldline, in explicit form.

Instead, we look for the most general {\em infinitesimal} gauge
transformation that leaves the form of the metric invariant.  We make
the ansatz $\delta x^\mu=:\xi^\mu$, and hence
\begin{equation}
\label{deltagansatz}
\delta g_{\mu\nu}=2\nabla_{(\mu}\xi_{\nu)},
\end{equation}
where
\begin{equation}
\xi^\mu:=\left(\xi^u,\xi^x,\xi^y,0\right),
\end{equation}
are functions of $(u,x,y)$. The general solution of $\delta
g_{xx}=\delta g_{xy}=0$ is
\begin{eqnarray}
\label{xiu}
\xi^u&=&A(u,y), \\
\label{xix}
\xi^x&=&\xi^x(u,x,y), \\
\label{xiy}
\xi^y&=&B(u,y)+
A_{,y}(u,y)\,S\int_0^x {G\over e^{2S f}R^2}\,dx'.
\end{eqnarray}
For $A=A(u)$ this reduces to the infinitesimal version of
(\ref{residualgaugeu})-(\ref{residualgaugex}). 

For simplicity, we now restrict the background metric to flat
spacetime in Bondi gauge, with $R=x$, $G=H=1$ and $b=f=0$. (\ref{xiy})
then simplifies to 
\begin{equation}
\xi^y=B(u,y)-{SA_{,y}(u,y)\over x}. 
\end{equation}
The resulting metric perturbation is 
\begin{eqnarray}
\delta G&=&A_{,u}+\xi^x_{,x},\\ 
\delta H&=&2(A_{,u}+\xi^x_{,u}), \\ 
\delta R&=&\xi^x+{xB_{,y}\over 2}-{SA_{,yy}-2yA_{,y}\over 2}, \\
\delta f&=&-{A_{,yy}\over 2x}+{B_{,y}+2yB\over 2S}, \\ 
\delta b&=&-{A_{,y}\over x^2}+{B_{,u}\over S}-{xA_{,uy}+\xi^x_{,y}\over x^2}.
\end{eqnarray}

For analysis, we relate the coordinates $(u,x,y)$ to the
cylindrical Minkowski coordinates defined by
\begin{equation}
t:=u+x, \qquad z:=-yx, \qquad \rho:=\sqrt{S}x,
\end{equation}
with $\varphi$ completing both coordinate systems. The inverse
transformation is
\begin{eqnarray}
u&=&t-\sqrt{\rho^2+z^2}, \\
x&=&\sqrt{\rho^2+z^2},\\
y&=&-{z\over \sqrt{\rho^2+z^2}},
\end{eqnarray}
and hence the coordinate basis vectors transform as
\begin{eqnarray}
\label{partialt}
\partial_t&=&\partial_u, \\
\label{partialz}
\partial_z&=&y\partial_u-y\partial_x-{S\over x}\partial_y, \\
\label{partialrho}
\partial_\rho&=&-\sqrt{S}\partial_u+\sqrt{S}\partial_x-{y\sqrt{S}\over x}\partial_y,
\end{eqnarray}
with $\partial_\varphi$ in both coordinate systems. 

From (\ref{xiu}-\ref{xix}) and (\ref{partialt}-\ref{partialrho}), the
residual gauge transformations expressed in this basis are
\begin{eqnarray}
\label{generalxi}
\xi&=&(A+\xi^x)\,\partial_t+\left(\sqrt{S}(\xi^x+yA_{,y})
-{xyB\over\sqrt{S}}\right)\partial_\rho \nonumber \\
&&+(SA_{,y}-xB-y\xi^x)\,\partial_z. 
\end{eqnarray}
$\xi$ is a regular vector field if and only if the coefficients of
$\partial_t,\partial_z,\partial_\rho$ are regular functions of
$(t,z,\rho)$. In particular, they must be finite and single-valued at
the origin $x=0$ and on the axis $S=0$.

We now focus on gauge transformations that map the $tz$-plane to
itself, as these include all that move the central worldline
along the symmetry axis. Setting the component
$\xi^\rho$ to zero is equivalent to
\begin{equation}
\xi^x={xyB\over S}-yA_{,y},
\end{equation}
and substituting this back into (\ref{generalxi}), we have 
\begin{equation}
\label{xiinztplane}
\xi=\left(A-yA_{,y}+{xyB\over S}\right)\partial_t
+\left(A_{,y}-{xB\over S}\right)\partial_z.
\end{equation}
The resulting metric perturbations are regular at the origin and on
the axis if and only if
\begin{equation}
A=T(u)+Z(u)\,y, \qquad B=-SZ'(u),
\end{equation}
where $T$ and $Z$ are arbitrary functions. The corresponding regular
gauge vector field is
\begin{equation}
\xi=\left[T(u)-z\,Z'(u)\right]\,\partial_t
+\left[Z(u)+x\,Z'(u)\right]\,\partial_z,
\end{equation}
which explains why we have named the free coefficients $T$ and
$Z$. The corresponding pure gauge metric perturbations are
\begin{eqnarray}
\delta H&=&2(T'(u)+zZ''(u)), \\
\delta G&=&T'(u), \\
\delta b&=&-Z''(u), 
\end{eqnarray}
with $\delta R=\delta f=0$. This family includes the three
Killing vector fields $\partial_t$, $\partial_z$ and
$z\,\partial_t+t\,\partial_z$ of Minkowski spacetime.

Restricting to $T(u)=0$ leaves us with
\begin{equation} 
\xi=Z'(u)\,z\,\partial_t+[Z(u)+x\,Z'(u)]\,\partial_z.
\end{equation}
This is equal to $Z(u)\partial_z$ at the origin, and so is the gauge
transformation that moves the origin along the symmetry axis. The
resulting metric perturbation is
\begin{equation}
\delta H=-2Z''(u)\,x\,y, \qquad \delta b=-Z''(u), 
\end{equation}
with the other metric coefficients unchanged. Hence an infinitesimal
non-inertial motion of the origin along the symmetry axis gives
exactly the $l=1$ regular metric perturbation (\ref{leq1perturbation})
that we already derived from the linearised Einstein equations.

From elementary flatness at the origin, the same statement must be
true for an arbitrary regular, twistfree axisymmetric
spacetime. Hence we have shown that $b(u,{\bf 0})=0$ is precisely the
gauge condition that forces the origin of our coordinate system to
move on a geodesic along the $z$-axis. We are imposing this as a
boundary condition on (\ref{dbdxeqn}) or (\ref{blhs1eqn}). Conversely,
by imposing an arbitrary value of $b(u,{\bf 0})$, we could accelerate
the origin during the evolution.


\section{Regularity of the metric at the origin}
\label{appendix:regularity}


Following common practice, Rinne \cite{RinnePhD} defines a scalar
field as regular if it is analytic in Cartesian coordinates
$(t,\xi,\eta,z)$ (which in turn we assume to give rise to a chart
without coordinate singularities). Transforming to the cylindrical
coordinates $(t,\rho,\varphi,z)$ defined by
\begin{equation}
\xi:=\rho\cos\varphi, \qquad \eta:=\rho\sin\varphi,
\end{equation}
with $t$ and $z$ unchanged, Rinne shows that a scalar field is regular
and axisymmetric if and only if it is an analytic function of
$(t,\rho^2,z)$, that is, analytic in $(t,\rho,z)$, with only even
powers of $\rho$.

Rinne then shows that an axisymmetric metric is regular, in the sense
that its components in the Cartesian coordinate basis are analytic
functions of the Cartesian coordinates, if and only if in the
cylindrical coordinate basis it takes the form
\begin{equation}
\left(\begin{array}{cccc}
{\cal A} & {\cal B} & \rho{\cal D} & \rho^2{\cal F} \\
{\cal B} & {\cal C} & \rho{\cal E} &  \rho^2{\cal G}\\
\rho{\cal D} & \rho{\cal E} & {\cal H}+\rho^2{\cal J} &  \rho^3{\cal K}\\
\rho^2{\cal F} & \rho^2{\cal G} &  \rho^3{\cal K} & \rho^2\left({\cal H}-\rho^2{\cal J}\right) \\
\end{array}\right),
\end{equation} with the coordinates in order $(t,z,\rho,\varphi)$,
and the ``Rinne coefficients'' ${\cal A}$, ${\cal B}$, ${\cal C}$,
${\cal D}$, ${\cal E}$, ${\cal F}$, ${\cal G}$, ${\cal K}$, ${\cal
  H}$, ${\cal J}$ regular, that is analytic functions of
$(t,z,\rho^2)$. Clearly, the metric is twist-free if and only if ${\cal
  F}={\cal G}={\cal K}=0$, and we now restrict to this case.

In the following, we define a scalar field and metric as
``Rinne-regular'' if they obey these definitions. We shall now ask
what a Rinne-regular metric and scalar field look like, first in
spherical polar coordinates retaining the regular time slicing $t$,
and secondly replacing $t$ by a retarded time $u$. 

In the first step, we change from the cylindrical coordinates
$(t,\rho,z,\varphi)$ to the spherical coordinates $(t,x,y,\varphi)$
defined by
\begin{equation}
\rho:=x\sqrt{1-y^2}, \qquad z:=-xy,
\end{equation}
or equivalently
\begin{equation}
x=\sqrt{\rho^2+z^2}, \quad y=-z/\sqrt{\rho^2+z^2},
\end{equation}
with $t$ and $\varphi$ unchanged.  Note that $x$ and $y$ are our
radial and angular coordinates, not the Cartesian coordinates
$(\xi,\eta,z)$ defined above.

With $\rho^2=x^2-z^2$, a function is axisymmetric and regular if and
only if it is an analytic function of $(t,x^2-z^2,z)$, but this is the
case if and only if it is an analytic function of $(t,x^2,z)$ or,
expressed entirely in spherical coordinates, of $(t,x^2,-xy)$. In
particular, this means it is even in $x$ at constant $z$, not constant
$y$: this will be important below.

By transforming to the coordinate basis with respect to
$(t,x,y,\varphi)$, and reparameterising the Rinne coefficients
as
\begin{eqnarray}
\label{calBredef}
{\cal B}&=&z{\cal D}-{\cal X}_1, \\
{\cal E}&=&2z{\cal J}-{\cal X}_2-z{\cal X}_3,\\
\label{calCredef}
{\cal C}&=&{\cal H}+(2z^2-\rho^2){\cal J}-2z{\cal
  X}_2+(\rho^2-z^2){\cal X}_3,
\end{eqnarray}
we can show that the Rinne-regular twist-free axisymmetric metric is
spherically symmetric if and only if ${\cal X}_1={\cal X}_2={\cal
  X}_3=0$.

Moreover, because the coefficients of the redefinitions
(\ref{calBredef}-\ref{calCredef}) themselves are Rinne-regular
functions, and because we cannot absorb them into redefinitions of
${\cal X}_{1,2,3}$ without making those singular, the reparameterised
twist-free axisymmetric is Rinne-regular if {\em and only if} the
``modified Rinne coefficients'' ${\cal A}$, ${\cal D}$, ${\cal H}$,
${\cal J}$, ${\cal X}_{1,2,3}$ are Rinne-regular.

In the second step, we change from the spherical coordinates
$(t,x,y,\varphi)$ to the retarded coordinates $(u,x,y,\varphi)$,
where the retarded time
$u$ is defined as
\begin{equation}
u(t,x,y):=t-h(t,x,y).
\end{equation}
Without loss of generality, we set $h(t,0,y)=0$, and we split $h$
into its odd and even in $x$ (at constant $z$) parts as
\begin{equation}
h(t,x,y):=x\,h_o(t,x^2,-xy)+x^2 h_e(t,x^2,-xy).
\end{equation}
For $u$ to be a non-trivial retarded time, we assume that $h_o\ne
0$. Hence $h$ is not Rinne-regular even if $h_o$ and $h_e$ are. The
simplest choice would be $u=t-x$, the standard retarded null
coordinate on Minkowski spacetime. In this case, we can invert the
definition of $u$ to give $t=u+x$, and so a Rinne-regular axisymmetric
function would be an analytic function of $(u+x,x^2,-xy)$. 

We now impose that the lines of constant $(u,y,\varphi)$ are null,
which is equivalent to $g_{xx}=g_{xy}=0$. These are linear equations
relating the modified Rinne coefficients, but their coefficients are
neither even nor odd in $x$ (at constant $z$), so they have
Rinne-regular solutions only if we impose their even and odd parts
separately, giving us four equations. Hence we see that beyond
spherical symmetry we need the two generic Rinne-regular functions
$h_e$ and $h_o$ in order to be left with five regular functions to
match to our metric coefficients $(G,H,R,f,b)$.

Which four of ${\cal A}$, ${\cal D}$, ${\cal H}$, ${\cal J}$,
${\cal X}_{1,2,3}$, $h_e$ and $h_o$ should we solve for? In spherical
symmetry the simple height function $h=x$ is already sufficiently
general. This suggests that we should always solve for four of the
seven modified Rinne coefficients, rather than the height function,
and that this solution should be regular also in the special case
$h=x$. These requirements force us to solve for ${\cal D}$,
${\cal H}$, ${\cal X}_1$ and ${\cal X}_2$.

We can now express our five metric coefficients in terms of the five
remaining arbitrary regular functions ${\cal A}$, ${\cal J}$, ${\cal
  X}_3$, $h_e$ and $h_o$. To avoid square roots and logarithms in the
solution, we solve not for $R$ and $f$ but for $R^4$ and
$e^{4Sf}$. The resulting expressions are not immediately
helpful, as they are explicit functions of $(t,x,y)$, not $(u,x,y)$,
and we do not have an explicit expression for $t(u,x,y)$.

However, we can expand $(G,H,R,f,b)$ in powers of $x$ at
constant $(u,y)$ by using the derivative operator
\begin{equation}
\left.{\partial\over \partial x}\right|_{u,y}=
\left.{\partial\over \partial x}\right|_{t,y}
-{u_{,x}(t,x,y)\over u_{,t}(t,x,y)}\left.{\partial\over \partial t}\right|_{x,y}
\end{equation}
to find the coefficients of the Taylor series. We are then in effect
inverting the height function order by order. The Taylor coefficients are
given in terms of ${\cal A}$, ${\cal J}$, ${\cal X}_3$, $h_e$ and
$h_o$ and their partial derivatives with respect to $(t,x^2,z)$, 
evaluated at $x=z=0$, which are finite and independent of each other. 

We do not give the series expressions here, but it is clear from their
construction that $G$, $H$, $R/x$, $f/x^2$ and $b$ are analytic
functions of $(x^2,-xy)$, at constant $u$, with the coefficients of
the Taylor series given explicitly in terms of the coefficients of the
Taylor series in $(x^2,-xy)$, at constant $t=u$, of ${\cal A}$,
${\cal J}$, ${\cal X}_3$, $h_e$. This confirms the ansatz that we
found to be consistent for expanding the field equations in powers of
$x$ at constant $u$ in Sec.~\ref{section:originexpansion}.



\end{document}